\documentclass[10pt,journal,compsoc]{IEEEtran}

\usepackage[nocompress]{cite}
\usepackage[pdftex]{graphicx}
\graphicspath{{./Figures/}}
\DeclareGraphicsExtensions{.pdf,.jpeg,.png}

\usepackage{amsmath,amssymb}
\usepackage{mathtools}
\usepackage{dsfont}

\usepackage{graphicx}
\graphicspath{Figures/}

\usepackage{tabularray}
\UseTblrLibrary{booktabs}

\usepackage{subcaption}
\usepackage{caption}
\usepackage{enumitem}

\newcommand\MYhyperrefoptions{bookmarks=true,bookmarksnumbered=true,
pdfpagemode={UseOutlines},plainpages=false,pdfpagelabels=true,
colorlinks=true,citecolor={black},
urlcolor=magenta,
linkcolor=blue,
filecolor=cyan,
pdftitle={Medical Image Segmentation Review: The success of U-Net},%
pdfsubject={Computer Vision, Medical Image Segmentation},%
pdfauthor={R. Azad, E. Khodapanah Aghdam, A. Rauland, Y. Jia, A. Haddadi Avval, A. Bozorgpour, S. Karimijafarbigloo, J.P. Cohen, E. Adeli, D. Merhof},%
pdfkeywords={Medical Image Segmentation, Deep Learning, U-Net, Convolutional Neural Network, Transformer}}%
\usepackage[\MYhyperrefoptions,pdftex]{hyperref}

\usepackage[nameinlink]{cleveref}

\usepackage{orcidlink}

\begin{document}
\bstctlcite{IEEEexample:BSTcontrol}

\title{Medical Image Segmentation Review: \\ The Success of U-Net}

\author{Reza~Azad\,\textsuperscript{\orcidlink{0000-0002-4772-2161}}, Ehsan~Khodapanah~Aghdam\,\textsuperscript{\orcidlink{0000-0002-2849-1070}},
Amelie~Rauland,
Yiwei~Jia\,\textsuperscript{\orcidlink{0000-0002-5824-8821}},
Atlas~Haddadi~Avval\,\textsuperscript{\orcidlink{0000-0002-3896-7810}},
Afshin~Bozorgpour\,\textsuperscript{\orcidlink{0000-0003-1857-1058}},
Sanaz~Karimijafarbigloo\,\textsuperscript{\orcidlink{0000-0002-6632-6121}},
Joseph~Paul~Cohen\,\textsuperscript{\orcidlink{0000-0002-1334-3059}},
Ehsan~Adeli\,\textsuperscript{\orcidlink{0000-0002-0579-7763}},
and~Dorit~Merhof\,\textsuperscript{\orcidlink{0000-0002-1672-2185}}%
\IEEEcompsocitemizethanks{\IEEEcompsocthanksitem R.~Azad and E.~Khodapanah~Aghdam contributed equally to this work.%
\IEEEcompsocthanksitem R.~Azad, A.~Rauland, Y.~Jia, and S.~Karimijafarbigloo are with the Institute of Imaging and Computer Vision, RWTH Aachen University, 52074 Aachen, Germany.%
\IEEEcompsocthanksitem E.~Khodapanah~Aghdam is with the Department of Electrical Engineering, Shahid Beheshti University, Tehran 1983969411, Iran.%
\IEEEcompsocthanksitem A.~Haddadi~Avval is with the School of Medicine, Mashhad University of Medical Sciences, Mashhad 9177899191, Iran.%
\IEEEcompsocthanksitem A.~Bozorgpour is with the Department of Computer Engineering, Sharif University of Technology, Tehran 1458889694, Iran.%
\IEEEcompsocthanksitem J.P.~Cohen is with the Center for Artificial Intelligence in Medicine \& Imaging, Stanford University, Palo Alto, California 94304, USA, and this work was done prior to joining Amazon.%
\IEEEcompsocthanksitem E.~Adeli is with Stanford University, Stanford, California 94305, USA.%
\IEEEcompsocthanksitem D.~Merhof is with the Faculty of Informatics and Data Science, University of Regensburg, 93053 Regensburg, Germany, and also with the Fraunhofer Institute for Digital Medicine MEVIS, 28359 Bremen, Germany.%
}
\thanks{(Corresponding author: D.~Merhof. E-mail: \href{mailto:dorit.merhof@ur.de}{dorit.merhof@ur.de}.)}\\ \protect
\thanks{This work has been submitted to the \textbf{IEEE} for possible publication. Copyright may be transferred without notice, after which this version may no longer be accessible.}
}


\IEEEtitleabstractindextext{%
\begin{abstract}
Automatic medical image segmentation is a crucial topic in the medical domain and successively a critical counterpart in the computer-aided diagnosis paradigm. U-Net is the most widespread image segmentation architecture due to its flexibility, optimized modular design, and success in all medical image modalities. Over the years, the U-Net model achieved tremendous attention from academic and industrial researchers. Several extensions of this network have been proposed to address the scale and complexity created by medical tasks. Addressing the deficiency of the naive U-Net model is the foremost step for vendors to utilize the proper U-Net variant model for their business. Having a compendium of different variants in one place makes it easier for builders to identify the relevant research. Also, for ML researchers it will help them understand the challenges of the biological tasks that challenge the model. To address this, we discuss the practical aspects of the U-Net model and suggest a taxonomy to categorize each network variant. Moreover, to measure the performance of these strategies in a clinical application, we propose fair evaluations of some unique and famous designs on well-known datasets. We provide a comprehensive implementation library with trained models for future research. In addition, for ease of future studies, we created an online list of U-Net papers with their possible official implementation. All information is gathered in \href{https://github.com/NITR098/Awesome-U-Net}{https://github.com/NITR098/Awesome-U-Net} repository.
\end{abstract}

}

\maketitle
\IEEEdisplaynontitleabstractindextext

\IEEEraisesectionheading{\section{Introduction} \label{sec:introduction}}

\IEEEPARstart{I}{mage} segmentation, defined as the partition of the entire image into a set of regions, plays a vital role in a wide range of applications. Medical image segmentation is a crucial example of this domain and offers numerous benefits for clinical use. Automated segmentation facilitates the data processing time and guides clinicians by providing task-specific visualizations and measurements. In almost all clinical applications the visualization algorithm not only provides insight into the abnormal regions in human tissue but also guides the practitioners to monitor cancer progression. Semantic segmentation as a preparatory step in automatic image processing technique can further enhance the visualization quality by modeling to detect specific regions which are more relevant to the task on hand (e.g., heart segmentation) \cite{antonelli2022medical}.

Image segmentation tasks can be classified into two categories: semantic segmentation and instance segmentation \cite{asgari2021deep,minaee2021image}. Semantic segmentation is a pixel-level classification that assigns corresponding categories to all the pixels in an image, whereas instance segmentation also needs to identify different objects within the same category based on semantic segmentation.
Designing segmentation methods to distinguish organ or lesion pixels requires task-specific image data to provide the appropriate critical details. Common medical imaging modalities for acquiring data are X-ray, Positron Emission Tomography (PET), Computed Tomography (CT), Magnetic Resonance Imaging (MRI), and Ultrasound (US) \cite{wu2021medical}.
Early traditional approaches to medical image segmentation mainly focused on edge detection, template matching techniques, region growing, graph cuts, active contour lines, machine learning, and other mathematical methods.
In recent years, deep learning has matured in diverse fields for solving many edge cases specific to the medical domain. Convolutional neural networks (CNNs) have successfully implemented feature representation extraction for images, thus eliminating the need for hand-crafted features in image segmentation, and their superior performance and accuracy make them the main choice in this field.

An initial attempt to model the semantic segmentation using a deep neural network was proposed in \cite{ciresan2012deep}. This approach passes the input images through the convolutional encoder to produce the latent representation. Then on top of the generated feature maps the fully connected layers are included to produce a pixel-level prediction. The main limitation of this architecture was the use of fully connected layers, which depleted the spatial information and consequently degraded the overall performance. Long et al. \cite{2014FCN} proposed Fully Convolutional Networks (FCNs) to address this limitation. The FCN structure applies several convolutional blocks consisting of the convolution, activation, and pooling layers on the encoder path to capture semantic representation, and similarly uses the convolutional layer along with the up-sampling operation in the decoding path to provide a pixel-level prediction. The main motivation underlying the successive up-sampling process on the decoding path was to gradually increase the spatial dimension for a fine-grained segmentation result.

Inspired by the architecture of FCNs and the encoder-decoder models, Ronneberger et al. develop the U-Net \cite{ronneberger2015u} model for biomedical image segmentation. It is tailored to practical use in medical image analysis and can be applied in a variety of modalities, including CT \cite{huang2018robust,yu2019liver,kazerouni2022diffusion,zhou2018unet++,zhang2019net}, MRI \cite{chen2018s3d,zhang2020dense,ibtehaz2020multiresunet,liu2019cu,baldeon2020adaresu}, US \cite{abraham2019novel,zhao2017improved,zhang2017image}, X-ray \cite{frid2018improving,waiker2020effective}, Optical Coherence Tomography (OCT) \cite{orlando2019u2,asgari2019u}, and PET \cite{zhong20183d,wang2022ica}.

FCN networks, specifically the U-Net, can efficiently exploit a limited number of annotated datasets by leveraging data augmentation (e.g., random elastic deformation) to extract detailed features of images without the need for new training data, resulting in good segmentation performance \cite{azad2022medical}. This superiority has made it a great success and has led to the extensive use of U-Net model in the field of medical segmentation. The U-Net network is composed of two parts. The first part is the contracting path that employs the downsampling module consisting of several convolutional blocks to extract semantic and contextual features. And in the second part, the expansive path applies a set of convolutional blocks equipped with the upsampling operation to gradually increase the spatial resolutions of the feature maps, usually by a factor of two, while reducing the feature dimensions to produce the pixel-wise classification score. The most significant and important part of U-Net is the skip connections which copy the outputs of each stage within the contracting path to the corresponding stages in the expansive path. This novel design propagates essential high-resolution contextual information along the network, which encourages the network to re-use the low-level representation along with the high-context representation for accurate localization.
This novel structure becomes the backbone in the field of medical image segmentation since 2015, and several variants of the model have been derived to progress the state of the art based on it.

\begin{figure}
	\centering
	\includegraphics[width=\columnwidth]{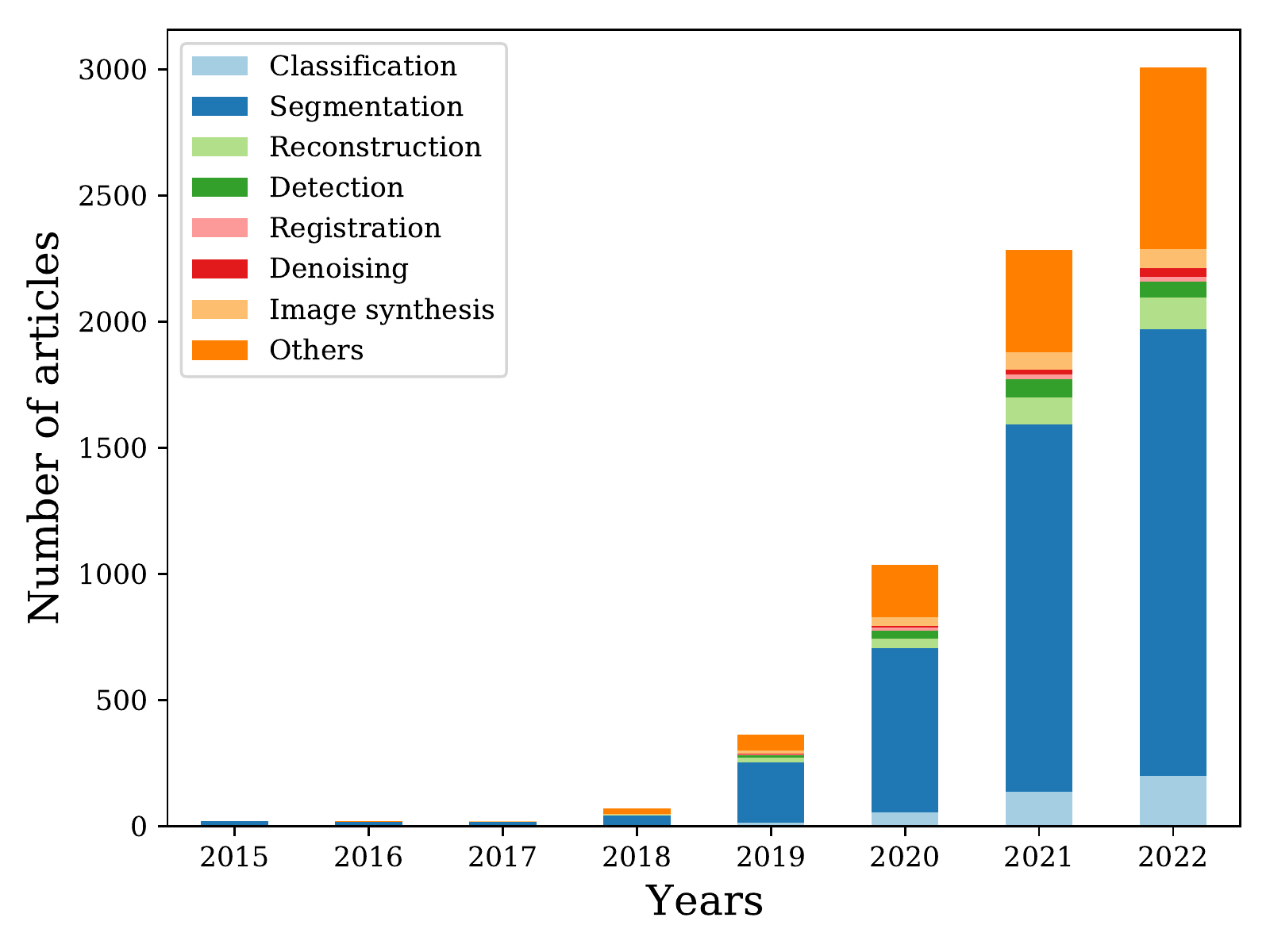}
	\caption{The number of research works published in the past decade using the U-Net model as their baseline to address various medical image analysis challenges. The visualization shows sumptuous attention from the research/industry community for this architecture, particularly the segmentation task which is the main objective of this review paper.}
	\label{fig:unetinchallenges}
\end{figure}

The auto-encoder design of U-Net makes it a unique tool for breaching  its structure in significant applications, e.g., image synthesis \cite{costa2017towards,wu2019uc,sun2022double}, image denoising \cite{reymann2019u,nasrin2019medical,lee2020mu}, image reconstruction \cite{guan2021dense,feng2020end}, and image super-resolution \cite{qiu2021progressive}. To provide more insight into the importance of the U-Net model in the medical domain, we provide \Cref{fig:unetinchallenges}, statistical information regarding the methods utilized U-Net model in their pipeline to address medical image analysis challenges. From \Cref{fig:unetinchallenges}, it is evident that U-Net influenced most of the diverse segmentation tasks in the medical image analysis domain with the extreme growth in publication numbers during the past decade and being bespeak for future remedies.

Our review covers the most recent U-Net-based medical image segmentation literature and discusses more than a hundred methods proposed until September 2022. We deliver a broad review and perspicuity on different aspects of these methods, including network architecture enhancements concerning vanilla U-Net, medical image data modalities, loss functions, evaluation metrics, and their critical contributions. According to the rapid developments in U-Net and its variants, we propose a summary of highly cited approaches in our taxonomy. we group the U-Net variants into the following categories:
\begin{enumerate}[leftmargin=*]
	\item \hyperref[sec:skip-connection]{Skip Connection Enhancements}
	\item \hyperref[sec:backbone-design]{Backbone Design Enhancements}
	\item \hyperref[sec:bottleneck]{Bottleneck Enhancements}
	\item \hyperref[sec:transformers]{Transformers}
	\item \hyperref[sec:rich-representation]{Rich Representation Enhancements}
\item \hyperref[sec:probablistic]{Probabilistic Design}
\end{enumerate}
Some of the key contributions of this review paper can be outlined as follows:
\begin{itemize}[leftmargin=*]
	\item This review covers the most recent literature on U-Net and its variants for medical image segmentation problems and overviews more than 100 segmentation algorithms proposed till September 2022, grouped into six categories.
	\item We provide a comprehensive review and insightful analysis of different aspects of U-Net-based algorithms, including the refinement of base U-Net architectures, training data modality, loss functions, evaluation metrics, and their critical contributions.
	\item We provide comparative experiments of some reviewed methods on popular datasets and offer codes and pre-trained weights on \href{https://github.com/NITR098/Awesome-U-Net}{GitHub}.
\end{itemize}

As a result, the remainder of the paper is organized as follows:  \Cref{sec:taxonomy} includes the taxonomy of review methods. \Cref{subsec:2dunet} and \Cref{subsec:3dunet} provide a a detailed insight into the basic 2D U-Net and 3D U-Net architectures, respectively. In \Cref{sec:unetextensions} we will cover U-Net extensions, overview at least five top-cited methods in each taxonomical branch, and highlights their key contribution. \Cref{sec:experiments} provides a comprehensive practical information such as the experimental datasets, training process, the loss functions, evaluation metrics, comparative results, and ablation studies. \Cref{sec:challenges} discusses the current challenges in the literature and future directions. Eventually, the last chapter provides the conclusion.

\section{Taxonomy} \label{sec:taxonomy}
\begin{figure*}[t]
	\centering
	\includegraphics[width=\textwidth]{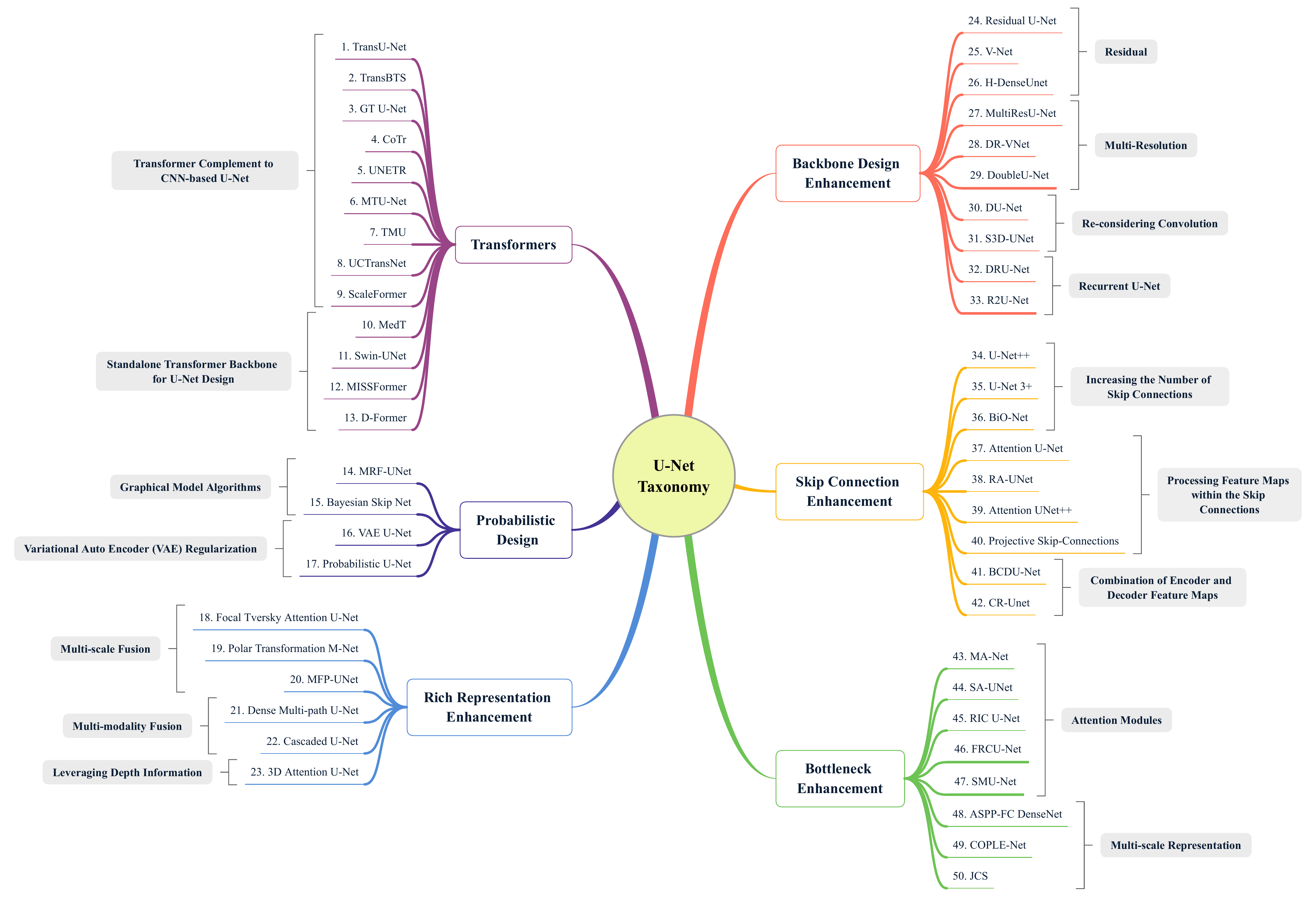}
	\caption{The proposed U-Net taxonomy categorizes different extensions of the U-Net model based on their underlying design idea. More specifically, our taxonomy takes into account the modular design of the U-Net model and shows where the improvement happens (e.g., skip connection). Due to the clarification and unity in the studies' denomination, we may utilize some brevities. In this case, each prefix number denotes 1. \cite{chen2021transunet}, 2. \cite{wang2021transbts}, 3. \cite{li2021gt}, 4. \cite{xie2021cotr}, 5. \cite{hatamizadeh2022swin}, 6. \cite{wang2022mixed}, 7. \cite{reza2022contextual}, 8. \cite{wang2022uctransnet}, 9. \cite{huang2022scaleformer}, 10. \cite{valanarasu2021medical}, 11. \cite{cao2021swin}, 12. \cite{huang2021missformer}, 13. \cite{wu2022d}, 14. \cite{brudfors2021mrf}, 15. \cite{klug2020bayesian}, 16. \cite{myronenko20183d}, 17. \cite{kohl2018probabilistic}, 18. \cite{abraham2019novel}, 19. \cite{fu2018joint}, 20. \cite{moradi2019mfp}, 21. \cite{dolz2018dense}, 22. \cite{lachinov2018glioma}, 23. \cite{islam2019brain}, 24. \cite{drozdzal2016importance}, 25. \cite{milletari2016v}, 26. \cite{li2018h}, 27. \cite{ibtehaz2020multiresunet}, 28. \cite{karaali2022dr}, 29. \cite{jha2020doubleu}, 30. \cite{jin2019dunet}, 31. \cite{chen2018s3d}, 32. \cite{kou2019microaneurysms}, 33. \cite{nasrin2019medical}, 34. \cite{zhou2019unet++}, 35. \cite{huang2020unet}, 36. \cite{xiang2020bio}, 37. \cite{oktay2018attention}, 38. \cite{jin2020ra}, 39. \cite{li2020attention}, 40. \cite{lachinov2021projective}, 41. \cite{azad2019bi}, 42. \cite{li2019cr}, 43. \cite{fan2020ma}, 44. \cite{guo2021sa}, 45. \cite{zeng2019ric}, 46. \cite{azad2021deep}, 47. \cite{azad2022smu}, 48. \cite{hai2019fully}, 49. \cite{wang2020noise}, 50. \cite{wu2021jcs}.}
	\label{fig:unet-taxonomy}
\end{figure*}

This section suggests a taxonomy that organizes different approaches presented in the literature to modify U-Net architecture for medical image segmentation.
Due to the modular design of U-Net, we proposed our taxonomy to cope with the inheritance design of U-net rather than the conceptual taxonomies offered in \cite{siddique2021u}. Furthermore, this property makes it difficult to fit each study into only one group so that a method may belong to several groups of divisions. \Cref{fig:unet-taxonomy} depicts our structure for taxonomy, and we think this taxonomy helps the field be organized and even motivational for future research. In \Cref{sec:unetextensions}, we will go through each concept of taxonomy. In the remainder of this section, we will first explain the naive 2D U-Net, and following that, we will introduce the 3D U-Net. Eventually, we will elaborate on the importance of the U-Net model from a clinical perspective. 

\subsection{ 2D U-Net} \label{subsec:2dunet}

Before recapitulating the U-Net structure in more detail, we will first consider the path that brings us to the U-Net architecture. The story begins with the EM segmentation challenge in 2012, where Ciresean et al. \cite{ciresan2012deep} were the first researchers who outperform the previous biomedical imaging segmentation methods using convolutional layers. The key factor that made them able to win the challenge was the availability of huge annotated data (CNN can learn comparatively better than the classical machine learning approach on large datasets \cite{o2019deep,hedjazi2017identifying}). However, access to the high amount of annotated data in biomedical tasks is always inherently challenging due to privacy concerns, the complexity of the annotation process, expert skill requirements, and the high price of taking images with biomedical imaging systems. The first step toward alleviating the need for large annotated data was proposed in \cite{ciresan2012deep}. This method used an image patching technique to not only increase the number of samples but also model the data distribution with small patches. Using this technique the CNN network learns the visual concept by simply deploying a sliding window. However, the sliding window usually brings more computation burden than its performance increase. Hence, there is always a trade-off between performance and computational complexity.

\begin{figure}
	\centering
	\includegraphics[width=\columnwidth]{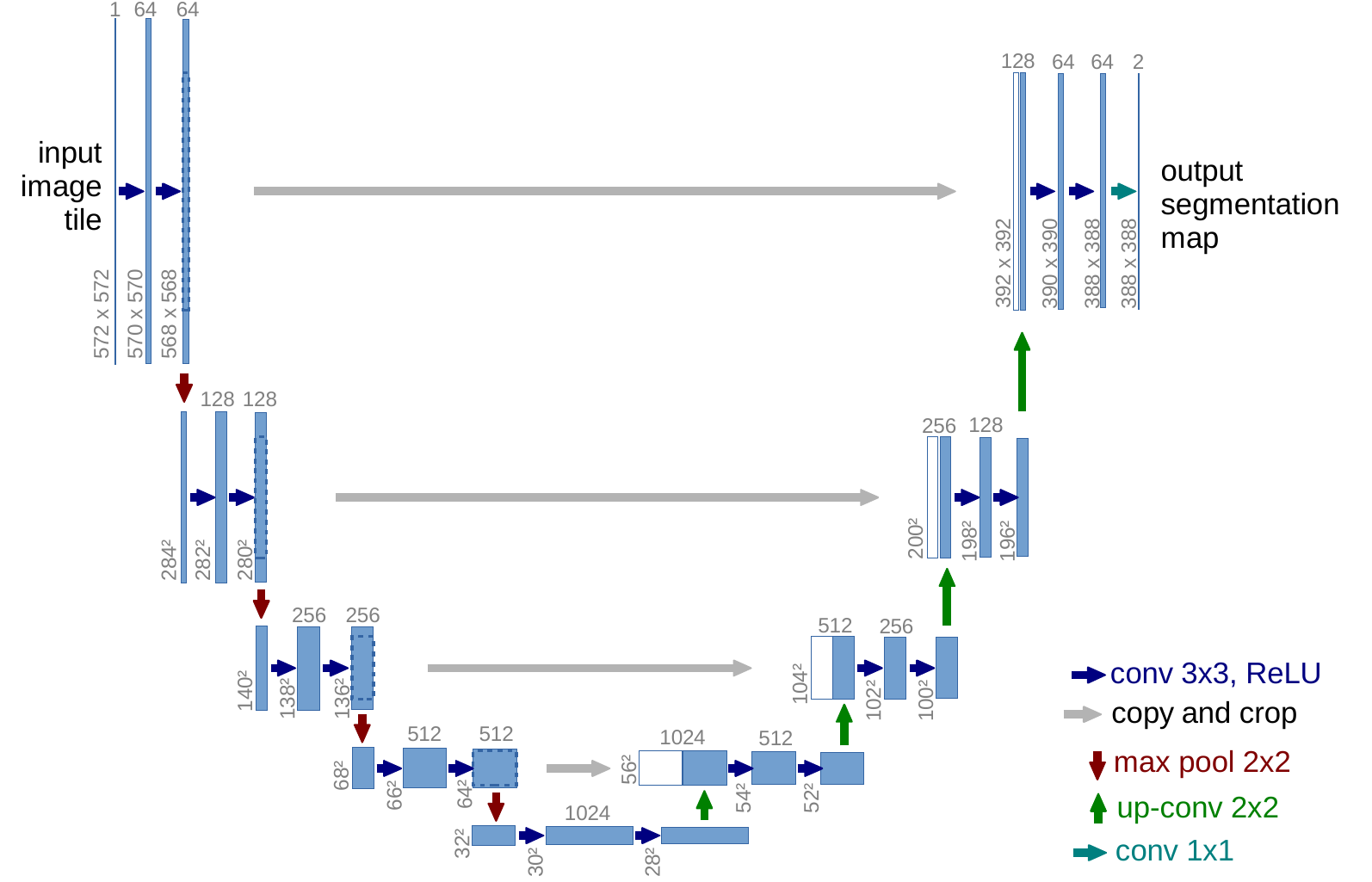}
	\caption{The initial 2D U-Net architecture that is designed to cope with semantic segmentation challenge. Figure from \cite{ronneberger2015u-arx}.}
	\label{fig:2dunet}
\end{figure}

In 2015, Ronnebreger et al. \cite{ronneberger2015u} proposed a new architecture with respect to Long et al.'s \cite{2014FCN} FCN framework in conjunction with \textit{ISBI cell tracking challenge}, where they won the competition by a large margin. \Cref{fig:2dunet} shows the structure of the U-Net model. Their proposed method is a cornerstone in a few attitudes those days. First, it is based on a fully convolutional network in an encoder-decoder design with insufficient data than the DNNs instinct with some intuitive data augmentation techniques. Second, their model was reasonably fast and outperformed other methods in the challenge.
The model architecture can be divided into two parts: The first part is the contracting path, also known as the encoder path, where its purpose is to capture contextual information. This path consists of repeated blocks, where each block contains two successive $3\times3$ convolutions, followed by a ReLU activation function and max-pooling layers. The max pooling layer is also included to gradually increase the receptive field of the network without imposing an additional computational burden.

The second part is expanding the path, also called the decoder path, where it aims to gradually up-sample feature maps to the desired resolution. This path consists of one $2\times2$ transposed convolution layer (up-sampling), followed by two consecutive $3\times3$ convolutions and a ReLU activation. The connection path between encoder and decoder paths (also known as a bottleneck) includes two successive $3\times3$ convolutions followed by a ReLU activation. The successive convolutional operations included in the U-Net model enables the network's receptive field size to be increased linearly. This process makes a network gradually learn coarse contextual and semantic representation in deep layers compared to shallow layers. Learning high-level semantic features makes the network slowly lose localization of extracted features, where this aspect is essential to reconstruct segmentation results. Ronnebreger et al. presented skip connections from the encoder path to the decoder path on the same scales to overcome this challenge. The existential reason for these skip connections is to impose localization information of extracted semantic features at the same stage from the encoder. To this end, the connection module concatenates low-level features coming from the encoder path with high-level representation derived from the decoding path to enrich localization information. Eventually, the network uses a $1\times1$ convolution to map the final representation to the desired number of classes. To mitigate the loss of contextual information in the missing image's border pixels, the U-Net model uses an overlap tile strategy. In addition, to deal with insufficient training data a typical data augmentation technique such as rotation, and gray-level intensities invariance, elastic deformation is utilized. It should be noted that elastic deformation is a common strategy to make the model resistant to deformations, a common variation in tissues. From a practical perspective, the original U-Net model outperformed a sliding-window convolutional network \cite{ciresan2012deep} in warping error terminology in the \textit{EM segmentation challenge} dataset \cite{ronneberger2015u}. This network also became a new state-of-the-art on two other cell segmentation datasets, \textit{PhC-U373} and \textit{DIC-Hela} cells, by a large margin of approximately $9\%$ and $31\%$ from the previous best methods in the \textit{ISBI Cell Tracking Challenge 2015} by reporting Intersection over Union (IoU) metric \cite{ronneberger2015u}.

\begin{figure*}[!ht]
	\centering
	\begin{subfigure}{0.106\textwidth} \centering
		\includegraphics[width=\linewidth, height=\linewidth]{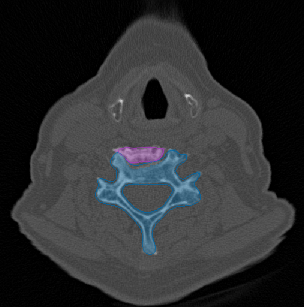}
	\end{subfigure}
	\hfill
	\begin{subfigure}{0.106\textwidth} \centering
		\includegraphics[width=\linewidth, height=\linewidth]{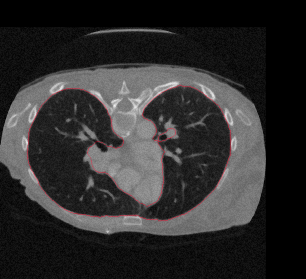}
	\end{subfigure}
	\hfill
	\begin{subfigure}{0.106\textwidth} \centering
		\includegraphics[width=\linewidth, height=\linewidth]{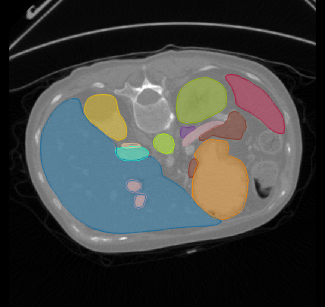}
	\end{subfigure}
	\hfill
	\begin{subfigure}{0.106\textwidth} \centering
		\includegraphics[width=\linewidth, height=\linewidth]{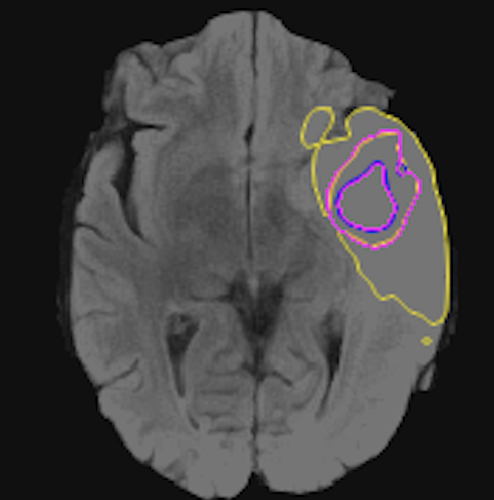}
	\end{subfigure}
	\hfill
	\begin{subfigure}{0.106\textwidth} \centering
		\includegraphics[width=\linewidth, height=\linewidth]{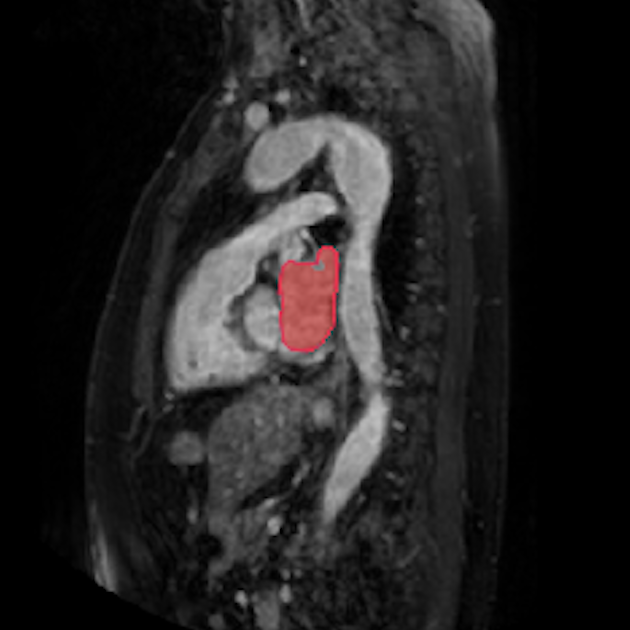}
	\end{subfigure}
	\hfill
	\begin{subfigure}{0.106\textwidth} \centering
		\includegraphics[width=\linewidth, height=\linewidth]{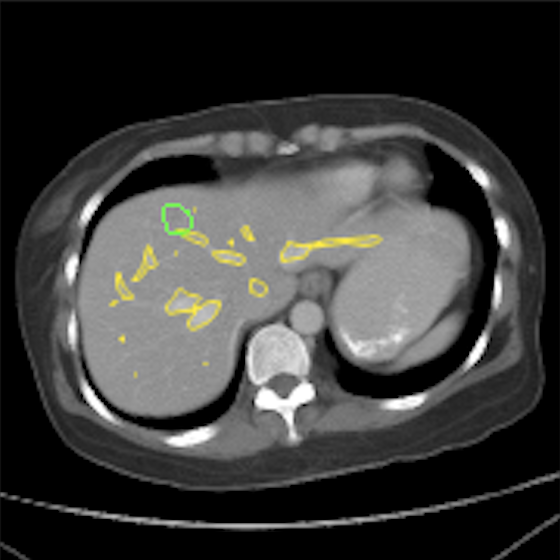}
	\end{subfigure}
	\hfill
	\begin{subfigure}{0.106\textwidth} \centering
		\includegraphics[width=\linewidth, height=\linewidth]{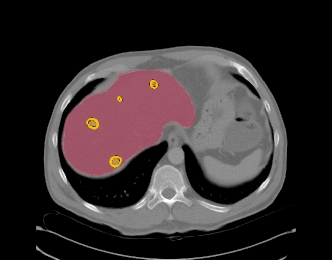}
	\end{subfigure}
	\hfill
	\begin{subfigure}{0.106\textwidth} \centering
		\includegraphics[width=\linewidth, height=\linewidth]{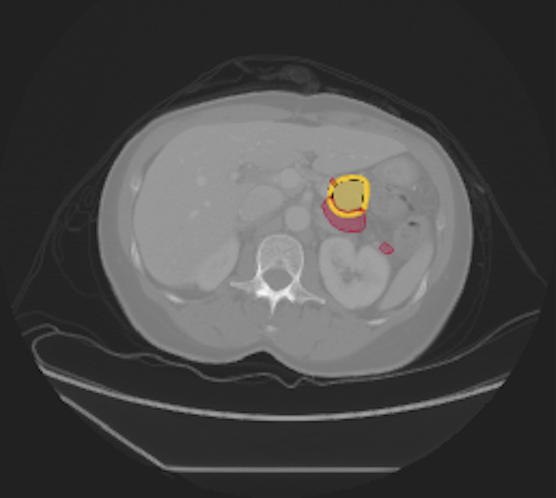}
	\end{subfigure}
	\hfill
	\begin{subfigure}{0.106\textwidth} \centering
		\includegraphics[width=\linewidth, height=\linewidth]{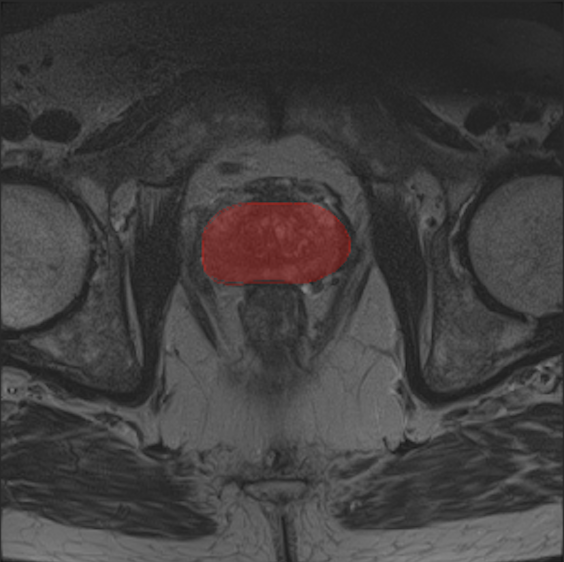}
	\end{subfigure}
	\\
	\begin{subfigure}{0.106\textwidth} \centering
		\includegraphics[width=\linewidth, height=1.2\linewidth]{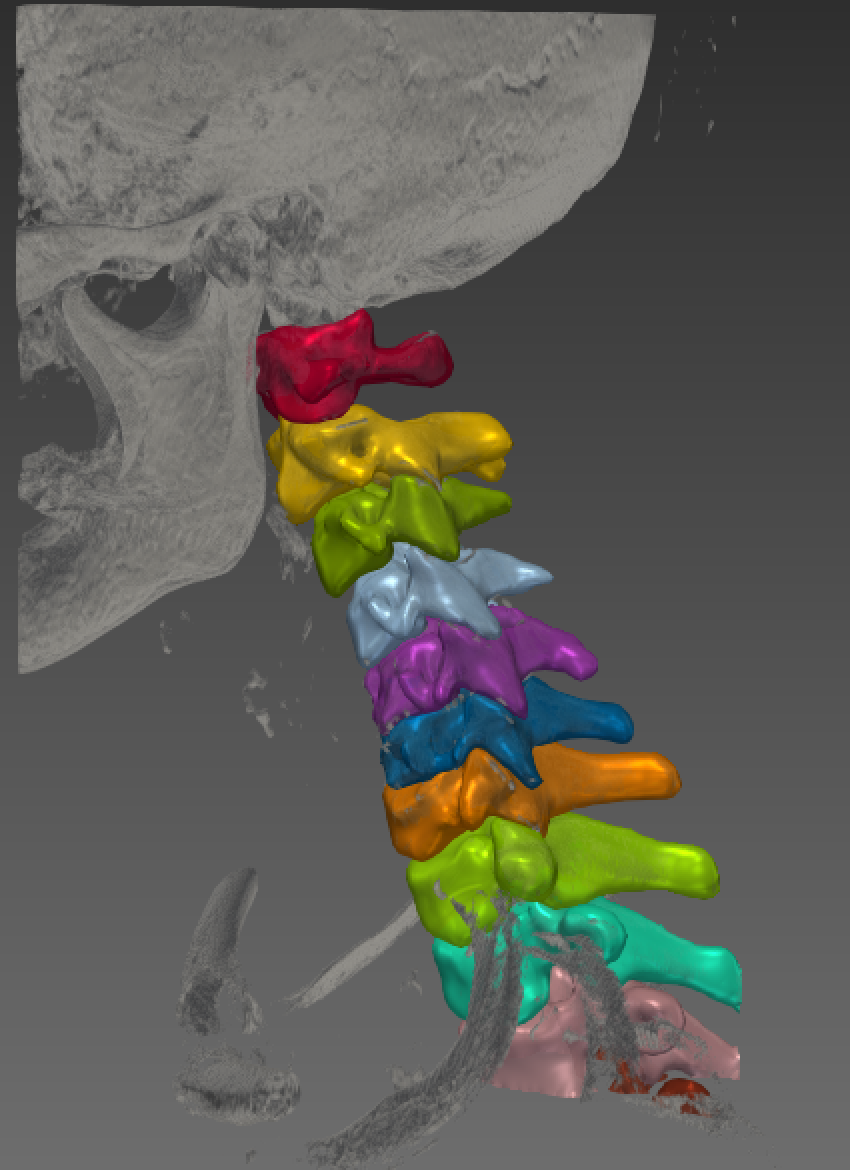}
	\end{subfigure}
	\hfill
	\begin{subfigure}{0.106\textwidth} \centering
		\includegraphics[width=\linewidth, height=1.2\linewidth]{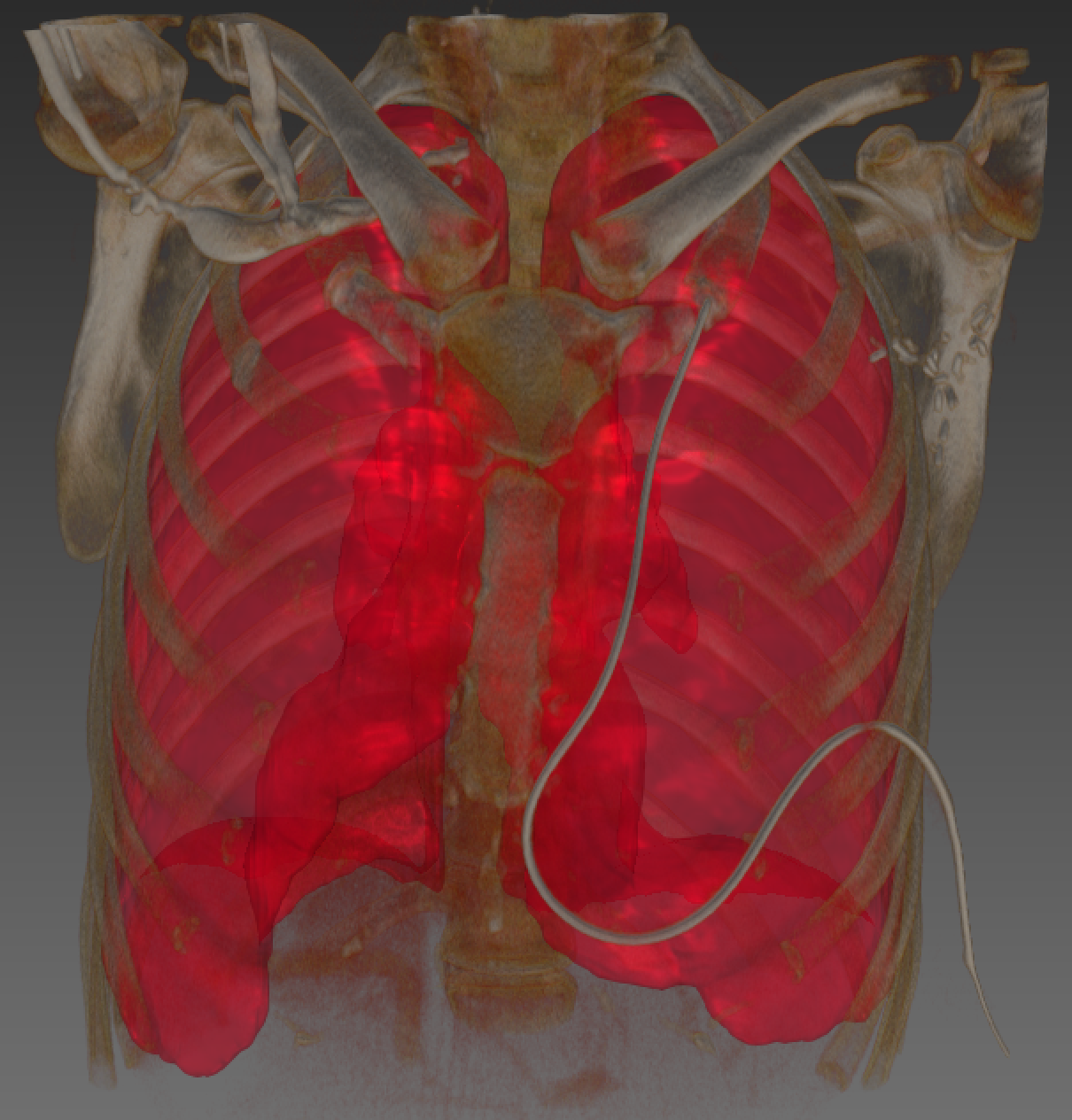}
	\end{subfigure}
	\hfill
	\begin{subfigure}{0.106\textwidth} \centering
		\includegraphics[width=\linewidth, height=1.2\linewidth]{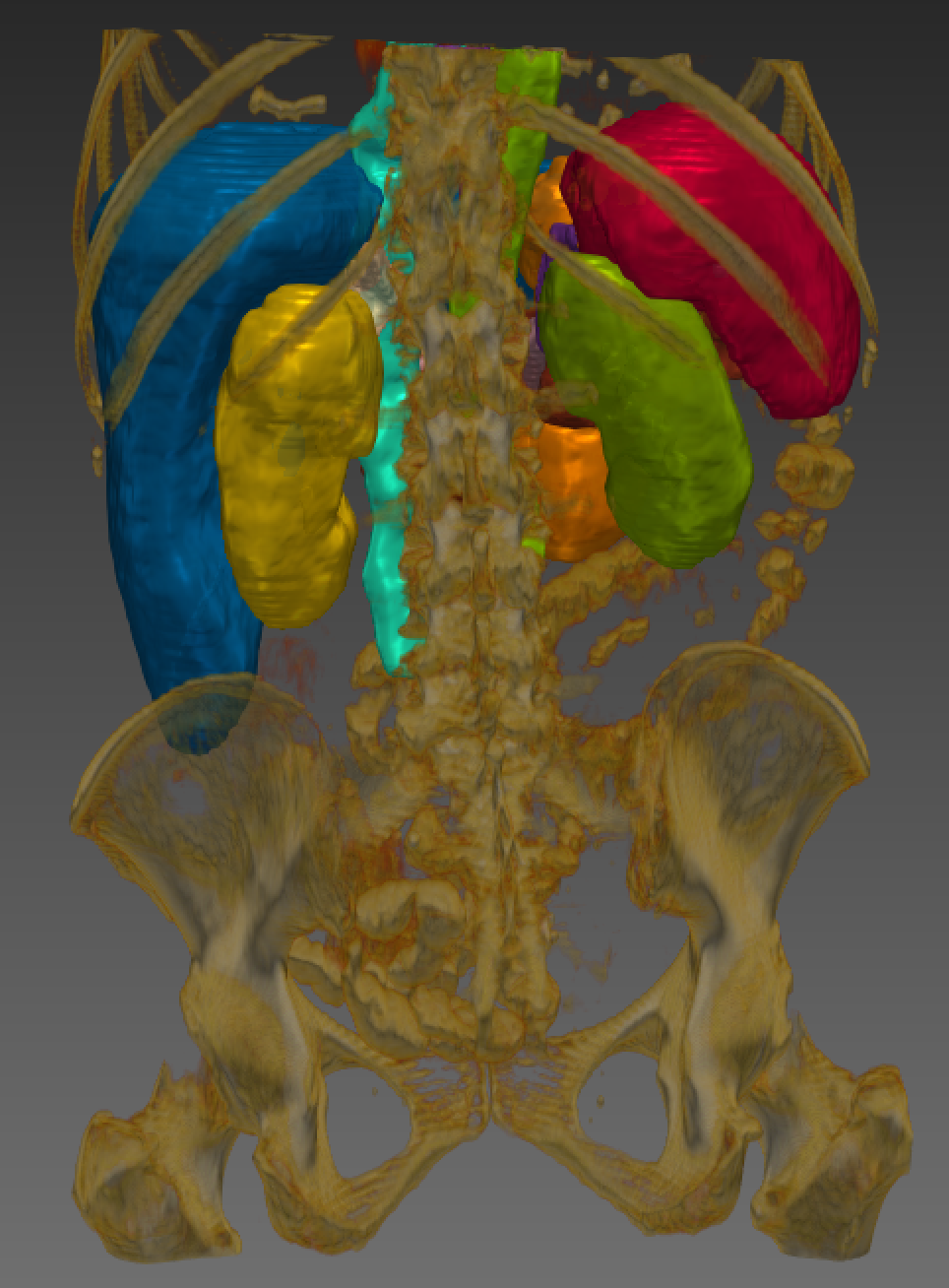}
	\end{subfigure}
	\hfill
	\begin{subfigure}{0.106\textwidth} \centering
		\includegraphics[width=\linewidth, height=1.2\linewidth]{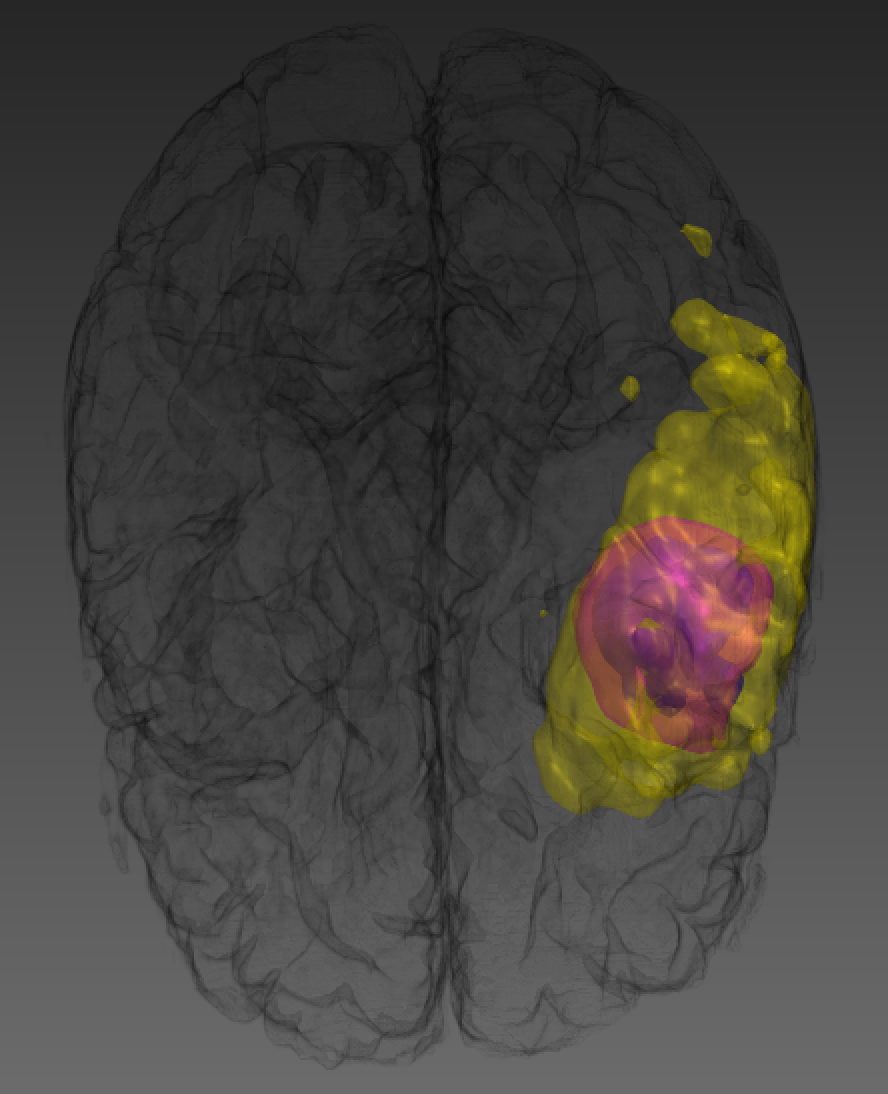}
	\end{subfigure}
	\hfill
	\begin{subfigure}{0.106\textwidth} \centering
		\includegraphics[width=\linewidth, height=1.2\linewidth]{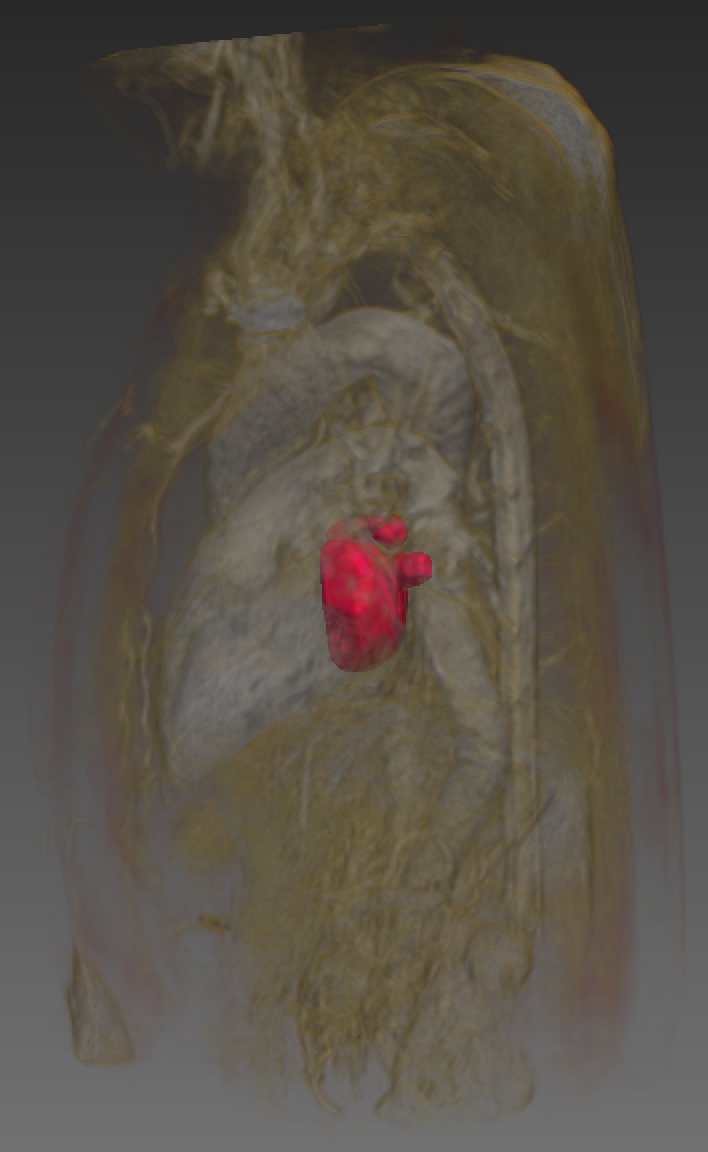}
	\end{subfigure}
	\hfill
	\begin{subfigure}{0.106\textwidth} \centering
		\includegraphics[width=\linewidth, height=1.2\linewidth]{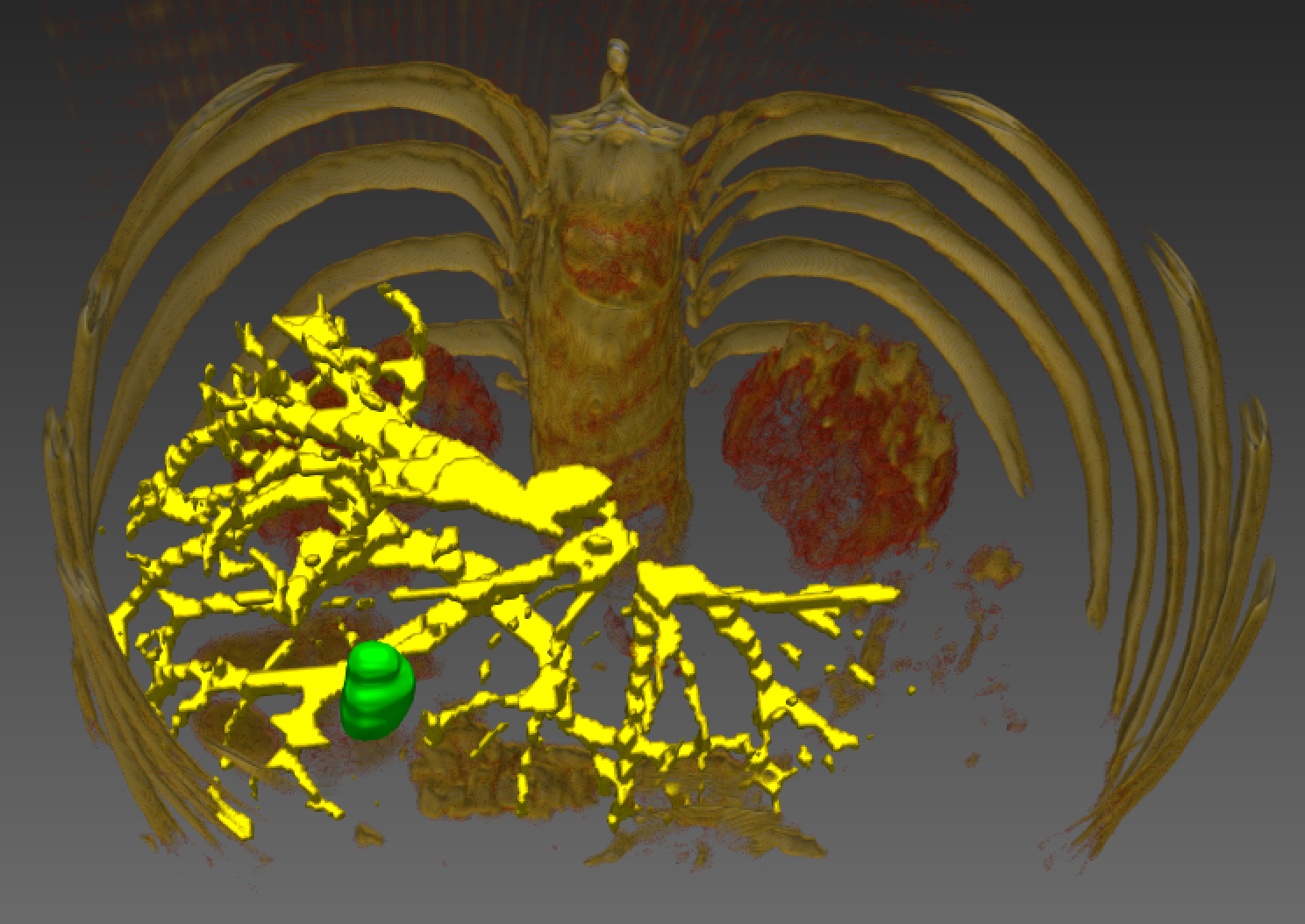}
	\end{subfigure}
	\hfill
	\begin{subfigure}{0.106\textwidth} \centering
		\includegraphics[width=\linewidth, height=1.2\linewidth]{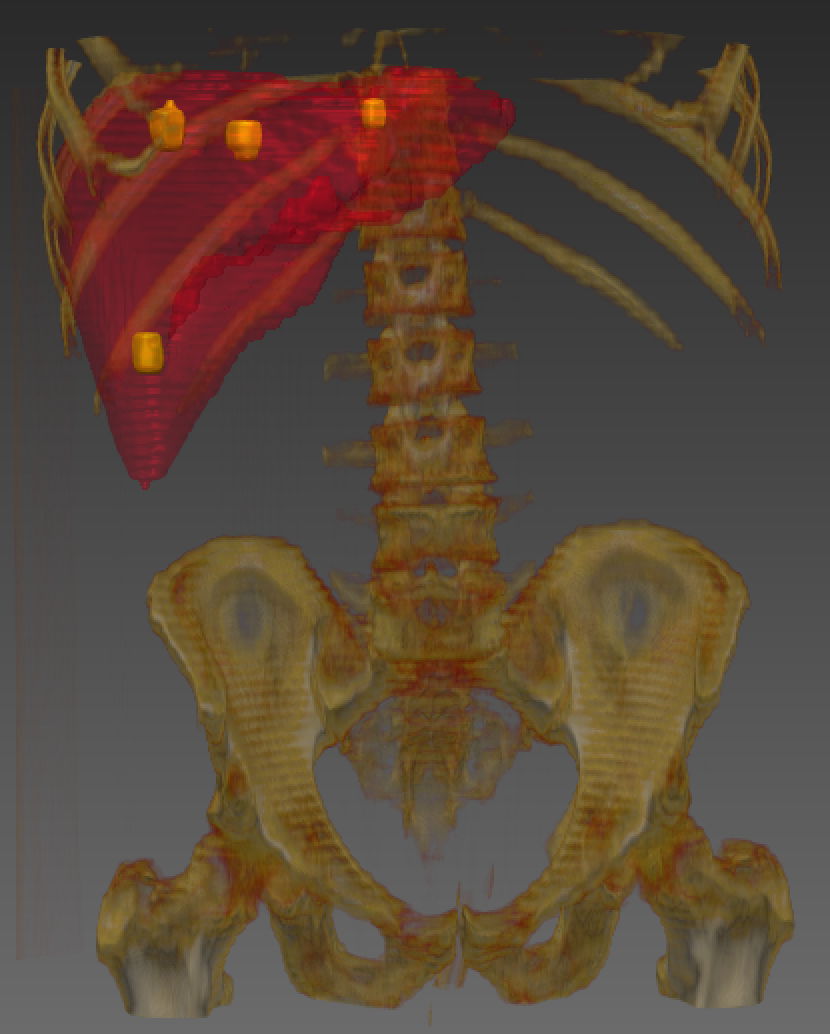}
	\end{subfigure}
	\hfill
	\begin{subfigure}{0.106\textwidth} \centering
		\includegraphics[width=\linewidth, height=1.2\linewidth]{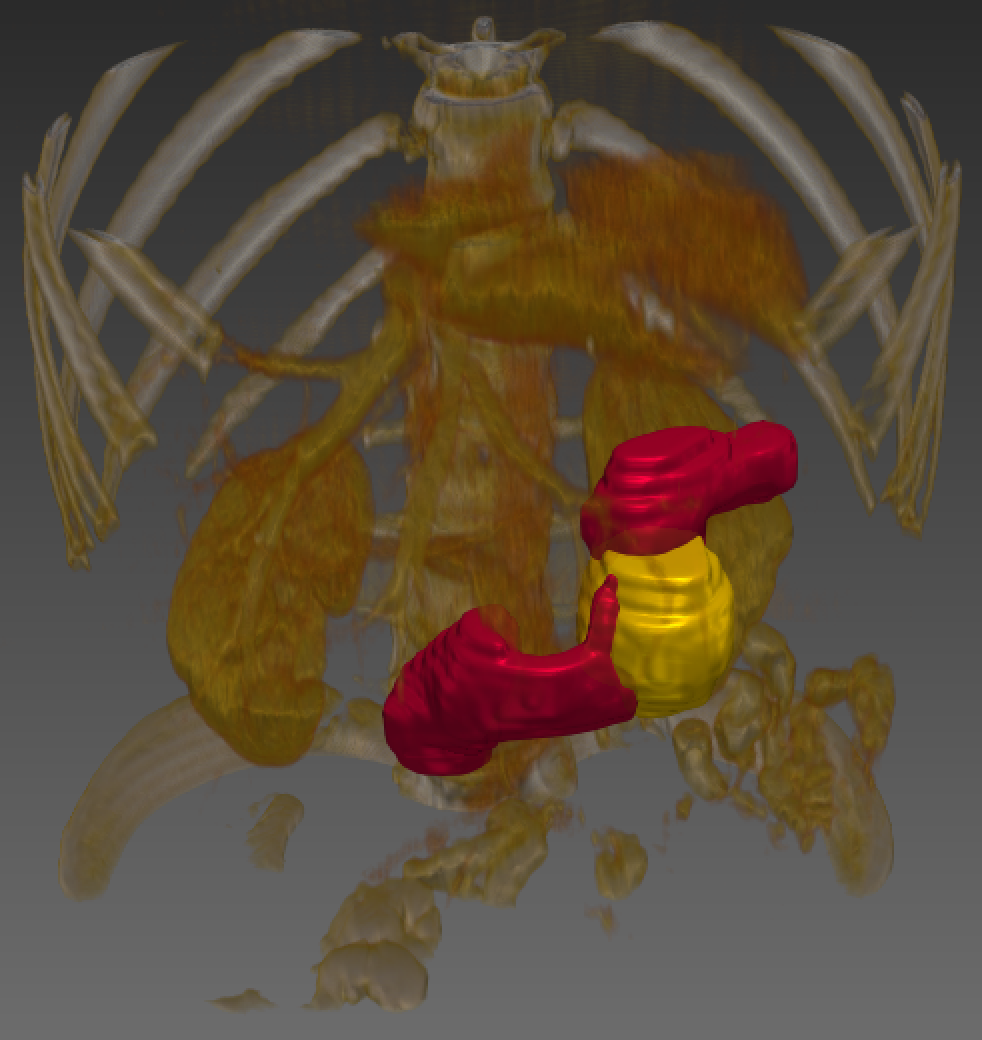}
	\end{subfigure}
	\hfill
	\begin{subfigure}{0.106\textwidth} \centering
		\includegraphics[width=\linewidth, height=1.2\linewidth]{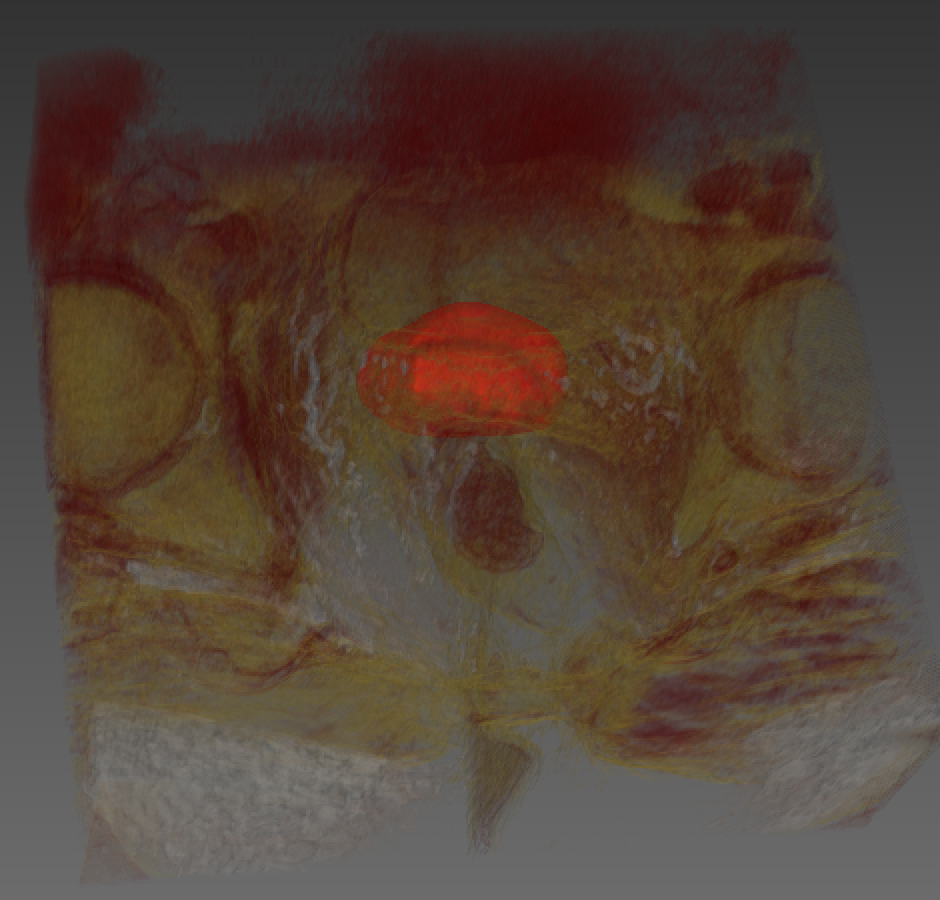}
	\end{subfigure}
	\caption{
		Sample of the 3D medical dataset and a single selected 2D frame, where the target area (e.g., organ) is highlighted using the annotation mask. 
		c.1) Cervical spine \cite{kaggle-cervical-spine-fracture-detection},
		c.2) Lung \cite{mader_2017}, 
		c.3) Fourteen abdominal organs \cite{landman2015miccai}, 
		c.4) Brain \cite{simpson2019large, menze2014multimodal}, 
		c.5) Heart \cite{simpson2019large}, 
		c.6) Hepatic vessel \cite{simpson2019large}, 
		c.7) Liver \cite{simpson2019large}, 
		c.8) Pancreas \cite{simpson2019large}, 
		c.9) Prostate \cite{simpson2019large}.
	}
	\label{fig:dataset-3d} 
\end{figure*}

\subsection{3D U-Net} \label{subsec:3dunet}
Due to the abundance and representation power of volumetric data, most medical image modalities are three-dimensional. So, {\c{C}}i{\c{c}}ek et al. \cite{cciccek20163d} proposed a 3D volumetric-based U-Net not only to pay attention to this need but also to overcome the time-consuming slice-by-slice annotation process for data. As it is noticeable that neighboring slices share the same information, there is no need for this much data redundancy. In \cite{cciccek20163d}, they replaced all 2D operations in U-Net architecture with the equivalent 3D companions and embedded a batch normalization layer for faster convergence after each 3D convolution layer. 3D U-Net was successfully applied to sparsely annotated three samples of Xenopus kidney embryos with reporting comparison results of IoU between 2D U-Net and 3D U-Net. To further support this we find the top 9/10 participants of the Kidney Tumor Segmentation (KiTS) 2021 challenge hosted by the Medical Image Computing and Computer Assisted Intervention (MICCAI) 2021 society \cite{heller2019kits19,heller2021state} challenge utilized a 3D U-Net and 1/10 utilized a 2D one \footnote{\url{https://kits21.kits-challenge.org/public/html/kits21_results.html\#}}.

\Cref{fig:dataset-3d} shows samples of 2D and 3D medical image segmentation challenges designed for different tasks. It can be seen that the 3D data provides more comprehensive information regarding the tissue and tumors, however, compared to the 2D data it has more computational cost.


\subsection{Clinical Importance and Effect of U-Net} \label{subsec:clinicalimportance}
\begin{figure*}[!t]
	\centering
	\includegraphics[width=\textwidth]{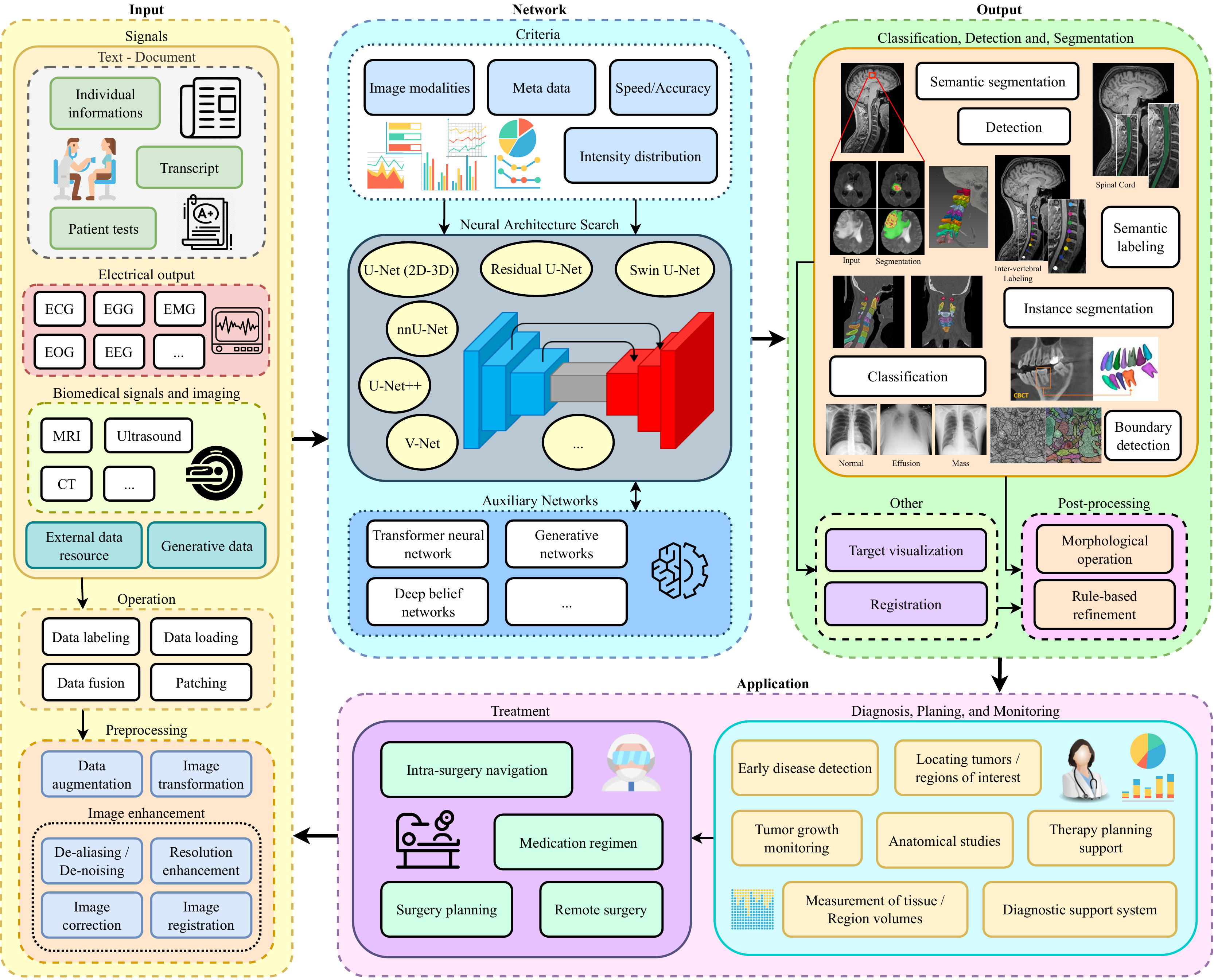}
	\caption{A detailed overview of the U-Nets core involvement in medical image analysis and clinical use. An illustration of how U-Nets are involved in clinical decisions is discussed in research papers. The first block deals with image acquisition, preparation, and pre-processing steps to provide the data in a common format for the deep neural network. The second step uses a neural architecture search algorithm to find an efficient architecture for the task at hand while the third block is designed to perform post operations to further refine the network output. Finally, the application block uses the software output to assist specialists with a certain action (e.g., tumor growth monitoring).}
	\label{fig:unet-pipeline}
\end{figure*}

During the start of the COVID-19 pandemic and the inevitable loss of healthcare and staff, the importance of utilizing artificial intelligence in images and test analysis was prompted. According to WHO, between January 2020 till May 2021, almost 80,000 to 180,000 healthcare and staff could have died from COVID-19 infection worldwide \cite{who2021healthcare}. Compensating for these skilled workforce losses, each country would incur a significant economic cost, and also transferring experiences among medical staff is a time-consuming process. In this direction, Michael et al. \cite{fitzke2021oncopetnet} applied a Large scale segmentation network to count specific cells in pathological images. They explicitly indicate that detecting cancerous cells from histopathological images is a challenging task that relies on the experiences of the expert pathologist. However, workflow efficiency can be increased with automatic system. Indeed the recent success of deep-learning-based segmentation methods, expansion of medical datasets and their easy accessibility, and facilitated access to modern and efficient GPUs, their applicability to specific image analysis problems of end-to-end users are eased. Semantic segmentation transforms a plain biomedical image modality into a meaningful and spatially well-organized pattern that could aid medical discoveries and diagnoses \cite{falk2019u,isensee2021nnu} and sometimes is beneficial to patients too as they may be able to avoid an invasive medical procedure \cite{aerts2014decoding}. Medical image segmentation is a vital component and a cornerstone in most clinical applications, including diagnostic support systems \cite{de2018clinically,bernard2018deep}, therapy planning support \cite{nikolov2018deep,mehrtash2017deepinfer}, intra-operative assistance \cite{hollon2020near}, tumor growth monitoring \cite{kickingereder2019automated,esmaeili2018direction}, and image-guided clinical surgery \cite{tonutti2017machine}. \Cref{fig:unet-pipeline} shows a general pipeline where the U-Net can be utilized in a clinical application to reduce experts' burden and accelerate the disease detection process. The entire end-to-end paradigm for using deep learning-based methods, especially U-Net, is an empirical struggle to fit this concept into everyday life \cite{isensee2021nnu}. Computer-Aided Diagnosis (CAD) can build from four main counterparts: Input, Network, Output, and Application. The input block could leverage different analyses on various available data like documented transcripts, diverse human body signals (EEG/ECG), and medical images. The multi-modal fusion of different data types could boost the performance of a pipeline for higher accuracy diagnosis. Based on specific criteria like Image modality, and data distribution, the network module could make decisions to choose one of the U-Net extensions which fit more to the setting. The output is a task-specific counterpart that the ultimate application block's decision could decide.

On the other hand, international image analysis competitions have a high demand for automatic segmentation methods, accounting for $70\%$ \cite{maier2018rankings} in the biomedical section, which universities of medical sciences primarily host or collaborate with them.
One of the advantages of deep learning competitions over conventional hypothesis-driven research is innate distinctions in the approach to problem-solving \cite{prevedello2019challenges}. Data competitions, by nature, encourage multiple individuals or groups to address a specific problem independently or collaboratively. According to Maier-Hein et al. \cite{maier2018rankings} of the 150 medical segmentation competitions before 2016 the majority used U-Net based models

Based on the points above and across-the-board of U-Net-based architectures, medical and clinical facilities could utilize these in real-world and commercial settings where nnU-Net \cite{isensee2021nnu} is one of these successful end-to-end designs. 

\section{U-Net Extensions} \label{sec:unetextensions}
U-Net is a ubiquitous network according to its approximately 48 thousand citations during its first release in 2015. This is evidence that it can handle diverse image modalities in broad domains and not only in medical fields. From our sight, the core advantage of U-Net is its modular and symmetric design, which makes it a suitable choice for broad modification and collaboration with diverse plug-and-play modules to increase performance. Therefore, by pursuing this cue, we infringe the Ronneberger et al. \cite{ronneberger2015u} network to modular improvable counterparts besides solid auxiliary modification for achieving SOTA or par with segmentation performances. In this respect, we offer our taxonomy (\Cref{fig:unet-taxonomy}) and divide the diverse variants of U-Net modifications into systematic categories as follows:
\begin{enumerate}[leftmargin=*]
	\item \hyperref[sec:skip-connection]{Skip Connection Enhancements}
	\item \hyperref[sec:backbone-design]{Backbone Design Enhancements}
	\item \hyperref[sec:bottleneck]{Bottleneck Enhancements}
	\item \hyperref[sec:transformers]{Transformers}
	\item \hyperref[sec:rich-representation]{Rich Representation Enhancements}
\item \hyperref[sec:probablistic]{Probabilistic Design}
\end{enumerate}

This taxonomy aims to provide comprehensive and practical information for both vendors and researchers. In the following parts of this section, each category will be extensively discussed along with relevant papers. 

\subsection{Skip Connection Enhancements} \label{sec:skip-connection}
Skip connections are an essential part of the U-Net architecture as they combine the semantic information of a deep, low-resolution layer with the local information of a shallow, high-resolution layer.
This section provides a definition of skip connections and explains their role in the U-Net architecture before introducing extensions and variants of the classic skip connection used in the original U-Net.
Skip connections are defined as connections in a neural network that do not connect two following layers but instead skip over at least one layer. Considering a two-layer network a skip connection would connect the input directly to the output, skipping over the hidden layer \cite{bishop2006pattern}.
In image segmentation, skip connections were first used by Long et al. in \cite{2014FCN}. At the time, the most common use of convolutional networks was for image classification tasks which only have a single label as output. In a segmentation task, however, a label should be assigned to each pixel in the image adding a localization task to the classification task.

Long et al. \cite{2014FCN} added additional layers to a usual contracting network using upsampling instead of pooling layers to increase the resolution of the output and obtain a label for every pixel. Since local, high-resolution information gets lost in the contracting part of the network it cannot be completely recovered when upsampling these volumes. To combine the deep, coarse semantic information with the shallow fine appearance information they add skip connections that connect up-sampled lower layers with finer stride with the final prediction layer.

In the original U-Net architecture by Ronneberger et al. \cite{ronneberger2015u} each level in the encoder path is connected to the corresponding same-resolution level in the decoder path by a skip connection to combine the global information describing what with the local information resolving where. 
The difference to the above approach is not only the higher number of skip connections but also the way in which the features are combined. Long et al. \cite{2014FCN} up-sampled feature maps from earlier layers to the output resolution and added them to the output of the final layer. Ronneberger et al. \cite{ronneberger2015u} concatenated the features of the corresponding encoder and decoder level and process them together by passing them through two convolutional layers and an up-sampling layer together.

Li et al. \cite{li2018h} conducted an ablation study on skip connections by training a dense U-Net with and without skip connections. The results clearly show that the network with the skip connections generalize better than the network without skip connections.

Over the following years, many variants and extensions of the original U-Net architecture were developed concerning the skip connections \cite{banerjee2022ultrasound,asadi2020multi}. Different types of extensions dealing with processing the encoder feature maps passed through the skip connections, combining the two sets of feature maps, and extending the number of skip connections will be presented in the following sections.

\subsubsection{Increasing the Number of Skip Connections}
In 2020 Zhou et al. \cite{zhou2019unet++} introduced the U-Net++ in which they redesign skip connections to be more flexible and therefore exploit multiscale features more effectively. Instead of restricting skip connections to only aggregate features that have the same scale in the encoder and decoder path, they redesign them in such a way that features of different semantic scales can be aggregated \cite{zhou2019unet++}.

They argue that there has been no proof so far that encoder and decoder feature maps at the same scale are the best match for feature fusion and therefore design a more flexible setup.
\begin{figure}
	\centering
	\includegraphics[width=\columnwidth]{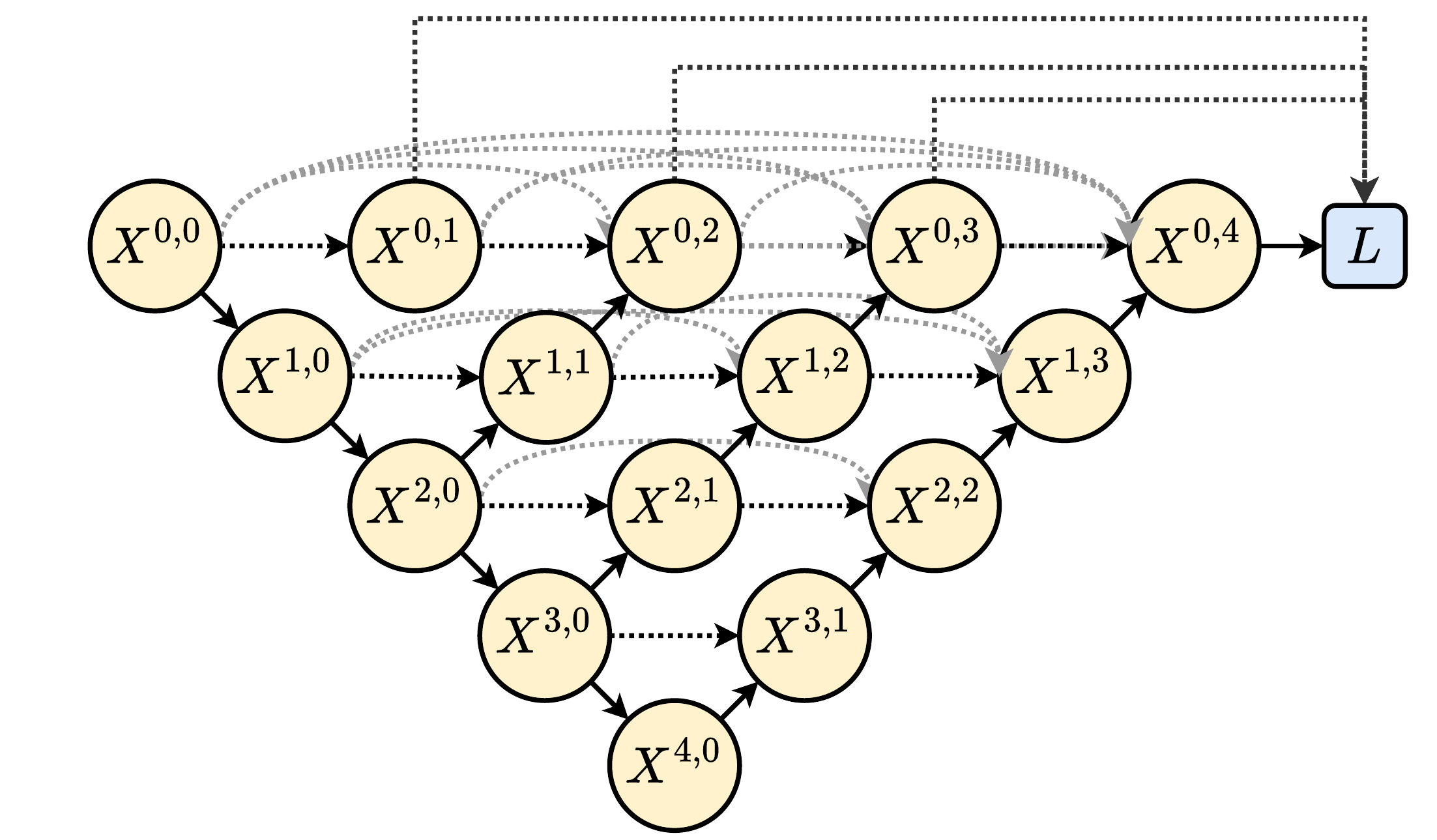}
	\caption{Architecture of the U-Net++ \cite{zhou2019unet++}. The U-Net++ uses a nested structure of convolutional layers $X^{i,j}$ (indicate $j$-th convolution layer in the $i$-th scale) connected through the skip connection paths to model multi-scale representation.}
	\label{fig:unet++}
\end{figure}
In their approach, they tackle two problems simultaneously. Since the optimal depth of a U-Net is unknown apriori and usually has to be determined through an exhaustive search, they incorporate U-Nets of different depths into one architecture. As can be seen in \Cref{fig:unet++} all the U-Nets share the same encoder but have their own decoder. Instead of only passing the same-scale encoder feature maps through the skip connections, each node in the decoder is also presented with the feature maps of the same-level decoders of the U-Nets with a lower depth. It can then be learned during training, which of the presented feature maps should ideally be used for the segmentation.

Huang et al. \cite{huang2020unet} take the dense skip connections introduced in the U-Net++ one step further by introducing full-scale skip connections in their architecture the U-Net3+. They argue that both the original U-Net with plain skip connections between same-level encoder and decoder nodes and the U-Net++ with the dense and nested skip connections do not sufficiently explore features from full scales making it challenging for the network to learn the position and boundary of an organ explicitly.
\begin{figure}
	\centering
	\includegraphics[width=\columnwidth]{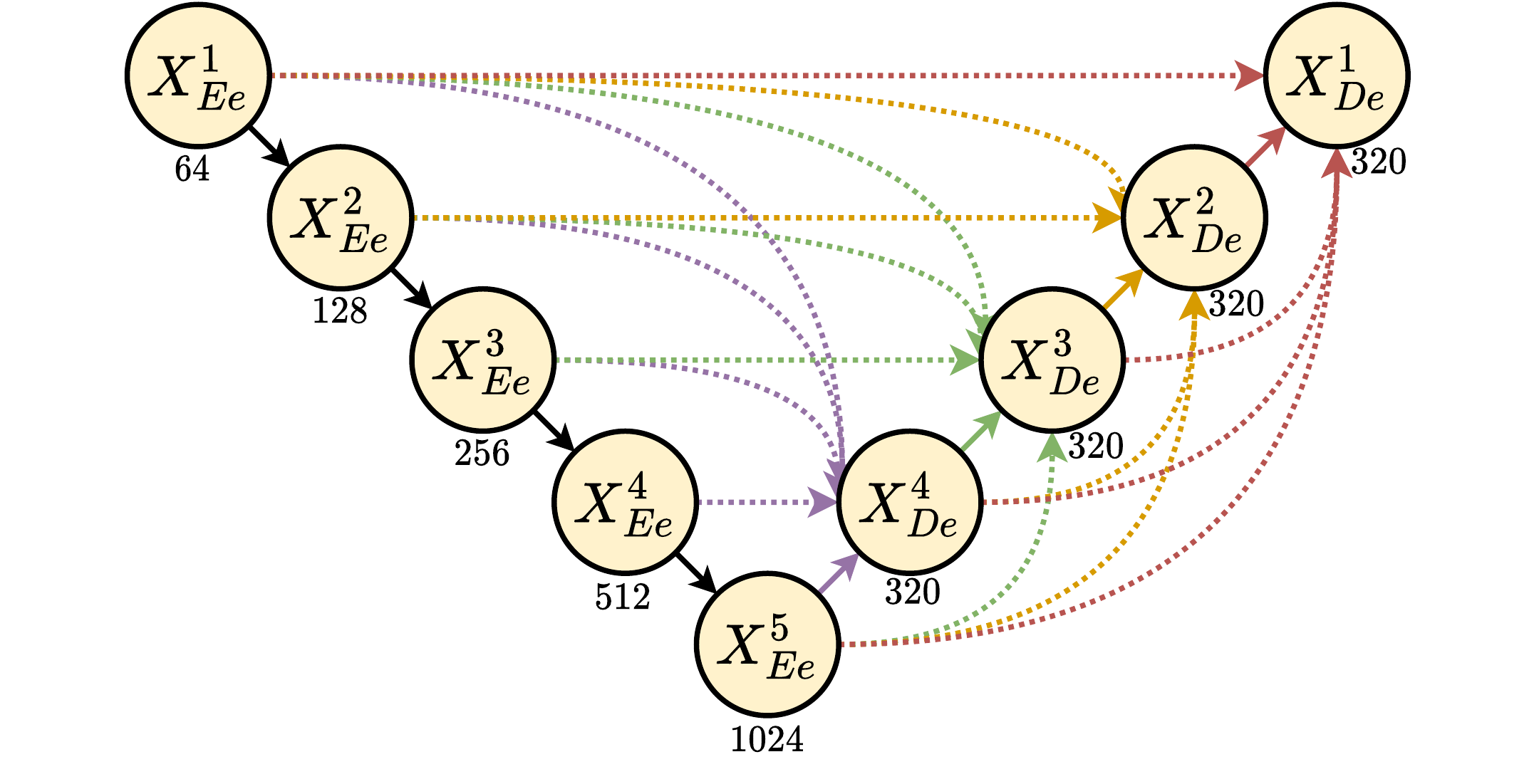}
	\caption{Architecture showing the full-scale skip connections of the UNet3+ \cite{huang2020unet}. $X_{Ee}^i$ and $X_{De}^i$ denote the encoder and decoder path feature maps at $i$-th scale, respectively. In addition, each subscript under each feature map symbol represents the number of feature maps to the corresponding block.}
	\label{fig:unet3+}
\end{figure}
To overcome this limitation they connect each decoder level with all encoder levels and all preceding decoder levels as can be seen in \Cref{fig:unet3+}. Since not all feature maps arriving at a decoder node through skip connections have the same scale, higher-resolution encoder feature maps will be downscaled using a max-pooling operation and lower-resolution feature maps coming from intra-decoder skip connections will be upsampled using bilinear upsampling. Additionally, apart from the up- or down-sampling operation, each skip connection is equipped with a $3\times 3$ convolutional layer calculating 64 output maps. The 64 feature maps arriving through each skip connection are stacked and the stack of feature maps is passed through another convolutional layer, followed by batch normalization and a ReLU activation before being further processed in the respective decoder node.

\begin{figure}
	\centering
	\includegraphics[width=\columnwidth]{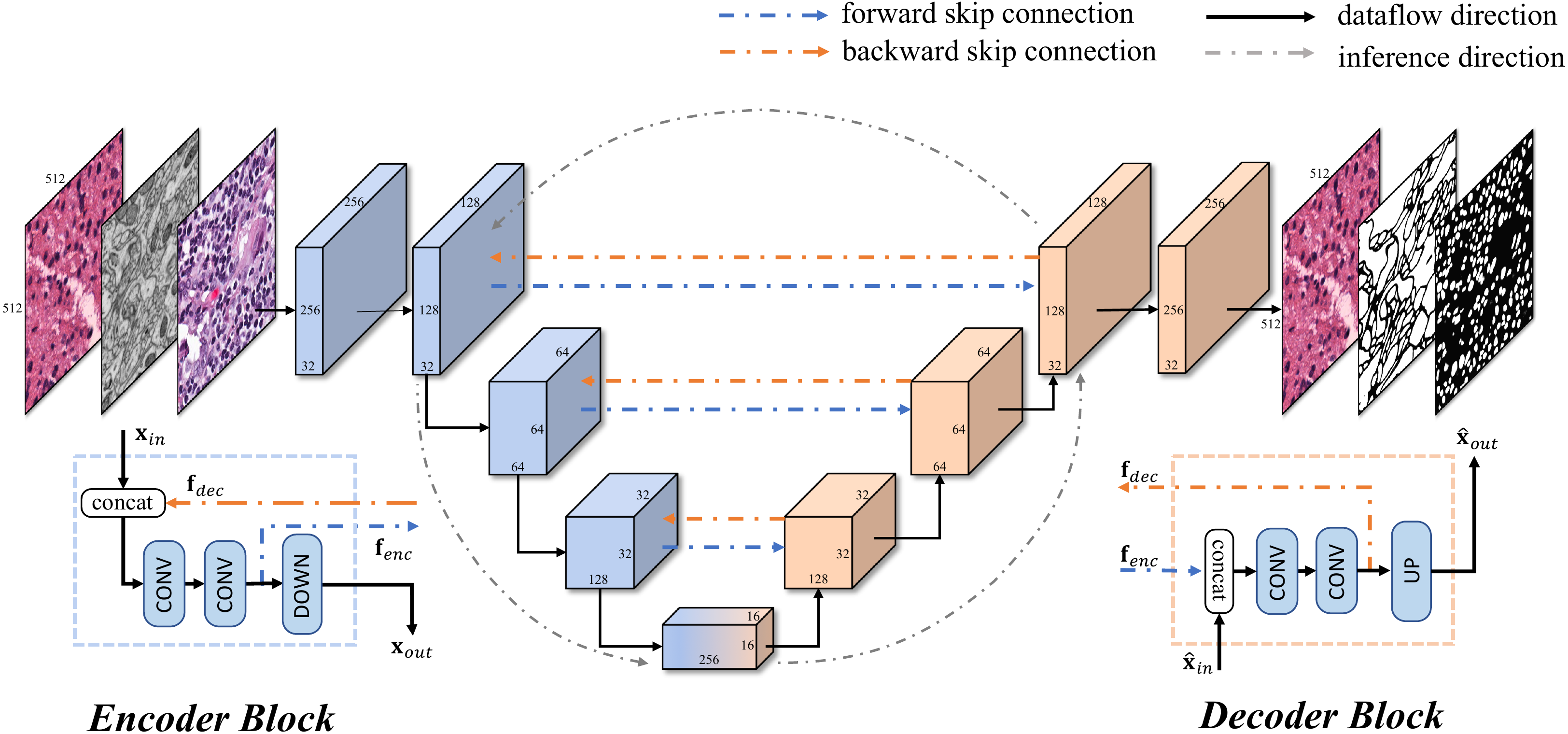}
	\caption{BiO-Net Architecture. Figure from \cite{xiang2020bio-arx}.}
	\label{fig:bio}
\end{figure}

Instead of increasing the number of forward skip connections, Xiang et al. \cite{xiang2020bio} add additional backward skip connections: Their Bi-directional O-Shape network (BiO-Net) is a U-Net architecture with bi-directional skip connections. 
This means that there are two types of skip connections:
\begin{enumerate}
	\item The forward skip connections are known from the original U-Net architecture, combining encoder and decoder layers at the same level. These skip connections preserve the low-level visual features from the encoder and combine them with the semantic decoder information.
	\item The backward skip connections pass decoded high-level features from the decoder back to the same level encoder. 
	The encoder can then combine the semantic decoder features with its original input and flexibly aggregate the two types of features. 
\end{enumerate}
Together these two types of skip connections build an O-shaped recursive architecture that can be traversed multiple times to receive improved performance (See \Cref{fig:bio}).
The recursive output of the encoder and decoder can be defined as follows:
\begin{align}
	\mathbf{x}_{out}^i&= \text{DOWN}(\text{ENC}([\text{DEC}([\mathbf{f}_{enc}^{i-1}, \mathbf{\hat{x}}_{in}^{i-1}]), \mathbf{x}_{in}^i])), \nonumber \\ 
	\mathbf{\hat{x}}_{out}^i&= \text{UP}(\text{DEC}([\text{ENC}([\mathbf{f}_{dec}^{i}, \mathbf{x}_{in}^{i}]), \mathbf{\hat{x}}_{in}^i])),
\end{align}
Here, $i$ represents the current inference iteration, $\text{UP}$ stands for an upsampling operation, $\text{DOWN}$ for a downsampling operation, $\text{DEC}$ and $\text{ENC}$ stand for a decoder and encoder level, respectively. An additional improvement was achieved when collecting decoded features from all iterations and feeding them to the last decoder stage together to calculate the final output. Although the recurrent training scheme might increase training time, this extension of the U-Net has the advantage that it does not introduce any additional parameters as claimed by the authors.

\subsubsection{Processing Feature Maps within the Skip Connections}
In the attention U-Net established by Oktay et al. \cite{oktay2018attention}, attention gates (AGs) are added to the skip connections to implicitly learn to suppress irrelevant regions in the input image while highlighting the regions of interest for the segmentation task at hand.
\begin{figure}
	\centering
	\includegraphics[width=\columnwidth]{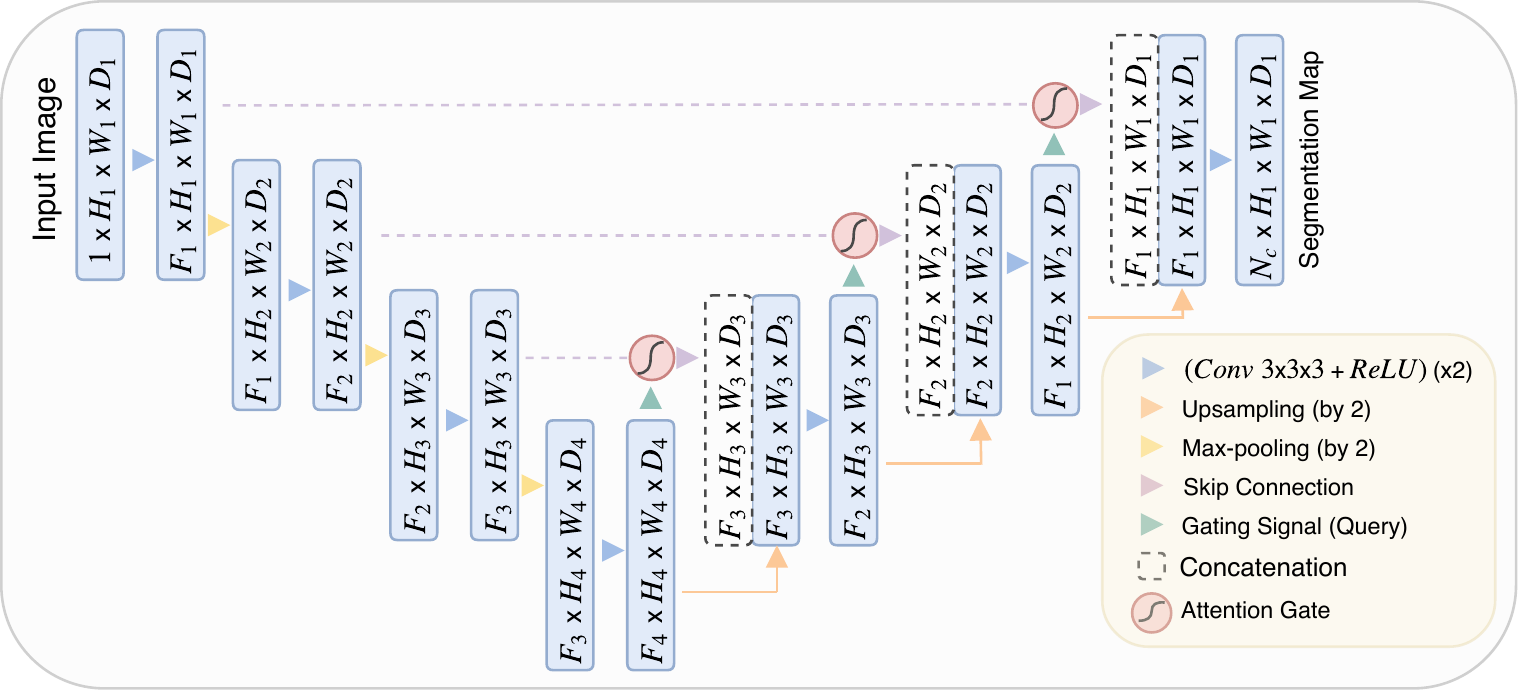}
	\caption{Attention U-Net Architecture. Figure from \cite{oktay2018attention}.}
	\label{fig:attention-unet}
\end{figure}
In biomedical imaging, when organs to be segmented show high inter-patient variation in terms of shape and size, a common approach is to use a cascaded network. The first network extracts a rough region of interest (ROI) including the organ to be segmented and the second network predicts the exact organ segmentation in this ROI. These approaches, however, suffer from redundant model parameters and high computational resources. Adding attention gates to the skip connections maintains a high prediction accuracy without the need for an external organ localization model. It is therefore trainable from scratch and introduces no significant computational overload and only a few additional model parameters. The output of an AG is the elementwise multiplication of the input feature maps with attention coefficients $\alpha_i \in [0, 1]$ as $\mathbf{\hat{x}}_{i, c}^l=\mathbf{x}_{i, c}^l\cdot \alpha_i^l$. For the computation of the attention coefficients both the input feature maps $x$, that have been passed through the skip connection from the encoder and the gating signal $g$ are analyzed. Here, the gating signal is collected from a coarser scale as can be seen in \Cref{fig:attention-unet} for adding contextual information. The applied additive attention is formulated as follows:
\begin{align}
	q_{\text{att}}^l&= \psi^T(\sigma_1(W_x^T\mathbf{x}_i^l+W_g^Tg_i+b_g))+b_\psi \nonumber \\
	\alpha_i^l&=\sigma_2(q_{\text{att}}^l(\mathbf{x_i^l}, g_i; \Theta_{\text{att}})),
\end{align}
where $\sigma_1$ and $\sigma_2$ are ReLU and sigmoid activations respectively, $W_x\in \mathbb{R}^{F_l\times F_{int}}$, $W_g\in \mathbb{R}^{F_g\times F_{int}}$ and $\psi \in \mathbb{R}^{F_{int}\times 1}$ are linear transforms and $b_g$ and $b_\psi$ are bias terms. Adding an AG to a skip connection, therefore, highlights the ROIs in the feature maps from the encoder path before they are concatenated with the feature maps of the decoder path. So in addition to adding higher resolution information, additional information on the location of the object(s) to be segmented is added, eliminating the need for cascaded multi-network approaches.

The attention U-Net++ by Li et al. combines the attention U-Net with the U-Net++ \cite{li2020attention}. Attention gates as described in \cite{oktay2018attention} are added to all the skip connections of the U-Net++ with its nested U-Nets and dense skip connections.
With similar motivation, Jin et al. \cite{jin2020ra} introduced a 3D U-Net with attention residual modules in the skip connections, called the RA-UNet.
The network was developed for the task of segmenting tumors in the liver. The main difficulties of this task lie in the large spatial and structural variability, low contrast between liver and tumor, and similarity to nearby organs.
The added attention residual learning mechanism in the skip connections improve the performance by focusing on specific parts of the image as claimed by the authors. The output of the attention module ($\mathbf{OA}$) in the RA-UNet structure is formulated as:
\begin{equation}
	\mathbf{OA}(\mathbf{x})=(1+\mathbf{S}(\mathbf{x}))\mathbf{F}(\mathbf{x}),
\end{equation}
where $\mathbf{S}(\mathbf{x})$ originates from the soft mask branch and has values in [0,1] to highlight important features and suppress noise and redundant features in the original feature maps $\mathbf{F}(\mathbf{x})$ passed through the trunk branch. The soft mask branch itself uses a residual encoder-decoder architecture to calculate its output.

To improve performance on the difficult task of the ovary and follicle segmentation from ultrasound images, Li et al. \cite{li2019cr} added spatial recurrent neural networks (RNNs) to the skip connections of a U-Net. Since there are usually many small follicles in an image, it is very likely that the neighboring follicles are spatially correlated. In addition, there might be a possible spatial correlation between the follicles and the ovary. As the max-pooling operation in the original U-Net brings a loss of spatially relative information the spatial RNNs should improve the segmentation results by learning multi-scale and long-range spatial contexts.

\begin{figure}
	\centering
	\includegraphics[width=\columnwidth]{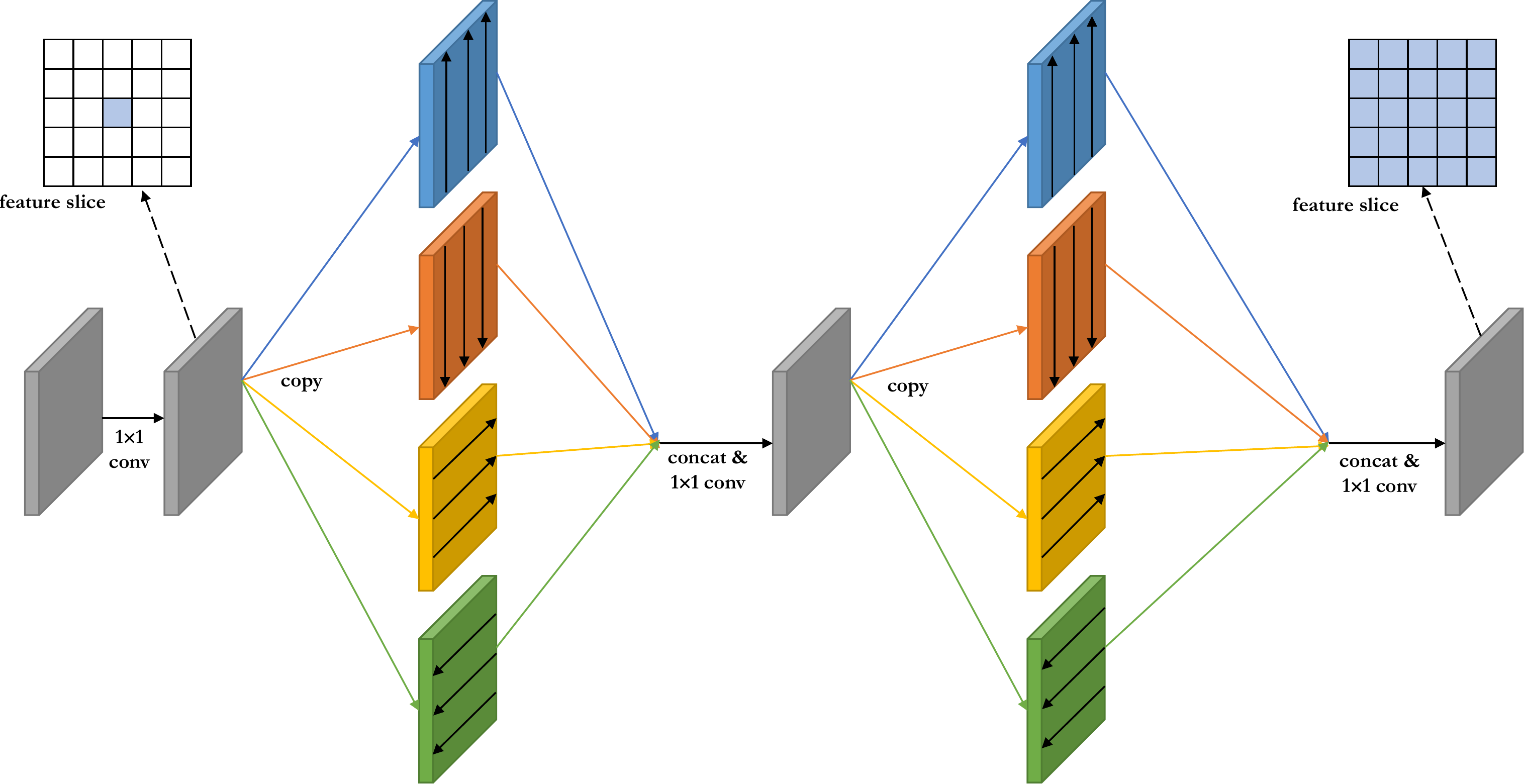}
	\caption{The spatial RNN module used in the skip connections of the CR-Unet by Li et al. in \cite{li2019cr}.}
	\label{fig:crUnet}
\end{figure}

Li et al. \cite{li2019cr} built the spatial RNNs from plain RNNs with a ReLU activation. Each spatial RNN module takes feature maps as input and produces spatial RNN features as output. It uses four independent data translations to integrate local spatial information in up, down, left, and right directions. The maps from each direction are concatenated and passed through a $1\times 1$ convolutional layer to produce feature maps where each point contains information from all four directions. The process is then repeated to extend the local spatial information to global contextual information. As can be seen in \Cref{fig:crUnet}, the final feature maps passed through the skip connection are a combination of the original encoder feature maps and the RNN features extracted from these maps. The authors claim that the architecture is especially strong at avoiding the segmentation of false positives and detecting and segmenting very small follicles. A limitation of the RNN modules is that they make training more difficult and computationally expensive. To compensate for this, Li et al. added deep supervision.

While most medical applications demand segmentations to be in the same dimension as the input image, there are also medical protocols that require segmentation of the image projection, e.g., Liefers et al. \cite{liefers2019dense} studied the retinal vessel segmentation as a $2$D $\rightarrow$ $1$D retinal OCT segmentation task. This adds the problem of dimensionality reduction to the segmentation. Lachinov et al. \cite{lachinov2021projective} introduced a U-Net with projective skip connections to handle $N$D $\rightarrow$ $M$D segmentations, where $M<N$.
\begin{figure}
	\centering
	\includegraphics[width=\columnwidth]{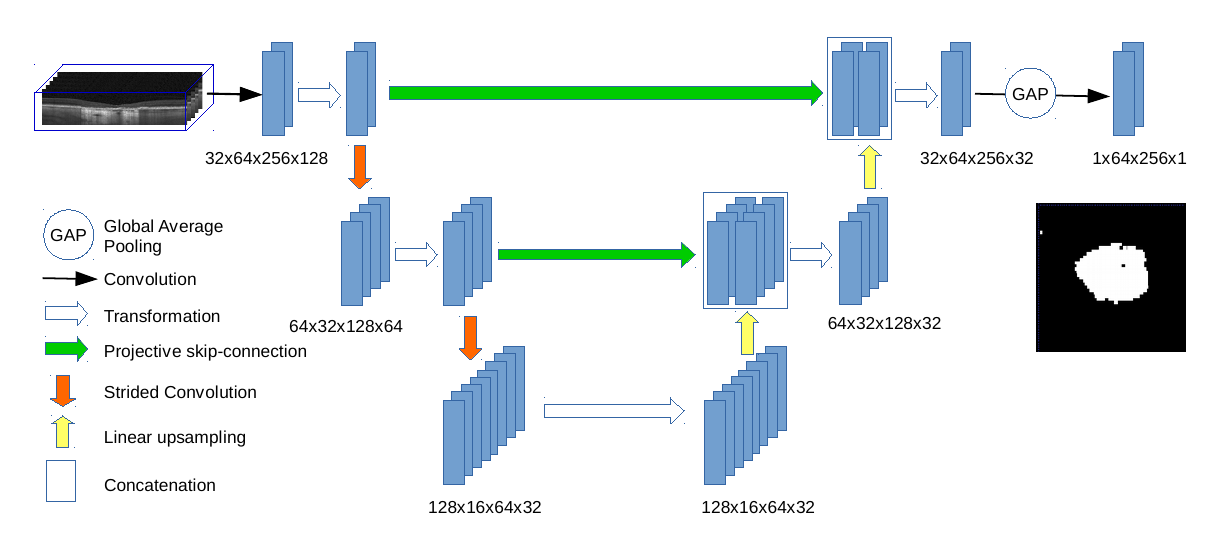}
	\caption{Architecture of U-Net with 3D input and 2D output segmentation. Figure from \cite{lachinov2021projective-arx}.}
	\label{fig:projective}
\end{figure}
The encoder is a classic U-Net encoder with residual blocks.
The decoder however only restores the input resolution for the $M$ dimensions of the segmentation. 
The remaining reducible dimensions $M< d \leq N$ are left compressed. This means that the sizes of the encoder and decoder feature maps no longer match which is why Lachinov et al. \cite{lachinov2021projective} introduce the projective skip connections.
The encoder feature maps passed along the projective skip connections are processed by an average pooling layer with varying kernel size so that the dimensions which are not present in the segmentation are reduced to the size they have in the bottleneck. 
This way they can be concatenated with the corresponding decoder feature maps. Global Average Pooling (GAP) and a convolutional layer are added after the last decoder level to calculate the final $M$D segmentation.
The overall architecture for $N = 3$ and $M = 2$ can be seen in \Cref{fig:projective}.
The third dimension is not upsampled to its original resolution in the decoder path and is finally reduced to one by the GAP.

\subsubsection{Combination of Encoder and Decoder Feature Maps}
\begin{figure}
	\centering
	\includegraphics[width=\columnwidth]{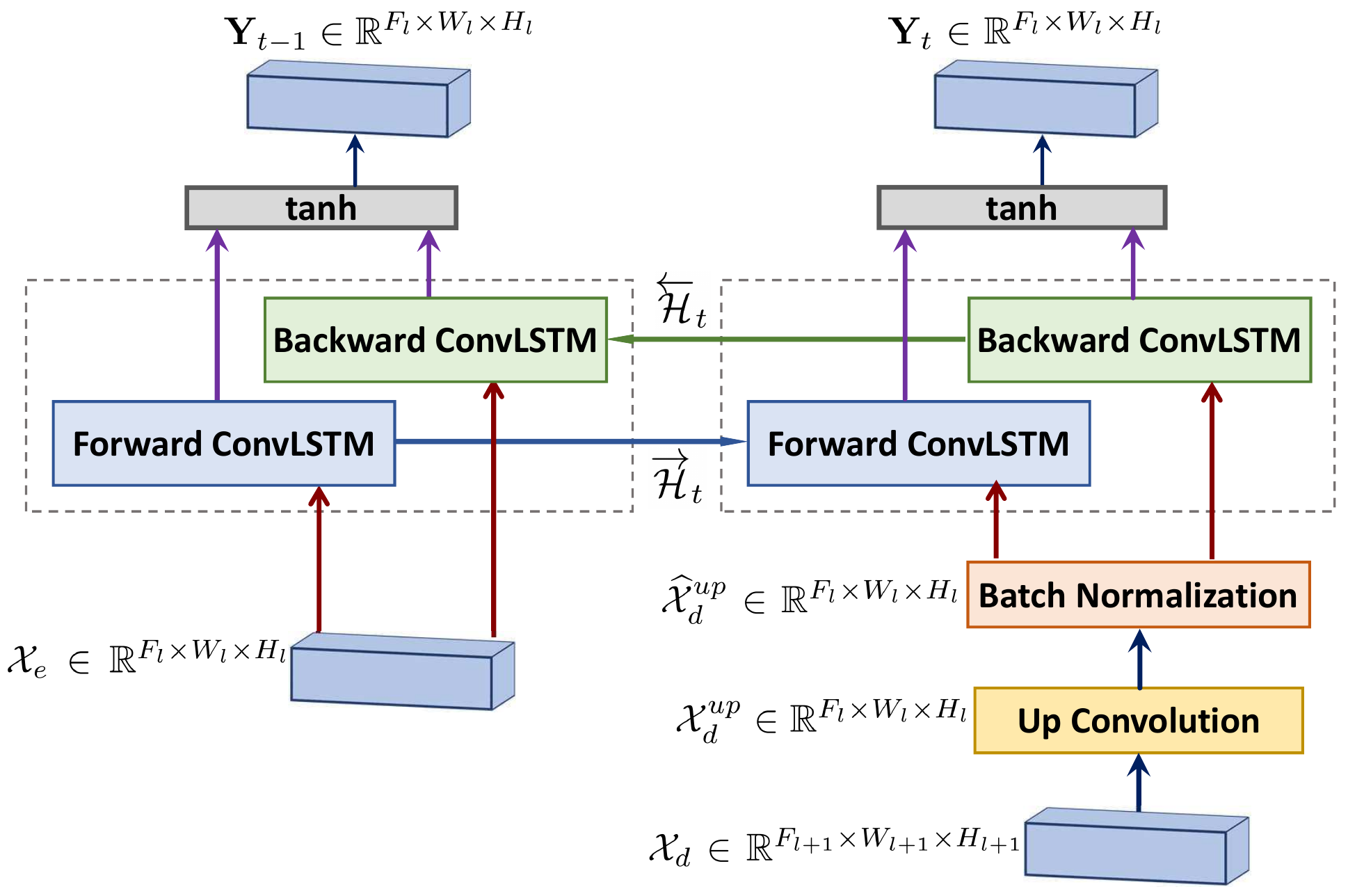}
	\caption{Bi-directional convolutional LSTM module for the combination of encoder feature maps passed through the skip connection and decoder feature maps from the previous up-convolutional layer. $\mathcal{X}_e$ and $\mathcal{X}_d$ represent the set of feature maps copied from the encoding part and the output of the previous scale's convolutional layer, respectively. $\mathbf{Y}_i$ indicates the output of the BConvLSTM module in a $i$-th block of a single skip connection. Figure from \cite{azad2019bi-arx}.}
	\label{fig:bcd-unet}
\end{figure}

Another extension of the classic skip connections is introduced in the BCDU-Net by Azad et al. \cite{azad2019bi} where a bi-directional convolutional long-term-short-term-memory (LSTM) module is added to the skip connections. Azad et al. argue that a simple concatenation of the high-resolution feature maps from the encoder and the feature maps extracted from the previous up-convolutional layer containing more semantic information might not lead to the most precise segmentation output. Instead, they combine the two sets of feature maps with non-linear functions in the bi-directional convolutional LSTM module. Ideally, this leads to a set of feature maps rich in both local and semantic information. The architecture of the bi-directional convolutional LSTM module used to combine the feature maps at the end of the skip connection can be seen in \Cref{fig:bcd-unet}.
It uses two ConvLSTMs, processing the input data in two directions in the forward and backward paths. The output will be determined by taking into consideration the data dependencies in both directions. In contrast to the approach by Li et al. \cite{li2019cr}, where only the encoder feature maps are processed by the RNN and then concatenated with the decoder features, this approach processes both sets of feature maps with the RNN.

\subsection{Backbone Design Enhancements} \label{sec:backbone-design}
Apart from adapting the skip connections of a U-Net it is also common to use different types of backbones in newer U-Net extensions. The backbone defines how the layers in the encoder are arranged and its counterpart is therefore used to describe the decoder architecture.

In the original U-Net by Ronneberger et al. \cite{ronneberger2015u} each level in the encoder consists of two $3\times 3$ convolutional layers with ReLU activation followed by a max pooling operation. The number of feature maps doubles at each level. Any 2D or 3D CNN image classifier can be used as an encoder in a U-Net adding its mirrored counterpart as the decoder. Dozens of studies modified the vanilla U-Net main blocks to broaden the receptive fields of convolution operations and extract rich, and fine-grained semantic representations for challenging multi-class problems, e.g., \cite{jha2020doubleu,li2022mfaunet,nguyen20213d,weng2021inet,lalonde2021capsules}. This section presents several prominent backbones used in the U-Net architecture and explains their benefits and downsides.

\subsubsection{Residual Backbone}
A very common backbone for the U-Net architecture is the ResNet initially developed by He et al. \cite{he2016deep}. Residual networks enable deeper network architectures by tackling the vanishing gradient problem that often occurs when stacking several layers in deep neural networks as well as a degradation problem that leads to first saturating and then degrading accuracy when adding more and more layers to a network. Residual building blocks, explicitly fit a residual mapping by adding skip connections and performing an identity mapping that is added to the output of the stacked layers.

In their implementation of a residual U-Net, Drozdzal et al. \cite{drozdzal2016importance} refer to the standard skip connections in the U-Net as long skip connections and the residual skip connections as short skip connections, as they only skip ahead over two convolutional layers. Using residual blocks as the backbone in a U-Net, Drozdzal et al. \cite{drozdzal2016importance} can build deeper architectures and find that the network training converges faster compared to the original U-Net. Milletari et al. \cite{milletari2016v} report the same findings in their 3D U-Net architecture using 3d residual blocks as the backbone.
\begin{figure}
\centering
\includegraphics[width=\columnwidth]{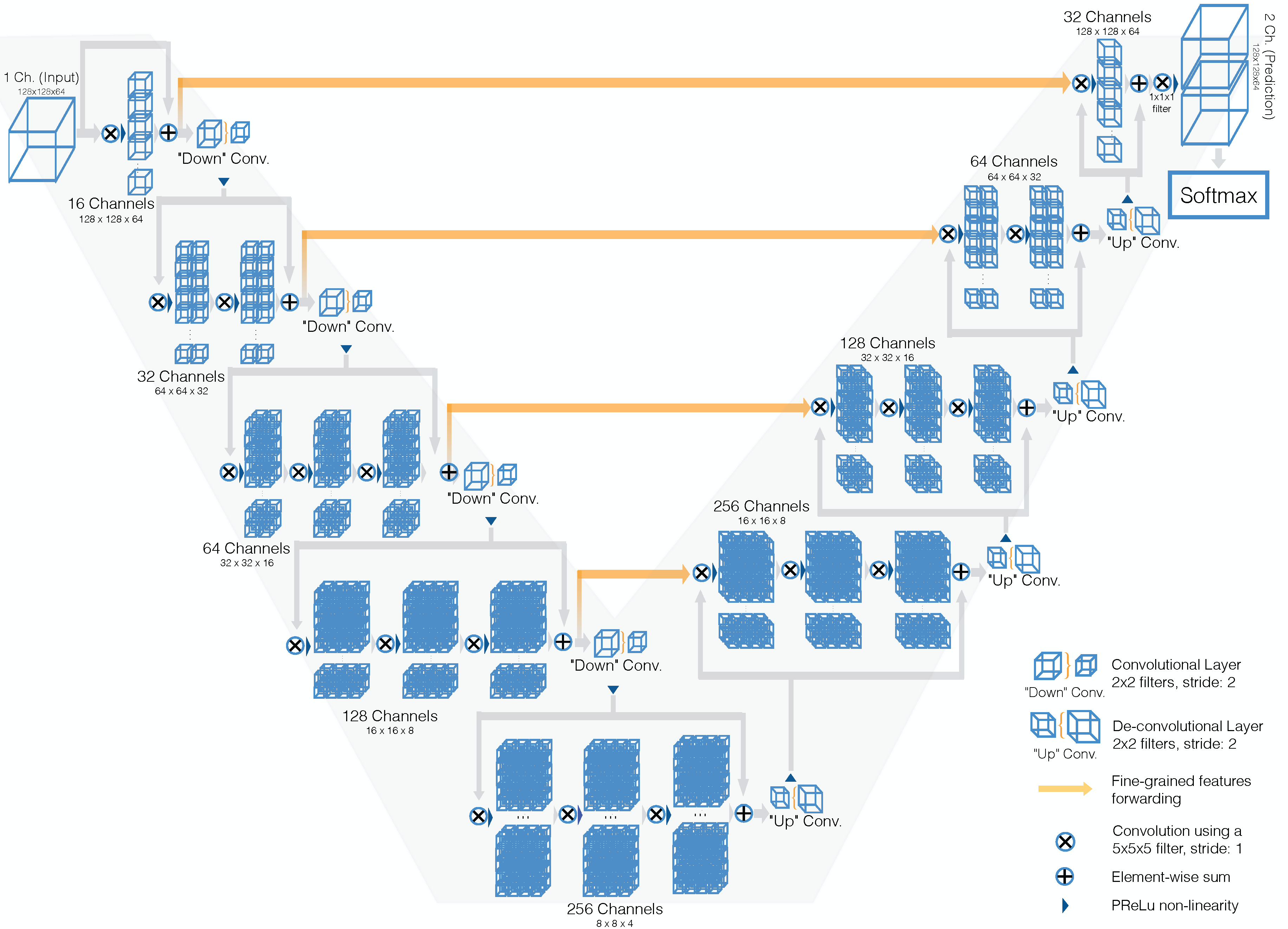}
\caption{Schematic representation of V-Net structure for volumetric biomedical image segmentation that comprises residual connections in each scale. In addition, V-Net utilizes the strided convolutional kernels rather than max-pooling layers to lessen the memory footprint in the training stage. Figure from \cite{milletari2016v-arx}.}
\label{fig:v-net}
\end{figure}

A prominent adaption of the backbone is to exchange all 2D convolutions with 3D convolutions to process an entire image volume as can often be found in medical applications. When processing a 3D image in a slice-wise fashion using 2D convolutions, the contexts on the z-axis can not be captured and learned by the network. Using fully convolutional architecture with 3D convolutions elevates this drawback and can fully leverage the spatial information along all three dimensions.

A drawback of using 3D convolutional layers as the backbone in a u-net is the high computational cost and GPU memory consumption which limits the depth of the network and the filter's size i.e. its field-of-view. Milletari et al. \cite{milletari2016v} fully convolutional volumetric, V-Net architecture uses 3D residual blocks (\Cref{fig:v-net}) as a backbone, thereby enabling fast and accurate segmentation in 3D images. The H-DenseUNet by Li et al. \cite{li2018h} uses two U-Nets, one with 2D-dense-blocks as the backbone and the other with 3D-dense-blocks as the backbone. This enables them to first extract deep intra-slice features and then learn inter-slice features in shallower volumetric architecture with a lower computational burden.

\subsubsection{Multi-Resolution blocks}
\begin{figure}
	\centering
    \begin{subfigure}{0.48\columnwidth}
        \includegraphics[width=\textwidth]{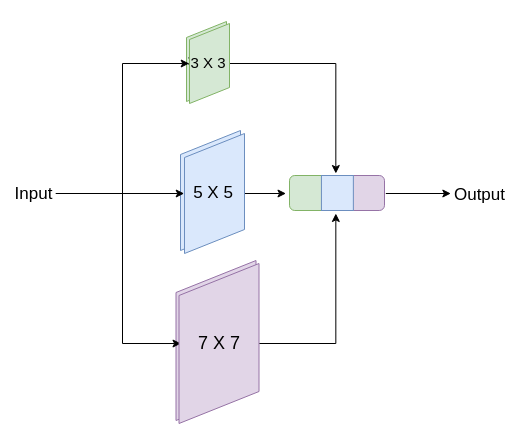}
        \caption{}
        \label{fig:multires_1}
    \end{subfigure}
    \hfill
    \begin{subfigure}{0.48\columnwidth}
        \includegraphics[width=\textwidth]{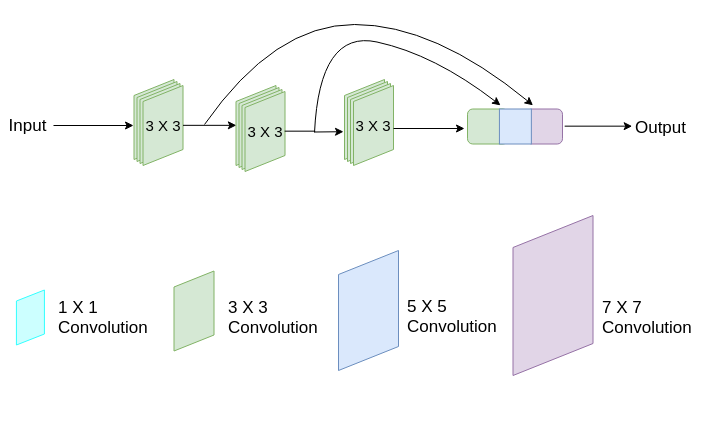}
        \caption{}
        \label{fig:multires_2}
    \end{subfigure}
    \hfill
    \begin{subfigure}{\columnwidth}
        \includegraphics[width=\textwidth]{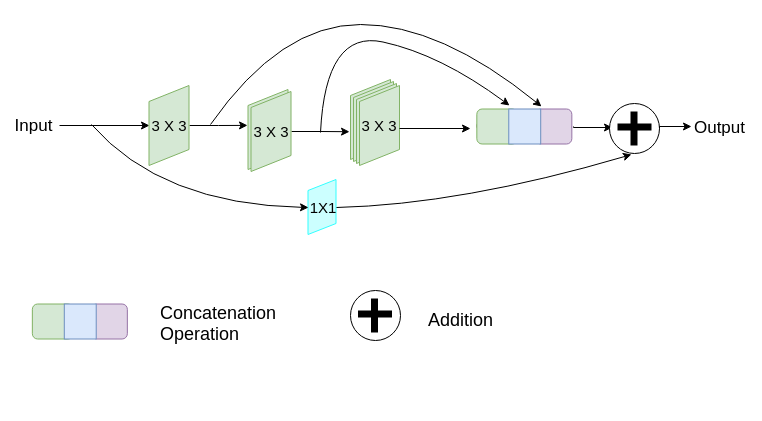}
        \caption{}
        \label{fig:multires_3}
    \end{subfigure}
	\caption{The MultiRes Block (c) is developed from the original inception Block (a) with three parallel convolutions with $3\times3$, $5\times5$, and $7\times7$ kernels by expressing the $5\times5$ and $7\times7$ convolutions as two and three consecutive $3\times3$ convolutions as can be seen in (b) and adding a residual skip connection. Figure from \cite{ibtehaz2020multiresunet-arx}.}
	\label{fig:multiresunet}
\end{figure}

To tackle the difficulty of analyzing objects at different scales, Ibtehaz et al. introduce the MultiResUNet with inception-like blocks as a backbone \cite{ibtehaz2020multiresunet}. Inception blocks, introduced by Szegedy et al. \cite{szegedy2016rethinking}, use convolutional layers with different kernel sizes in parallel on the same input and combine the perceptions from different scales before passing them deeper into the network. The two following convolutions with $3\times3$ kernels in the classical U-Net resemble one convolution with a $5\times5$ kernel. For incorporating a multi-resolution analysis into the network, $3\times3$ and $7\times7$ convolutions should be added in parallel to the $5\times 5$ convolution. This can be achieved by replacing the convolutional layers with inception-like blocks. Adding the additional convolutional layers increases the memory requirement and computational burden. Ibtehaz et al., therefore, formulate the more expensive $5\times5$ and $7\times7$ convolutions as consecutive $3\times3$ convolutions. The final \textit{MultiRes block} is created by adding a residual connection. The evolution from the original inception block to the MultiRes block can be seen in \Cref{fig:multiresunet}. Instead of keeping an equal number of filters for all consecutive convolutions, the number of filters is gradually increased to further reduce the memory requirements.
In the final architecture, the two consecutive $3\times3$ convolutions from the original U-Net are replaced by one MultiRes block, leading to faster convergence, improved delineation of faint boundaries, and higher robustness against outliers and perturbations.

Another well-known backbone for U-Net extensions is the DenseNet introduced by Huang et al. in \cite{huang2017densely}. Similarly to residual networks, the DenseNet also aims at fighting the vanishing gradient problem by creating skip connections from early layers to later layers. The DenseNet maximizes the information flow by connecting all layers with the same feature map size with each other. This means that every layer obtains concatenated inputs from all preceding layers. Contrary to what one might expect, a dense net actually requires fewer parameters compared to a traditional CNN because it does not have to relearn redundant feature maps and can therefore work with very narrow layers with e.g. only 12 filters and can learn multi-resolution features. The direct connection from each layer to the loss function implements implicit deep supervision which helps train deeper network architectures without vanishing gradients.

Karaali et al. \cite{karaali2022dr} utilized Dense Residual blocks in the U-Net-like representation for retinal vessel segmentation. To this end, they were inspired by DenseNet \cite{huang2017densely}, and ResNet \cite{he2016deep} to design a Residual Dense-Net (RDN) block. In their architecture the first sub-block comprises successive batch Normalization, ReLu, Convolution, and Dropout counterparts, which employs the dense connectivity pattern as in \cite{huang2017densely}. The following sub-block applies a residual connectivity pattern. Using a DenseNet-like backbone helps the U-Net architecture learn more relevant features using fewer parameters. The residual connectivity smooths the information flow across the layers to facilitate the optimization step.

\subsubsection{Re-considering Convolution}
\begin{figure}
	\centering
	\includegraphics[scale=0.3]{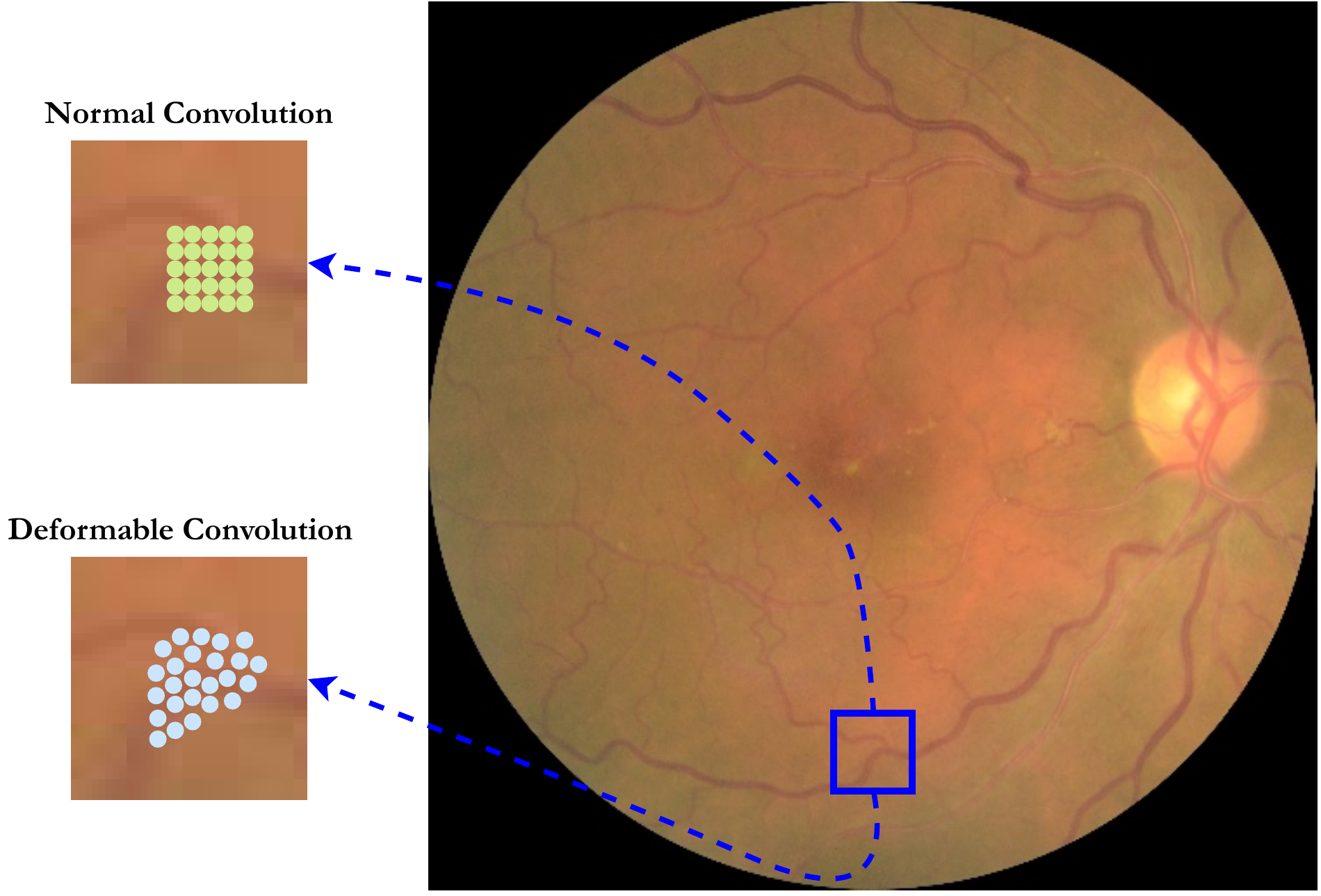}
	\caption{One exemplary deformable convolutions with their respective normal convolution with a $5\times5$ kernel.}
	\label{fig:dunet}
\end{figure}

This direction aims to reduce the computational burden of the naive convolution operation by re-considering the alternative convolutional operations. Jin et al. \cite{jin2019dunet} exchange each $3\times3$ convolutional layer in the original U-Net with a deformable convolutional block for the accurate segmentation of retinal vessels. Their architecture is named DUNet. The deformable convolutional blocks are inspired by the work on deformable convolutional networks by Dai et al. \cite{dai2017deformable} and should adapt the receptive fields to adjust optimally to different shapes and scales of complicated vessel structures in the input features. In deformable convolutions, offsets are learned and added to the grid sampling locations normally used in the standard convolution. One exemplary illustration of adjusted sampling locations for a $5\times 5$ kernel can be seen in \Cref{fig:dunet}.

In a classic convolution the kernel sampling grid $G$ would be defined as:
\begin{equation}
	G = {(-2, -2), (-2, -1), ..., (2, 1), (2, 2)}.
\end{equation}
Considering this grid, every pixel $m_0$ in the output feature map $\mathbf{y}$ can be calculated as:
\begin{equation}
	\mathbf{y}(m_0)=\sum_{m_i\in G}\mathbf(w)(m_i)\cdot \mathbf{x}(m_0+m_i)
\end{equation}
from the input $\mathbf{x}$. In the deformable convolution, an offset $\Delta m_i$ is added to the grid locations.
\begin{equation}
	\mathbf{y}(m_0)=\sum_{m_i\in G}\mathbf(w)(m_i)\cdot \mathbf{x}(m_0+m_i+\Delta m_i) 
\end{equation}\\
Every deformable convolutional block consists of a convolutional layer, to learn the ideal offsets from the input. A deformable convolution layer applying the convolution with the adapted sampling points followed by batch normalization and ReLU activation. Since the calculated offset $\Delta m_i$ is usually not an integer, the input value at the sampling point is determined using bilinear interpolation. Exchanging the simple convolutions with deformable convolutions helps the network adapt to different shapes, scales, and orientations but comes at a higher computational burden because an additional convolutional layer per block is needed to determine the offsets of the sampling grid.

\begin{figure}
	\centering
	\includegraphics[scale=0.5]{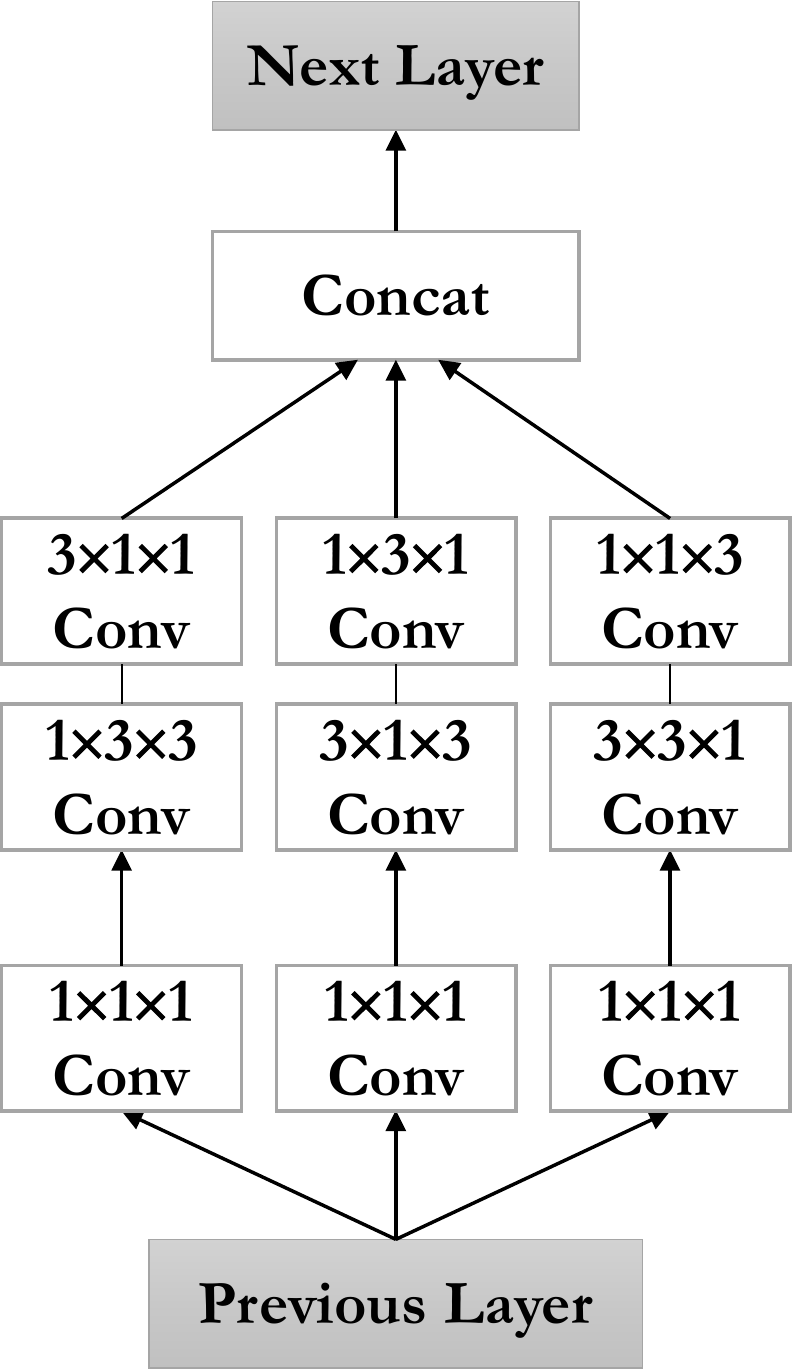}
	\caption{Architecture of one separable 3D convolutional block \cite{chen2018s3d}. Using parallel design the architecture performs the 3D convolution in three paths with less computation burden. }
	\label{fig:separable}
\end{figure}

When segmenting from 3D images it is important to make use of the full spatial information from the volumetric data. However, this is not possible with 2D convolutions and 3D convolutions are computationally very expensive. To address this problem, Chen et al. \cite{chen2018s3d} used separable 3D convolutions as the backbone of the U-Net. Each 3D convolutional block in the original U-Net is replaced by an S3D block which can be seen in \Cref{fig:separable}.
The 3D convolution is divided into three branches where each branch represents a different orthogonal view so that the input is processed in axial, sagittal, and coronal views. Additionally, a residual skip connection is added to the separated 3D convolution. Using separable 3D convolutions as the backbone of the U-Net, Chen et al. \cite{chen2018s3d} can take into consideration the full spatial information from the volumetric data in the U-Net architecture without the extremely high computational burden of standard 3D convolution.

\subsubsection{Recurrent Architecture}
Recurrent neural networks (RNN) are used frequently to process sequential data such as in speech recognition. Liang et al. \cite{liang2015recurrent} were among the first groups to design a recurrent convolutional neural network (RCNN) for images recognition. Although the input image, in contrast to sequential data, is static, the activity of each unit is modulated by the activities of its neighboring units because the activities of RCNNs evolve over time. By unfolding the RCNN through time, they can obtain arbitrarily deep networks with a fixed number of parameters.

Using these RCNN blocks as the backbone of the U-Net architecture enhances the ability of the model to integrate contextual information. Alom et al. \cite{alom2019recurrent} used RCNN blocks as a backbone in their RU-Net architecture, ensuring better feature representation for segmentation tasks.

\begin{figure}[ht!]
	\centering
	\begin{subfigure}[][][c]{0.48\columnwidth}
		\centering
		\includegraphics[scale=0.5]{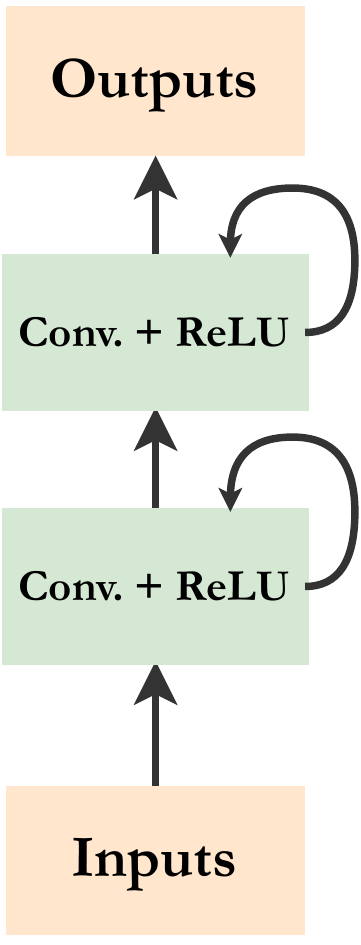}
		\caption{Architecture of double recurrent convolutional block.}
		\label{fig:runet1}
	\end{subfigure}
	\hfill
	\begin{subfigure}[][][c]{0.50\columnwidth}
		\centering
		\includegraphics[scale=0.5]{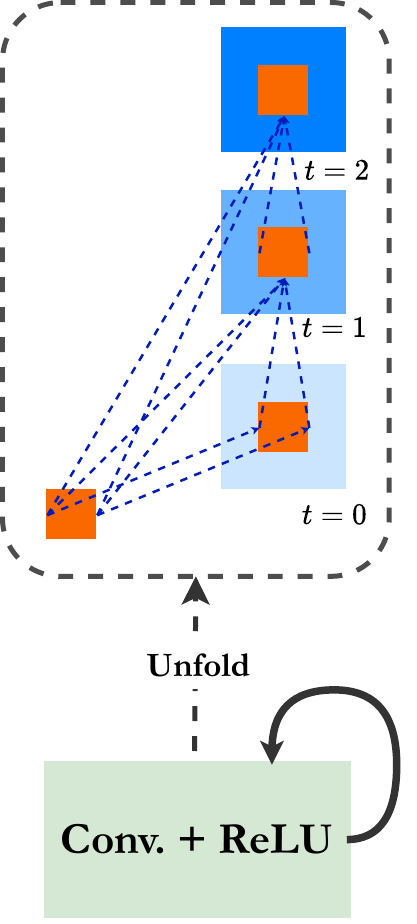}
		\caption{A single unfolded recurrent convolutional block for $t=2$.}
		\label{fig:runet2}
	\end{subfigure}
	\caption{Recurrent Convolutional Mechanism introduced by Alom et al. \cite{alom2019recurrent}, namely RU-Net.}
\end{figure}
\Cref{fig:runet1} shows a recurrent convolutional unit, they used as a backbone.
\Cref{fig:runet2} shows one of the two sub-blocks in \Cref{fig:runet1} unfolded for $t=2$, which is also the unfolding parameter chosen in their experiments. Adding additional residual connections to the separate recurrent convolutional blocks enables deeper networks and results in their R2U-Net architecture.

\subsection{Bottleneck Enhancements} \label{sec:bottleneck}
The U-Net architecture can be separated into three main parts: the encoder (contracting path), the decoder (expanding path), and the bottleneck which lies between the encoder and decoder. The bottleneck is used to force the model to learn a compressed representation of the input data which should only contain the important and useful information needed to restore the input in the decoder. To this end, various modules are designed in multiple studies \cite{zhang2022multi,azad2021deep} to recalibrate and highlight the most discriminant features. In the original U-Net, the bottleneck consists of two $3\times 3$ convolutional layers with ReLU activation. More recent approaches however have extended the classic bottleneck architecture to improve performance.

\subsubsection{Attention Modules}
\begin{figure}
	\centering
	\includegraphics[width=\columnwidth]{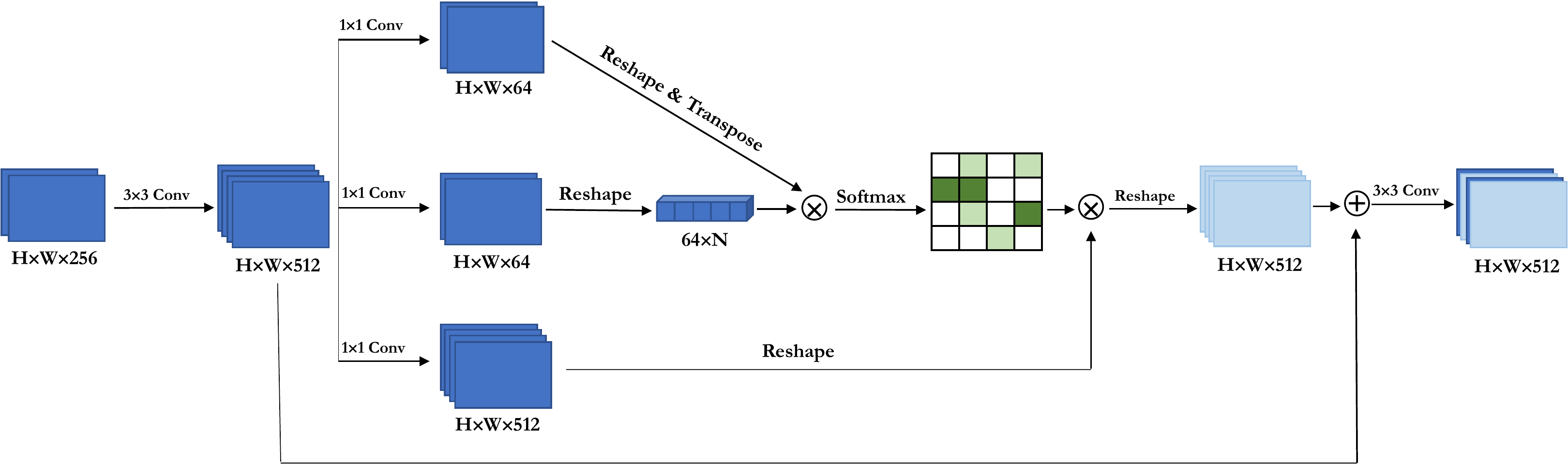}
	\caption{Architecture of the position-wise attention block introduced in \cite{fan2020ma}.}
	\label{fig:MAnet}
\end{figure}

Several works apply attention modules in the bottleneck of their U-Net architecture.
Fan et al. used a position-wise attention block (PAB) in their MA-Net to model spatial dependencies between pixels in the bottleneck feature maps with self-attention \cite{fan2020ma}.
The architecture of the PAB can be seen in \Cref{fig:MAnet}.
The feature maps passed into the bottleneck at the end of the encoder path are first processed by a $3\times 3$ convolutional layer.
The resulting outputs are then processed by three individual $1\times 1$ convolutional layers producing $A$, $B$, and $C$. 
$A$ and $B$ are reshaped to form two vectors.
A matrix multiplication of these two vectors passed through a softmax function yields the spatial feature attention map $P \in \mathbb{R}^{N\times N}$ in which the positions $p_{i, j}$ encode the influence of the $i^{th}$ position on the $j^{th}$ position in the feature map.
Subsequently, a matrix multiplication is performed between the reshaped $C$ and the spatial feature attention map $P$, and the resulting feature maps are multiplied with the input $I'$ before being passed through a final $3\times 3$ convolutional layer.
The final output $O$ is therefore defined as follows: 
\begin{equation}
	O_i = \alpha \sum_{i=1}^N(P_{ji}C_i)+I'_j
\end{equation}
$\alpha$ is set to zero at the beginning of training and it is learned to assign more weight during the training process.
Considering that the final output is the weighted sum of the feature maps across all positions and the original feature maps, it has a global contextual view and can selectively aggregate rich contextual information.
Intra-class correlation and semantic consistency are improved because the PAB can consider long-range spatial dependency between features in a global view.

Guo et al. also add a spatial attention module to the bottleneck of their SA-UNet architecture \cite{guo2021sa}. The spatial attention module should enhance relevant features and compress unimportant features in the bottleneck. In their approach the input feature maps are passed through an average pooling and a max pooling layer in parallel. Both pooling operations are applied along the channel dimension to produce efficient feature descriptors. The outputs are then concatenated and passed through a $7 \times 7$ convolutional layer and sigmoid activation to obtain a spatial attention map.
By multiplying the spatial attention map with the original input features, the inputs can be weighted based on their importance for the segmentation task at hand.
The attention module only adds 98 parameters to the original U-Net and is therefore computationally very lightweight.

In another work, Azad et al. \cite{azad2022smu} utilized the idea of a texture/style matching mechanism in the U-Net bottleneck for brain tumor segmentation. In their design, an attention agent is designed to distill the informative information from a full modality (four MRI modalities, T1, T2, Flair and T1c) into a missing-modality network (only Flair). Further information regarding the missing-modality task can be found in \cite{azad2022medical}. A deep frequency attention module is proposed in \cite{azad2021deep} to perform a frequency recalibration process on the U-Net bottleneck. This attention block aims to recalibrate the feature representation based on the structure and shape information rather than texture representation to alleviate the texture bias in object recognition.

\subsubsection{Multi-Scale Representation}
\begin{figure}
	\centering
    \begin{subfigure}[t][][t]{0.22\columnwidth}
        \centering
        \includegraphics[scale=0.5]{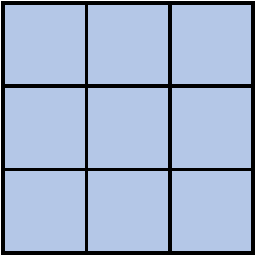}
        \caption{} \label{fig:atrous1}
    \end{subfigure}
    \hfill
    \begin{subfigure}[t][][t]{0.42\columnwidth}
        \centering
        \includegraphics[scale=0.5]{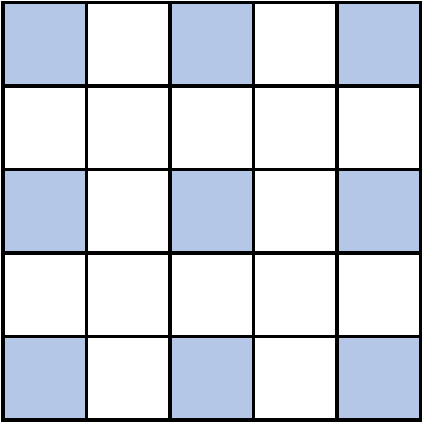}
        \caption{} \label{fig:atrous2}
    \end{subfigure}
    \hfill
    \begin{subfigure}[t][][t]{0.32\columnwidth}
        \centering
        \includegraphics[width=\textwidth]{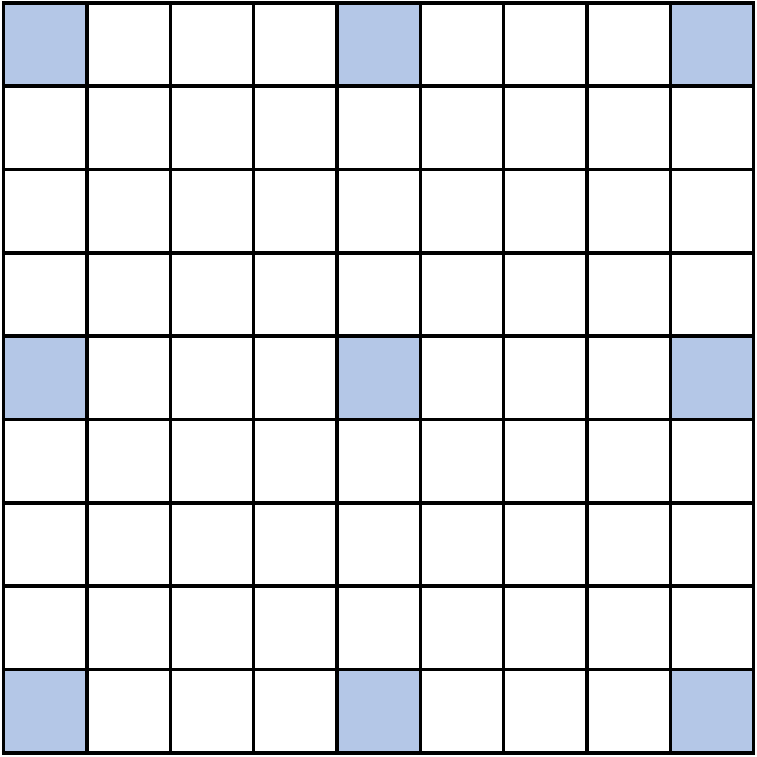}
        \caption{} \label{fig:atrous3}
    \end{subfigure}
	\caption{$3\times 3$ convolutional kernel with different atrous sampling rates $r$ of (a) $r=1$, (b) $r=2$ and (c) $r=4$.}
	\label{fig:atrous}
\end{figure}

The aim of this direction is to enhance the bottleneck design by including multi-scale feature representation, e.g. atrous convolution. The atrous convolutions are performed like standard convolutions, but with convolutional kernels with inserted holes in them.
The holes are defined by setting the weight of the convolutional kernel to zero at the corresponding locations and the pattern for doing so is defined by the atrous sampling rate $r$.
Considering a sampling rate $r$, this introduces $r-1$ zeros between consecutive filter values.
A $k\times k$ convolutional kernel is thereby enlarged to a $k+(k-1)*(r-1)\times k+(k-1)*(r-1)$ filter.
This way the receptive field of the layer is expanded without introducing any additional network parameters to be learned.\\
\Cref{fig:atrous} shows a $3\times 3$ kernel with atrous sampling rates $r$ of $r=1$, $r=2$ and $r=4$.
When the objects to be segmented are of very different sizes it is important for the network to extract multiscale information.
Combining the ideas of spatial pyramid pooling and atrous convolutions, the feature maps in the bottleneck of the U-Net can be resampled in parallel by atrous convolutions with different sampling rates and then combined to obtain rich multiscale features.

Hai et al. \cite{hai2019fully} use atrous spatial pyramid pooling (ASPP) in the bottleneck of a U-Net architecture for the segmentation of breast lesions. The final feature maps of the encoder are passed in parallel through a $1\times 1$ convolutional layer and three atrous $3\times 3$ convolutional layers with atrous sampling rates of 6, 12 and 18 respectively. These four processed groups of feature maps are concatenated together with the original feature maps passed to the bottleneck and processed by a final $1\times 1$ convolution before being passed to the decoder.

Wang et al. make use of ASPP in the bottleneck as well in their COPLE-Net for the segmentation of pneumonia lesions from CT scans of COVID-19 patients \cite{wang2020noise}. Here, four atrous convolutional layers with dilation rates of 1, 2, 4, and 6 respectively are used to process the bottleneck feature maps to capture multi-scale features for the segmentation of small and large lesions.

Similarly, Wu et al. \cite{wu2021jcs} proposed a multi-task learning paradigm, JCS, for COVID-19 CT image classification and segmentation. JCS \cite{wu2021jcs} is a two branches architecture, which utilizes a Group Atrous (GA) module, in its segmentation branches bottleneck for feature modification. GA first applies $1\times 1$ convolution operation to expand the channels of the feature map. Then the feature map is divided into four equal sets. Utilizing the atrous convolutions with different rates on these sets results in more global feature maps with diverse receptive fields. To fully extract more discriminant features from the final feature map, JCS adopts a squeeze and Excitation (SE) \cite{hu2018squeeze} block as an attention mechanism for recalibrating channel-wise convolution features.

\begin{figure*}[th!]
	\centering
	\begin{subfigure}[t][][c]{0.48\textwidth}
		\includegraphics[width=\textwidth]{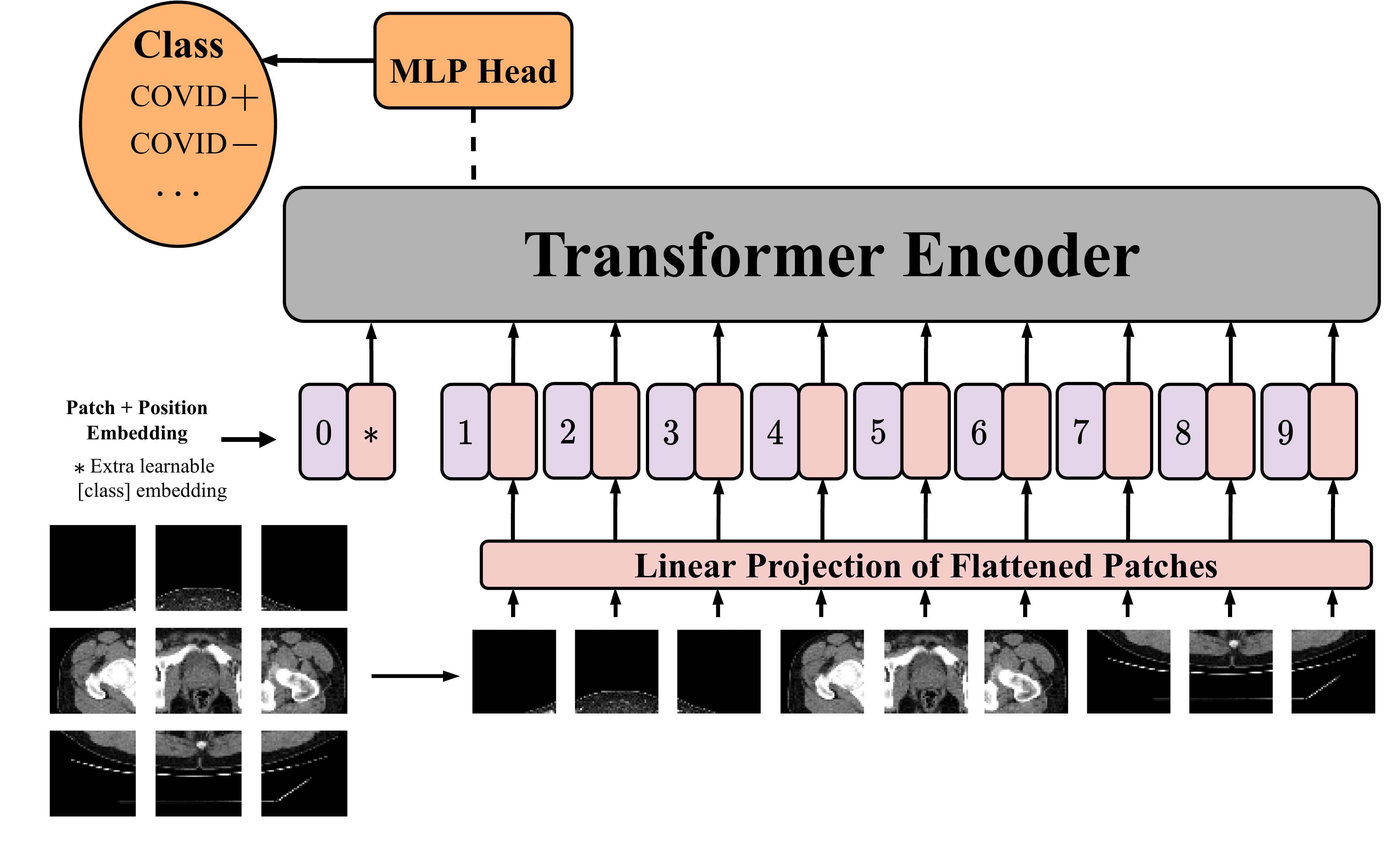}
		\caption{Overview of preliminary Vision Transformer (ViT) structure proposed by Dosovitskiy et al. \cite{dosovitskiy2020image}. ViT splits an image into fixed-size patches with a non-overlapping regime and then linearly embeds each of them to a 1D vector space, afterward adds position embeddings, and feeds the resulting sequence of vectors to a standard Transformer encoder. In order to perform classification, naive ViT uses the standard approach of adding an extra learnable \textit{classification token} to the sequence. However, this token is not practical in the segmentation literature, although the MLP head (from the dashed line in the above figure) is omitted in segmentation tasks.}
		\label{fig:vit}
	\end{subfigure}
	\hfill
	\begin{subfigure}[t][][c]{0.48\textwidth}
		\includegraphics[width=\textwidth]{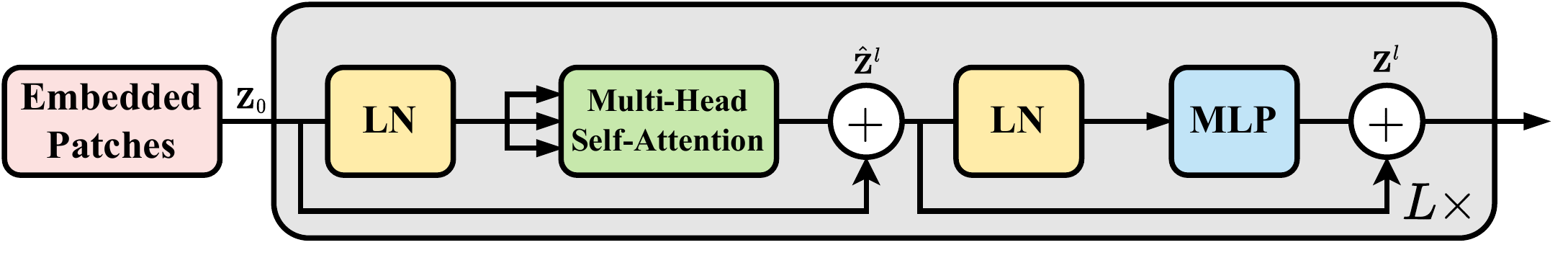}
		
		\includegraphics[width=\textwidth,height=0.7\textwidth]{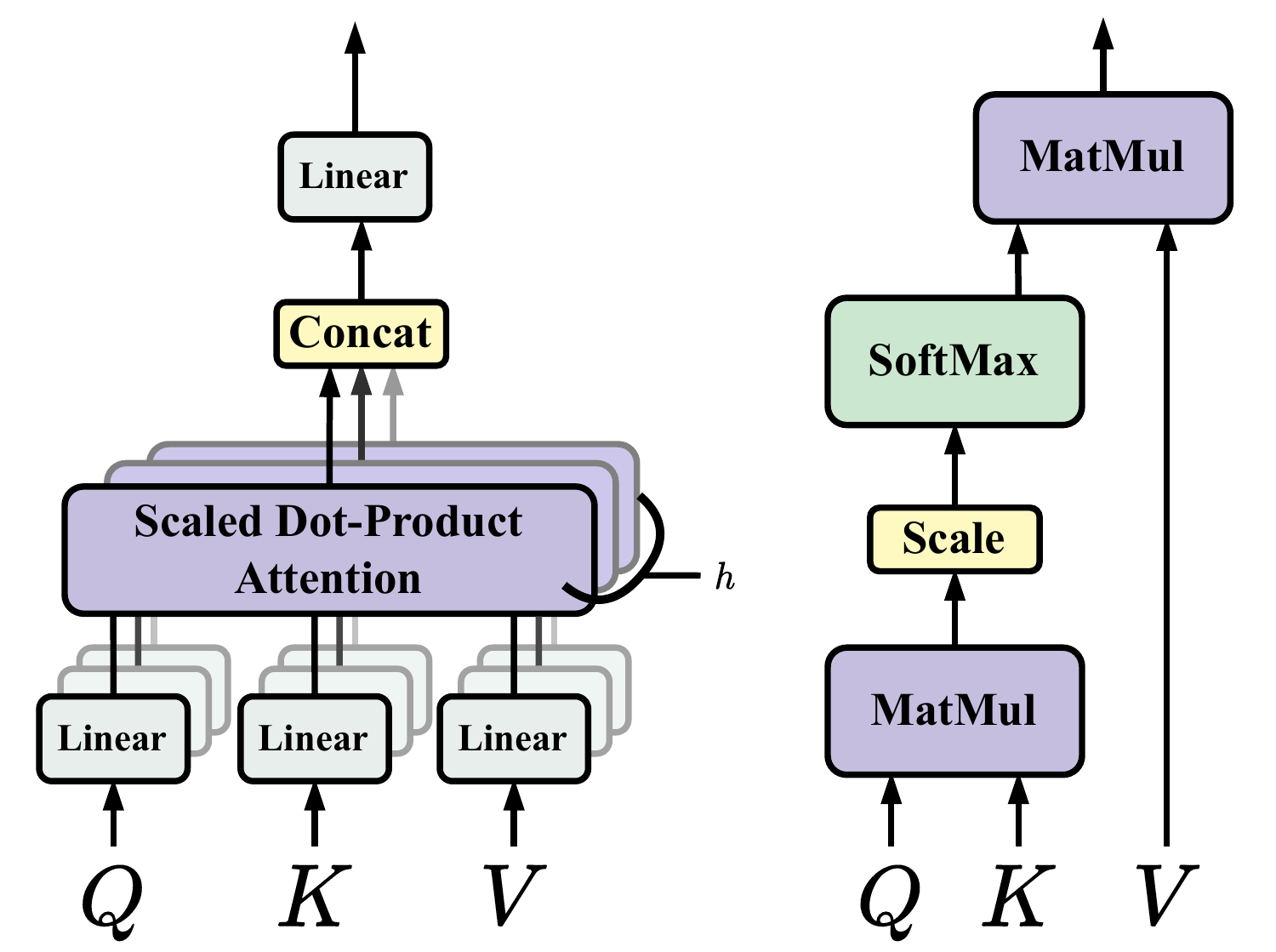}
		\caption{Up: Transformer Encoder block visualization, down-left: Multi-Head Self-Attention (MHSA) block, down-right: Attention mechanism block feed by the query (Q), key (K), and value (V) representations learned from the input embedded patches. LN and MLP denote the Layer Normalization \cite{ba2016layer} and Multi-Layer Perceptron operations, respectively.}
		\label{fig:transformer-encoder}
	\end{subfigure}
	\hfill
	\caption{The above figure portrays the ViT \cite{dosovitskiy2020image} pipeline with all of its detailed counterparts from a top-down view.}
	\label{fig:transformer}
\end{figure*}

\subsection{Transformers} \label{sec:transformers}
Inspired by the recent success of the Transformer models in Natural Language Processing (NLP), these models were further extended to perform vision recognition tasks. More specifically, the Vision Transformer model was introduced by Dosovitskiy et al. \cite{dosovitskiy2020image} to alleviate the deficiency of CNNs in capturing the long-range semantic dependencies.
Before going deeper into transformer-based methods, it might be practical to review the concept of vision transformers and the mechanism of self-attention utilized in these networks.

Contrary to the Transformers in NLP tasks  \cite{vaswani2017attention}, the computer vision tasks usually contains more than one dimensional data (e.g., 2D image, 3D video) which needs to be prepared for the transformer model. Hence, ViT's pipeline starts with image sequentialization (see \Cref{fig:vit}) process to prepare the tokenized sequence for the encoder module. From now on, the words \textbf{patch} and \textbf{token} will be used interchangeably.

If $x\in\mathbb{R}^{H\times W \times D \times C}$ is a volumetric 3D image with a $\left(H,W,D\right)$ spatial resolutions and $C$ input channels, first the $x$ is dividing into $N=\frac{H \times W \times D}{P^3}$ flattened uniform, non-overlapping patches $x_{p}^{i}\in \mathbb{R}^{N\times(P^3.C)}$ , $i \in \{1,\cdots,N\}$ with $(P,P,P)$ spatial resolution for each patch, therefore each patch is representing by a 1D sequence with a length of $1\times (P^3.C)$. Afterward, a linear layer applies on top of the sequence to map them to a $K$ dimensional embedding space. In order to retain the positional information of patches, a 1D learnable positional encoding $\mathbf{E}_{\text{pos}} \in \mathbb{R}^{N\times K}$ adds to patch embedding as follows:
\begin{align}
	\mathbf{z}_0=\left[x_{\text{class}}; x_p^1\mathbf{E};x_p^2\mathbf{E};\cdots;x_p^N\mathbf{E}\right] + \mathbf{E}_{\text{pos}}, \label{eq:patch-embedding}
\end{align}
where $\mathbf{E}$ denotes patch embedding operation and the class token,$x_{\text{class}}$, omitable in segmentation tasks. In the next step, the embedded patch feed to the stack of Transformer encoder blocks ($L \times$) containing the Multi-Head Self-Attention (MSHA), Multi-Layer
Perceptron (MLP), and Layer Normalization \cite{ba2016layer} sub-blocks to generate the latent representation. The following formulations show the mathematical process in Transformer encoder:
\begin{align}
	\hat{\mathbf{z}}^{l}&=\text{MSA}(\text{LN}(\mathbf{z}^{l-1}))+\mathbf{z}^{l-1} \notag \\ 
	\mathbf{z}^{l} & = \text{MLP}(\text{LN}(\hat{\mathbf{z}}^{l})) + \hat{\mathbf{z}}^{l}, \label{eq:multi-head-attention-block}
\end{align}
where $l \in \{1,\cdots,L\}$, $\hat{\mathbf{z}}^{l}$, and $\mathbf{z}^{l}$ denote output of MHSA operation and MLP function, respectively.

From \Cref{fig:transformer-encoder} the MHSA block comprises $h$ parallel Self-Attention sub-blocks that perform the attention (Scaled Dot-Product Attention) $h$ times with different \textbf{Q}eury ($Q$), \textbf{K}ey ($K$), and \textbf{V}alue ($V$) matrices from the input 1D sequence, $\mathbf{z}^{l} \in \mathbb{R}^{N\times K}$. The attention function is a mapping operation between query and key-value pairs to an output that measures the similarity between two components in $\mathbf{z}$ as:
\begin{align}
	\text{SA}(\mathbf{z})=\text{SoftMax}(\frac{QK^{T}}{\sqrt{K_h}})V,\label{eq:attention-eq}
\end{align}
where $\sqrt{K_h}$ denotes a normalization factor to preserve the attention matrix (\Cref{eq:attention-eq}) from the possible gradient vanishing or exploding through the training. Furthermore, the output of MHSA derives from the concatenation of multiple heads:
\begin{align}
	\text{MHSA}=\text{Linear}([\text{SA}_1(\mathbf{z});\text{SA}_2(\mathbf{z});\cdots;\text{SA}_h(\mathbf{z})]).
\end{align}
So far, we have briefly introduced the ViT pipeline and the related mathematics. In the next sections, we will discuss the integration of the Transformer into the U-Net structure in medical segmentation. We categorized the presence of Transformers in U-shaped networks into two sub-categories: (a) Transformer as a complement to CNN-based U-Net-like structures and (b) U-shaped standalone Transformer architectures.

\begin{figure}[!h]
	\centering
    \includegraphics[width=\columnwidth]{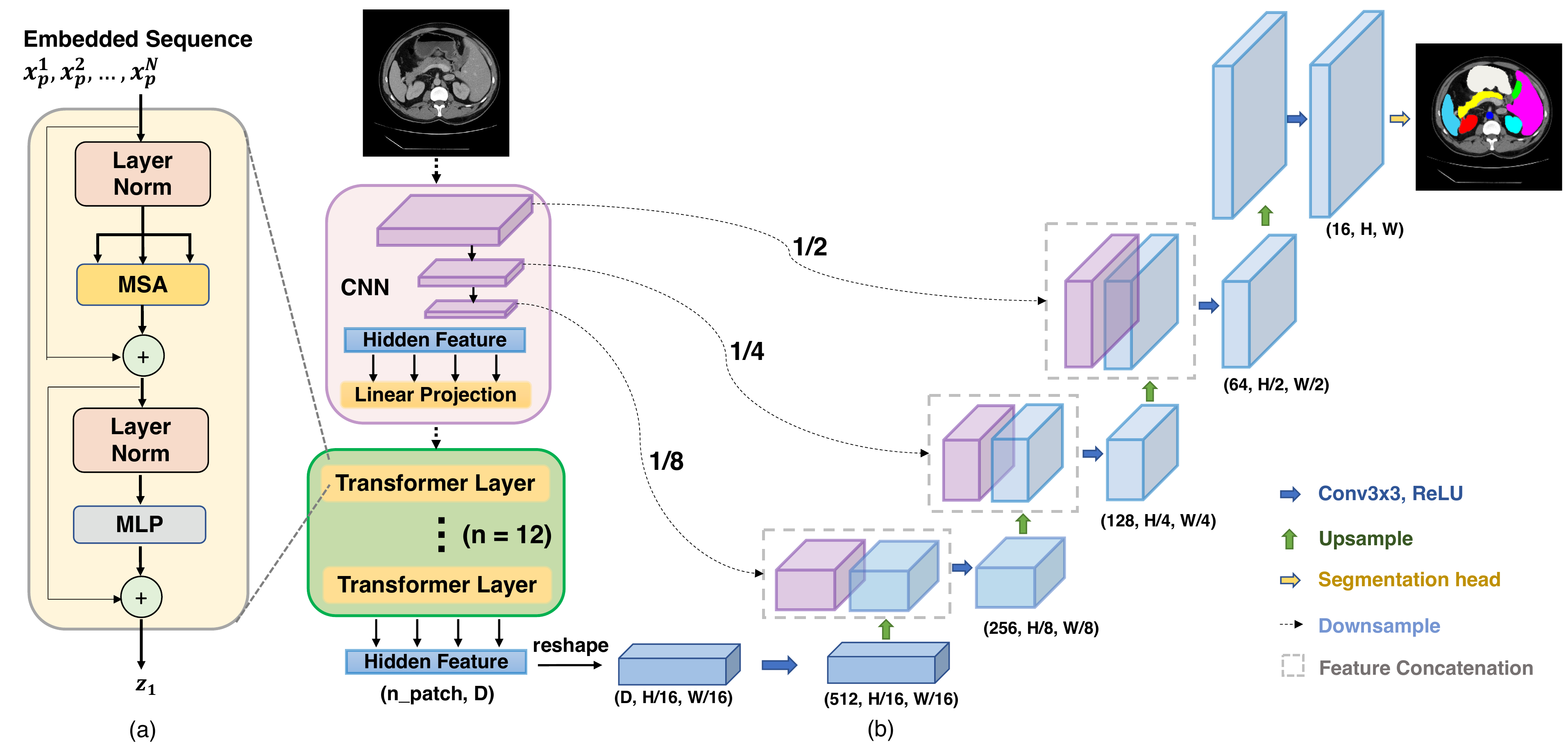}
    \caption{The overview architecture of the TransUNet. The Transformer layers are employed in the encoder part. The schematic of the Transformer is shown on the left. Figure from \cite{chen2021transunet}.}
    \label{fig:transunet}
\end{figure}	

\subsubsection{Transformer Complement to CNN-based U-Net} \label{sec:hybrid-transformer}
The success of convolutional neural networks (CNN) in diverse dense prediction tasks in the vision domain, e.g., segmentation, is noticeable. Their performance is underlined in their multi-scale representation and ability to capture local semantic and texture information. However, the local representation derving from the CNN architecture might not be robust enough to capture geometrical and structural information existing in the medical data. Therefore, there is a need for a mechanism to capture inter-pixel long relations to extend the performances of the existing CNN-based U-Net variants suffering from the limited receptive field of convolutional operations. Chen et al. \cite{chen2021transunet} proposed one of the first studies that utilized the Vision Transformer (ViT) in the U-Net structure to compensate for the U-Net's disability in long-range modeling dependencies, namely TransUNet (See \Cref{fig:transunet}). The stacked Transformers in the encoder path feed with the tokenized paths from abstract features extracted within the primal input to extract global contexts. The decoder path upsamples the encoded features combined with the high-resolution CNN feature maps to enable precise localization. Chen et al. clarified that the naive Transformer is not fit for downstream tasks like segmentation well due to its 1D functionality for capturing the interaction of the tokenized information. Therefore, they proposed this complementary Transformer design with U-Net, which they conducted several ablation studies to prove their superiority within the conventional attention collaborated networks such as Attention U-Net \cite{oktay2018attention} on Synapse \cite{synapse2015ct} and ACDC \cite{bernard2018deep} datasets.

TransUNet is a 2D network that processes the volumetric 3D medical image slice-by-slice, and due to its seminal ViT adaptation for its building blocks, it relies on pre-trained ViT models on large-scale image datasets. These restrictions made Wang et al. \cite{wang2021transbts} point them out and propose TransBTS as a U-Net-like architecture, modeling local and global information in spatial and slice/depth dimensions. While Transformer's computational complexity is quadratic and the volumetric 3D data is large, on the other hand, ViT's fixed-size tokenization process \cite{dosovitskiy2020image} discards the local structural information, TransBTS utilizes the 3D CNN backbone for its encoder and decoder path to capture local representation across spatial and depth dimensions and unleash from the high computational burden for the Transformer counterpart in overall. The essential key points in the amalgamation of the Transformer with the encoded low-resolution with high-level representation flow come from CNN blocks, are linear projection and feature mapping blocks, where the input/output signals reshape and downsample to be compatible for their usage. This hybrid network captures the local and global information from 3D data and demonstrates the improved performance within two Brain Tumor Segmentation (BraTS) 2019-2020 \cite{menze2014multimodal,bakas2017advancing,bakas2018identifying} datasets over the previous CNN U-Net structures.

Li et al. \cite{li2021gt} proposed the GT U-Net structure to address the low performance of previous segmentation methods in fuzzy boundaries while keeping the computational complexity low within the hybrid structure of CNN and the Transformer in a U-Net-like paradigm. Their method was applied to the private orthodontist tooth X-ray images and DRIVE dataset \cite{staal2004ridge}. All the main counterparts of U-Net are based on Group Transformer (GT) to dispense the quadratic computational complexity within these successive parallel convolution, Multi-Head Self-Attention (MHSA), convolution modules in each stage to gradually increase receptive filed and extracting long local-dependencies. So far, the presence of a Transformer in the segmentation tasks is crucial because if a network wants to provide an efficient prediction mask, it should be able to minimize the miss-classifying of the background and foreground pixels that leads to a reduction in False Positives (FP). Therefore learning long-range contextual features is as essential as fuzzy boundaries resulting from object overlappings or variation in exposure of the medical imaging devices. To mitigate the occurrence of this miss predicting in boundary levels, GT U-Net utilizes a Fourier descriptor loss term within binary cross entropy to impose the prior shape knowledge.

Xie et al. \cite{xie2021cotr} addressed the computational complexity that restrains the multi-scale functionality of conventional Self-Attention (SA) and proposed the hybrid CoTr architecture for volumetric medical image segmentation. The whole network is a U-Net-like structure with CNN-based 3D residual blocks for encoder and decoder paths with the amalgamation of Deformable Transformer (DeTrans) for multi-scale fusion, besides the conventional skip connections from the encoder to the decoder for better localization information and faster convergence. TransUNet suffers from parameters overload within MHSA, which treats all image tokenization positions equally. Therefore, CoTr instantiates the deformation concept from \cite{dai2017deformable,zhu2021deformable} into the deformable self-attention mechanism in Transformer to decrease the computation complexity and prepare the ground for using Transformer to process multi-scale and high-resolution feature maps. MS-DMSA layer is a deformable transformer instead of MHSA that focuses on only a small set of key sampling locations around a reference path. CoTr demonstrates the competitive results in a score-parameter trade-off on The Multi-Atlas Labeling Beyond the Cranial Vault (BCV) \cite{landman2015miccai} dataset.

UNETR \cite{hatamizadeh2022unetr} is a 3D segmentation network that directly utilizes volumetric data incorporating ViT solely at the encoder stage to capture global multi-scale contextual information in a 3D volumetric style which is usually of paramount importance in medical image segmentation domain. The architecture follows the U-shaped structure of \cite{ronneberger2015u} with skip connections carrying successive 3D convolution operations to the 3D CNN-based decoder. Using a CNN-based decoder is since transformers can not capture spatial localization information well despite their excellent capability of learning global information. Analogous to U-Net, Hatamizadeh et al. \cite{hatamizadeh2022unetr} uses the different stages of Transformer in the encoder to pass the flow from the contracting path to extracting path, and the multi-resolution contextual information (after reshaping the embedded sequence to a proper tensor shape and applying convolution operations) merges with CNN-based decoder to improve the segmentation mask prediction. UNETR produces uniform, non-overlapping patches from volumetric data and applies a linear projection to project patches into a constant embedding dimensional space throughout the Transformer layers. Their ablation studies depict that they outperformed the TranUNet \cite{chen2021transunet}, TransBTS \cite{wang2021transbts}, and CoTr \cite{xie2021cotr} on BCV \cite{landman2015miccai}, and MSD \cite{simpson2019large} datasets on an average of 1\% margin in the dice score metric.

In computer vision tasks, neighboring information of a specific region tends to be more correlated than far regions. To this end, Wang et al. \cite{wang2022mixed} proposed the MT-UNet network utilized with the Mixed Transformer Module (MTM) to capture long-range dependencies wisely concerning the most neighboring contextual information. Another critical point is that the ViT with Self Attention (SA) calculates the intra-tokens affinities, ignoring the inter-tokens connections dispensed through the other dimensions, especially in medical images. Therefore, MTM consists of an External Attention (EA) counterpart in itself to address this concern. MTM is used in conjunction with a U-Net-like structure accompanied by CNN blocks. CNN blocks are used to not only reduce the computational overhead by downsampling the input feature maps but also introduce a structure prior to the model in the case of small medical datasets. MT-UNet performs well on Synapse \cite{landman2015miccai} and ACDC \cite{bernard2018deep} datasets in comparison with TranUNet \cite{chen2021transunet}.

Azad et al. \cite{reza2022contextual} proposed a contextual attention network, namely TMU, for adaptively synthesizing the U-Net produced local feature with the ViT's global information for enhanced overlap boundary areas in medical images. TMU is two branches pipeline, wherein the first stream utilizes a U-Net-like block without a segmentation head (Resnet backbone \cite{he2016deep}) to extract high semantic features and object-level boundary heatmap interaction representation. In the next branch, the ViT-based Transformer module applies to non-overlap input images to extract long-range dependencies. Whereas the objective of segmentation differs from one subject to another data, as mentioned before, TMU aims to merge the local and global information adaptively. To do so, Azad et al. proposed a contextual attention mechanism to produce image-level contextual information and highlight the most discriminative regions within importance coefficients delivered by attention weights from Transformer. This paradigm not only revealed the efficiency of the boundary information as a prior and adaptive collaboration of local and long-range dependencies but also outperforms the conventional hybrid and solely CNN-based methods on SegPC challenge dataset \cite{gupta2018pcseg,gehlot2020ednfc,gupta2020gcti} and skin lesion segmentation datasets \cite{codella2018skin,codella2019skin,mendoncca2013ph}.

\begin{figure}[!htb]
    		\includegraphics[width=\columnwidth]{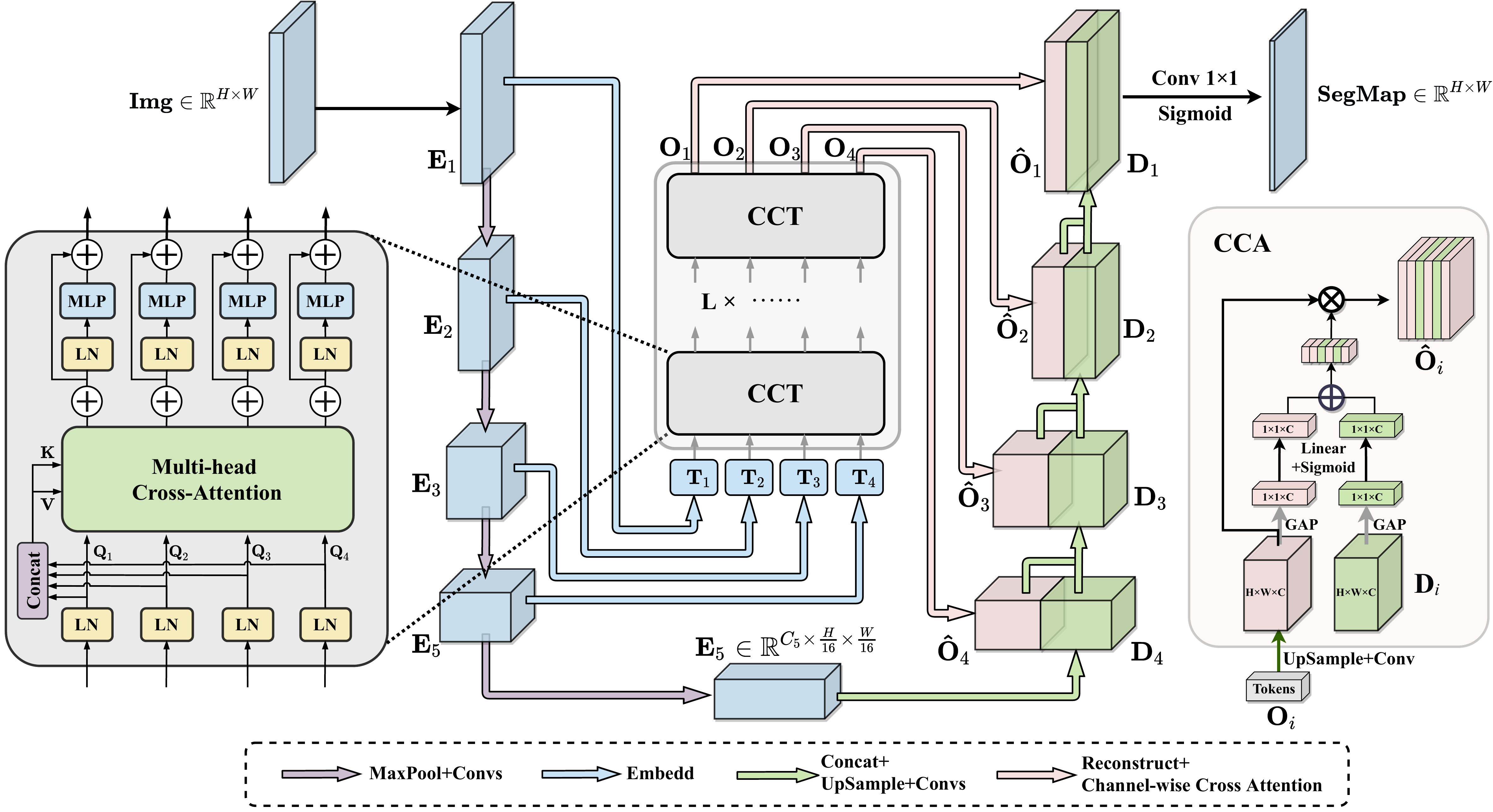}
    		\caption{Overview of the UCTransNet architecture. The original skip connections of U-Net architecture are replaced with the proposed CTrans Module including Channel-wise Cross fusion Transformer (CCT) and Channel-wise Cross Attention (CCA). CCT with the Multi-head Cross-Attention module is illustrated on the left. Figure from \cite{wang2022uctransnet-arx}.}
    		\label{fig:uctransnet}
\end{figure}
Skip connections in the U-Net-based model are used to transfer high spatial information from the encoder to the decoder for accurate localization, while the successive downsampling operations suffer from the loss of spatial information. However, Wang et al. \cite{wang2022uctransnet} studied the effectiveness of the preliminary U-Net skip connections and stated that the naive skip connections suffer from the highly semantic gap such as semantic gaps among multi-scale encoder features and between the encode-decoder stages. They proposed UCTransNet \cite{wang2022uctransnet} that alleviates these mentioned issues from the channel perspective with an attention mechanism, namely Channel Transformer (CTrans). CTrans is a modification for skip connections in a U-Net-based pipeline and consists of two sub-counterparts Channel Cross fusion with Transformer (CCT) and Channel-wise Cross-Attention (CCA), for aggregating multi-scale features adaptively and guiding the fused multi-scale channel-wise features to decoder effectively, respectively (see \Cref{fig:uctransnet}). CCT aims to fuse multi-scale encoder features to adaptively compensate for the semantic gap between different scales with the advantage of long-range dependency modeling in the Transformer. CCT tokenized feature maps at each stage within patch sizes from a multiple of $\frac{1}{2}$, preserving the channel dimensions. From \Cref{fig:uctransnet}, the proposed CCT module accompanies tokenized feature maps as a query and concatenated four tokens come from stages as key and value matrixes. With the use of instance normalization \cite{ulyanov2016instance} operation for gradient smoothing through the process, the primary distinction between the CCT and Self Attention (SA) is that the attention operation applies on the channel axis rather than the patch axis. Afterward, to rectify the gap between the encoder and decoder's inconsistent feature representation, the output tokens of CCT pass through CCA to apply a better fusion step and lessen the ambiguity with the decoder feature. The UCTransNet network performs SOTA Dice results on the GlaS \cite{sirinukunwattana2017gland}, MoNuSeg \cite{kumar2017dataset,kumar2019multi} and Synapse \cite{landman2015miccai} datasets in comparison with TransUNet \cite{chen2021transunet}.

\begin{figure}[!h]
    \includegraphics[width=\columnwidth]{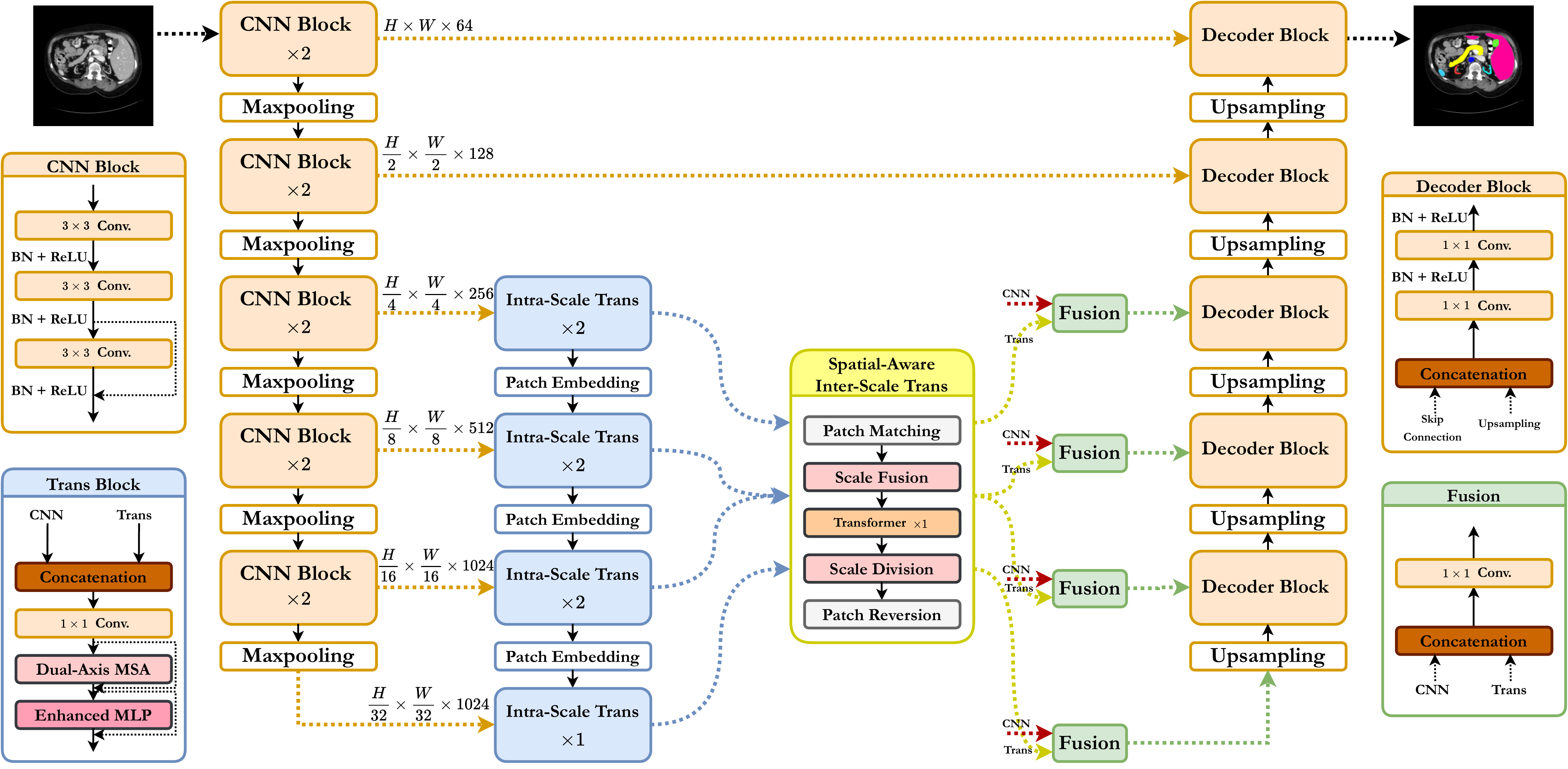}
    \caption{Overview of the ScaleFormer \cite{huang2022scaleformer} architecture. The input image is first passed through the CNN Block to extract the local details. The output features of the last several layers in the CNN block are then fed into the Intra-Scale Transformers to model the global information for each scale. Spatial-Aware Inter-Scale Transformer fuses the outputs of Inter-scale Transformer modules to enable the interaction among different scales. Finally, the decoder block performs upsampling and concatenation with features of corresponding scales to produce the segmentation prediction.}
    \label{fig:scaleformer}
\end{figure}

Similar to UCTransNet, Huang et al. \cite{huang2022scaleformer} addressed the inconsistency between local and global features in inter and intra-scales in conventional architectures (hybrid / standalone) \cite{chen2021transunet,cao2021swin,huang2021missformer,yan2022after} and proposed a ScaleFormer backbone based on a U-Net-like structure which during this study is the SOTA method in 2D modality. Their innovations through their design are to couple CNN-based features within long-range contextual features in each scale effectively within lightweight Dual-Axis MSA captures attention in a row/column-wise manner. In addition,  ScaleFormer \cite{huang2022scaleformer} make a bridge with a spatial-aware inter-scale Transformer to interact with the target regions' multi-scales features to surpass the shape, location, and variability of organs' limitation. ScaleFormer utilizes ResNet \cite{he2016deep} variant backbone, basic ResNet-34 blocks for CNN-feature extractor, and in each stage, scale-wise intra-scale transformer (Dual-Axis MSA) couples with the local features to highlight both detailed local-spatial and long-spatial affinities in each scale. To alleviate the deficiencies of previous methods in capturing sufficient information from multiple scales by a hierarchical encoder, spatial-aware inter-scale Transformers merge these features adaptively to strengthen the ScalFormer in effectively segmenting various-scale organs. From \Cref{fig:scaleformer}, the inter-scale Transformer is a computation-efficient design by applying successive point-wise convolutions followed by average-pooling on row/column-wise query and key matrices of the Transformer. This pipeline embraces the whole operations in a single block rather than multi blocks for row and column on input data \cite{ho2019axial}. The spatial-aware inter-scale Transformer is a conventional Transformer with a difference in interaction calculation of tokens cue. To be more precise, each input token to this Transformer first reshapes to its 2D representation. Every 2D representation of specific tokens in each scale concatenates with their successive 2D patch feature map in the following scales and then flattens to its 1D representation by producing a master token for that specific token and applying the standard Transformer calculation to it. Afterward, the enhanced representations are aggregated in the decoder path with local-level CNN features on the same stage in each skip connection to each decoder block. ScaleFormer proves its functionality through multiple datasets \cite{landman2015miccai,kumar2017dataset,kumar2019multi,bernard2018deep} by outperforming TransUNet \cite{chen2021transunet}, Swin-Unet\cite{cao2021swin}, MISSFormer \cite{huang2021missformer} and AFTer-UNet \cite{yan2022after}.

\Cref{fig:transunet,fig:scaleformer} show sample CNN-Transformer U-shaped structures that Transformer is an add-on to a U-Net-like network to model the long-range contextual information in diverse stages of U-Net from encoder-decoder to skip connection and bottleneck. The Swin UNETR \cite{hatamizadeh2022swin} is a modification to the original UNETR \cite{hatamizadeh2022unetr} that the 3D Vision Transformer replaced by the Swin Transformer in encoder path. There are still other studies such as \cite{azad2022transnorm,chen2021transattunet} noteworthy to review, but due to the limitation of the paper, we excluded them. 

\begin{figure*}[th!]
	\centering
	\begin{subfigure}[][][c]{0.48\textwidth}
		\includegraphics[width=\textwidth]{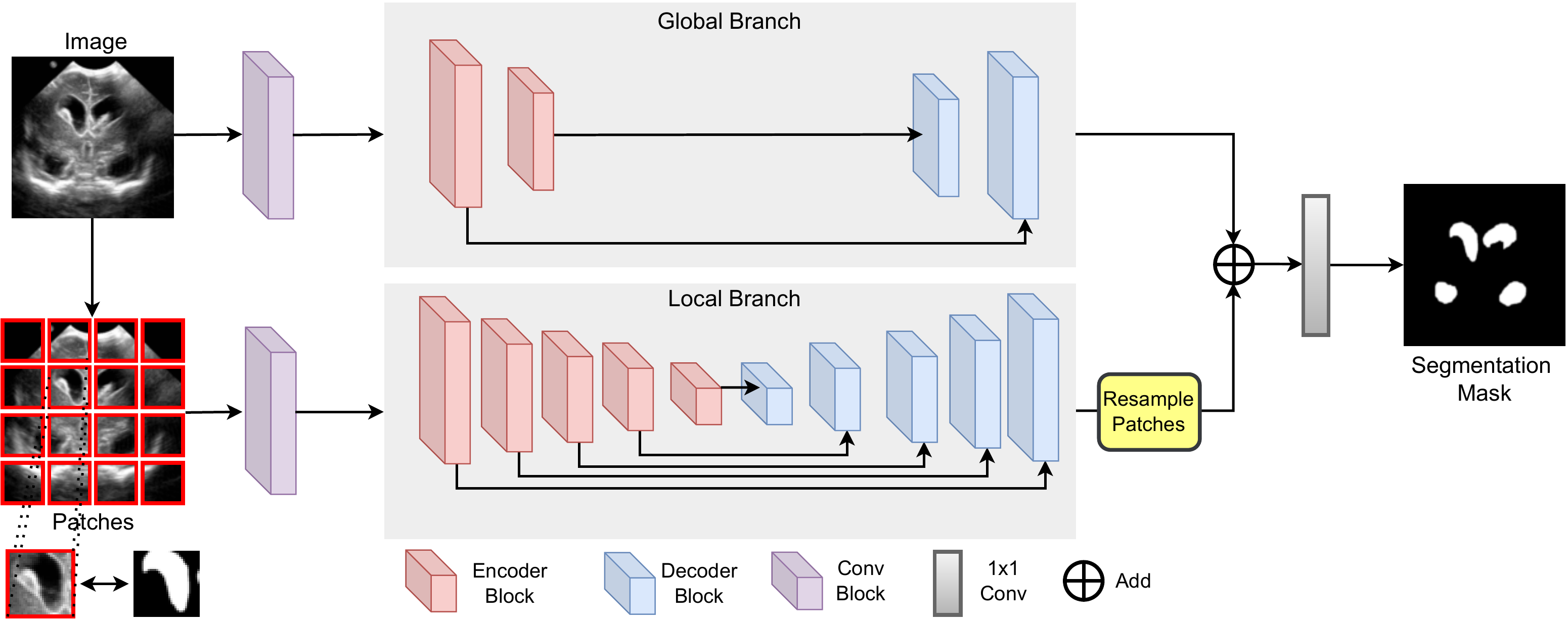}
		\caption{
			Overview of the MedT architecture. The network uses the LoGo strategy for training. The upper global branch utilizes the first fewer blocks of the transformer layers to encode the long-range dependency of the original image. In the local branch, the images are converted into small patches and then fed into the network to model the local details within each patch. The output of the local branch is re-sampled relying on the location information. Finally, a $1 \times 1$ convolution layer fuses the output feature maps from the two branches to generate the final segmentation mask. Figure from \cite{valanarasu2021medical-arx}.
		}
		\label{fig:medt}
	\end{subfigure}
	\hfill
	\begin{subfigure}[][][c]{0.48\textwidth}
		\includegraphics[width=\textwidth]{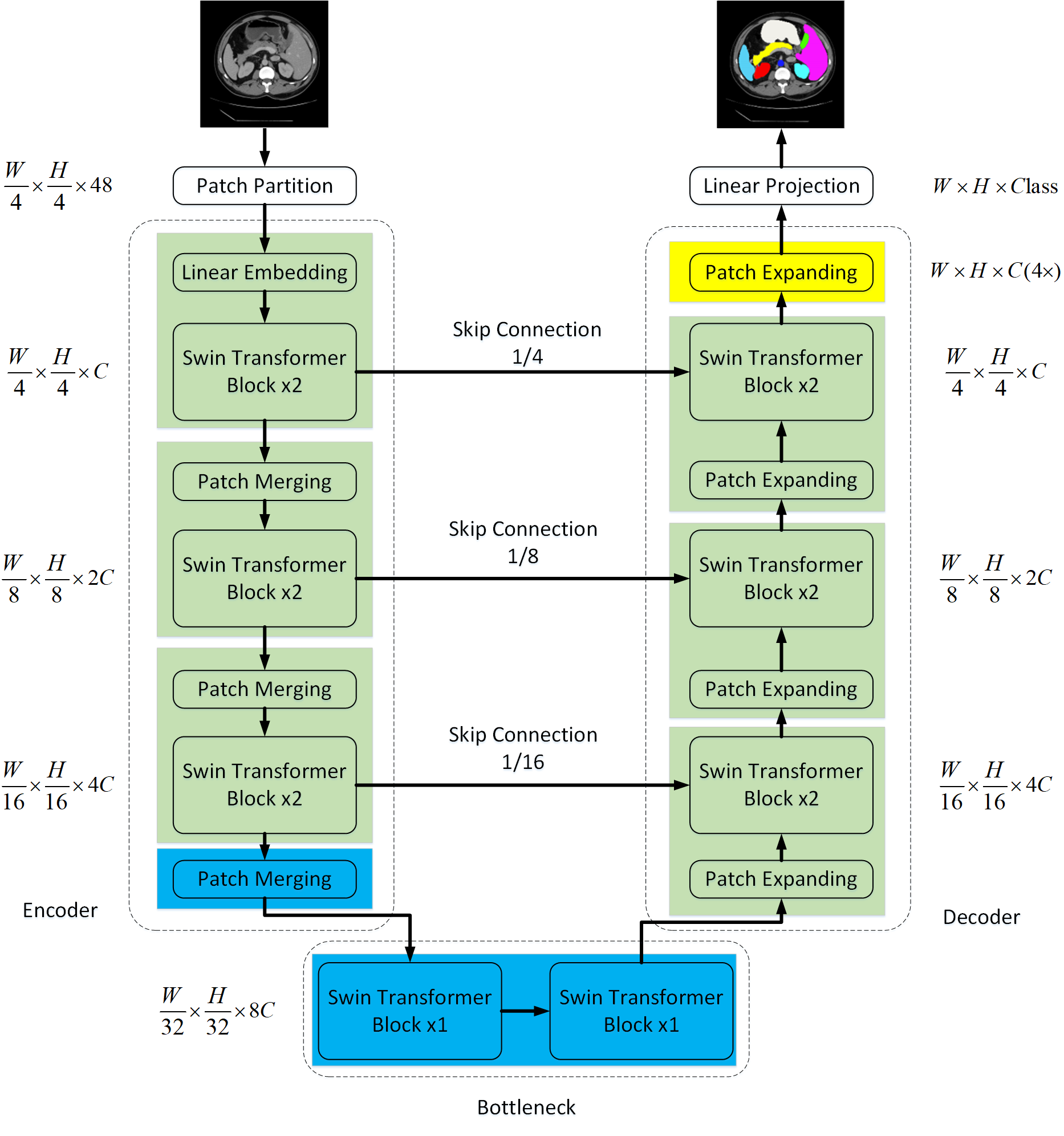}
		\caption{
			The architecture of the Swin-Unet follows the U-Shape structure. It contains the encoder, the bottleneck, and the decoder part which are built based on the Swin transformer block. The encoder and the decoder are connected with skip connections. Figure from \cite{cao2021swin}.
		}
		\label{fig:swin-unet}
	\end{subfigure}
	\hfill
	\begin{subfigure}[][][c]{0.48\textwidth}
		\includegraphics[width=\textwidth]{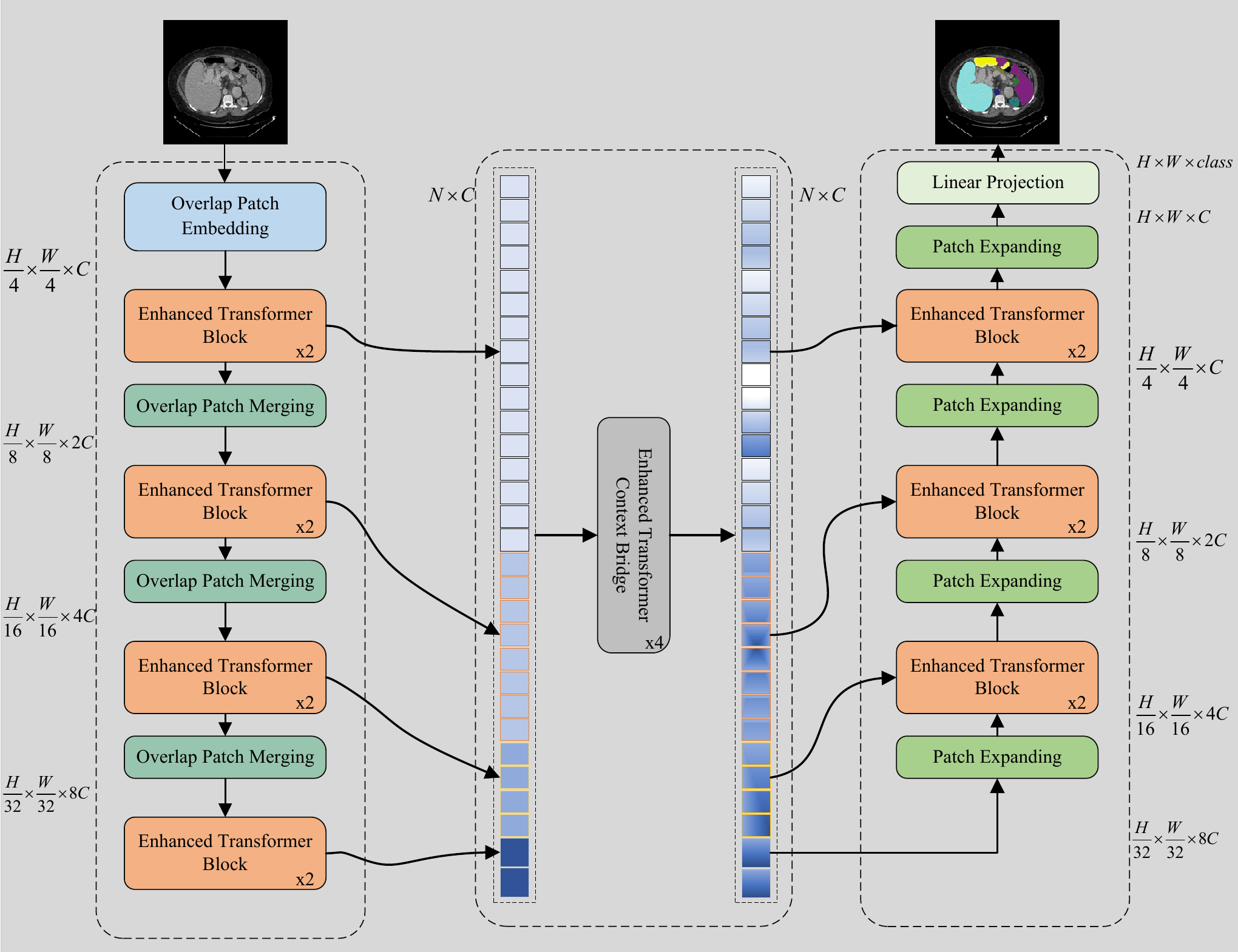}
		\caption{
			Overview of the MISSFormer architecture.
			The network is composed of a hierarchical encoder, a decoder, and an Enhanced Transformer Context Bridge. The encoder and decoder are constructed based on the enhanced transformer blocks and modules for patch processing. The outputs of each stage within the encoder are fused and passed through the bridge to model the local and global dependencies of different scales. Figure from \cite{huang2021missformer}.
		}
		\label{fig:missformer}
	\end{subfigure}
	\hfill
	\begin{subfigure}[][][c]{0.48\textwidth}
		\includegraphics[width=\textwidth]{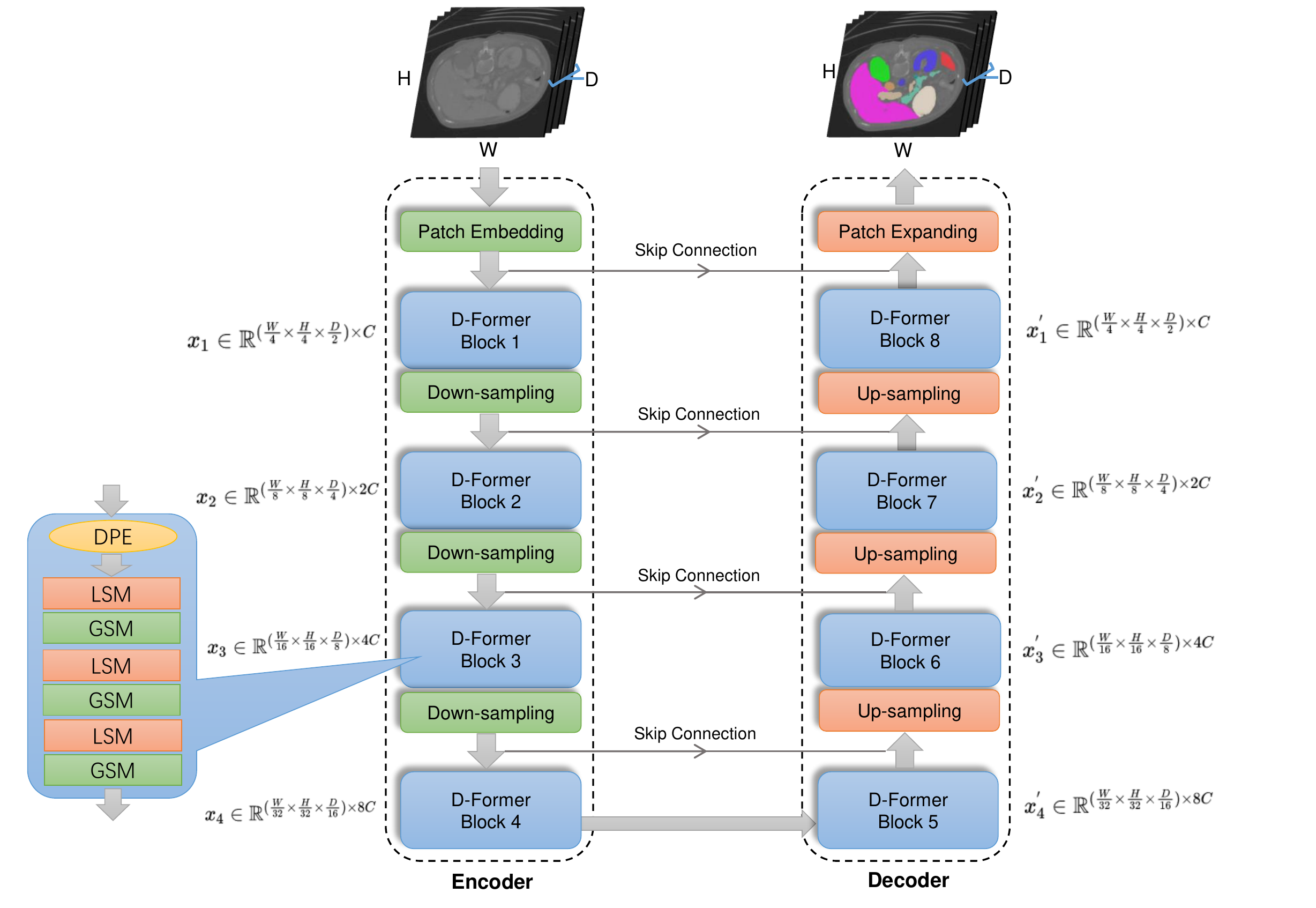}
		\caption{Overview of the D-Former architecture. One dynamic position encoding block (DPE), multiple local scope modules (LSMs), and global scope modules (GSMs) constitute each D-Former block. Figure from \cite{wu2022d}.}
		\label{fig:d-former}
	\end{subfigure}
	\caption{Overview of architectures mentioned in \Cref{sec:standalone-transformer}}
	\label{fig:standalone-transformer}
\end{figure*}

\begin{figure*}[th!]
	\centering
	\begin{subfigure}[t][][c]{0.3\textwidth}
		\includegraphics[width=\textwidth]{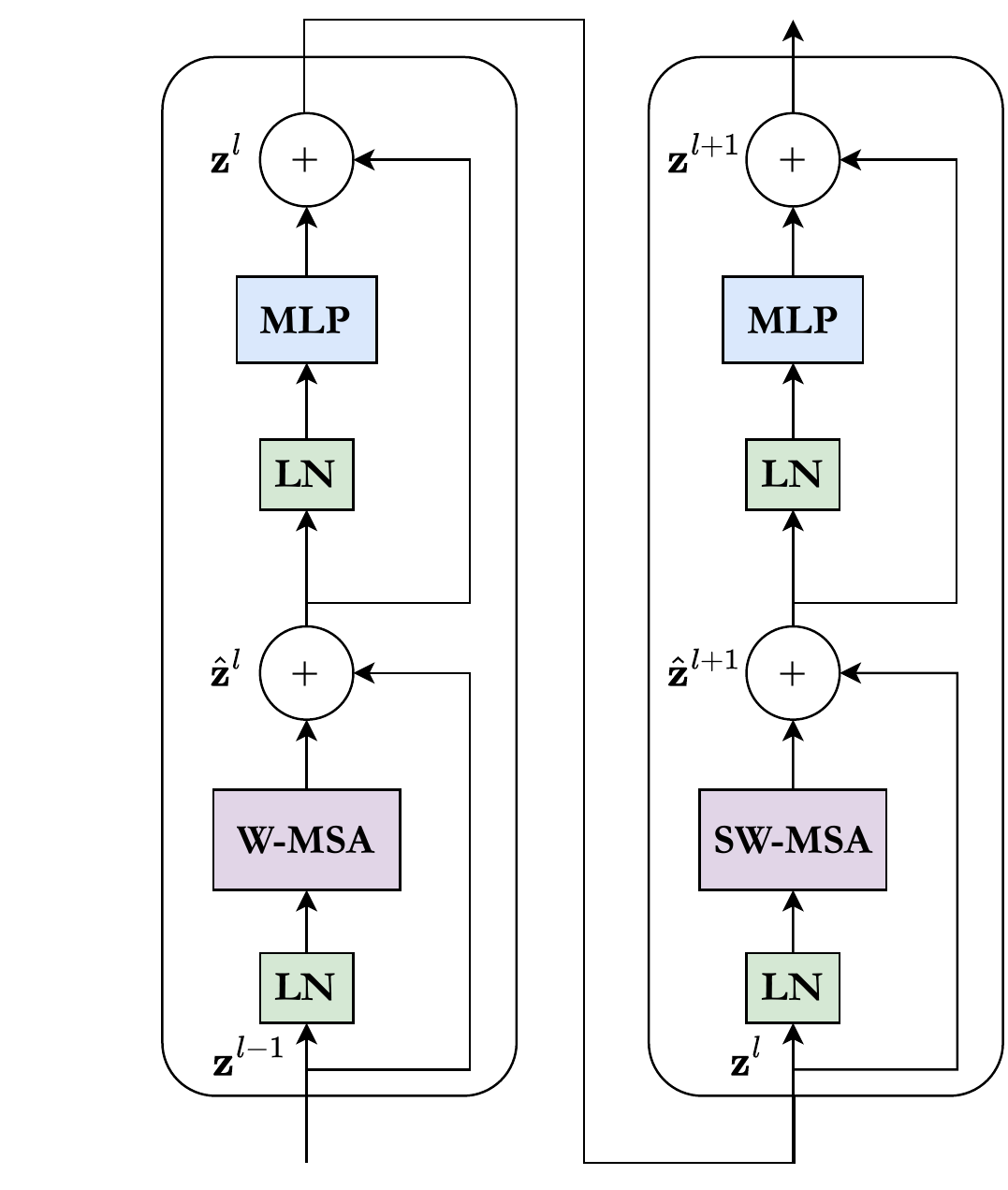}
		\caption{The calculations through the Swin block are:\\ $\hat{\mathbf{z}}^l = \text{W-MSA}(\text{LN}(\mathbf{z}^{l-1}))+\mathbf{z}^{l-1}$,\\$\mathbf{z}^l=\text{MLP}(\text{LN}(\hat{\mathbf{z}}^l))+\hat{\mathbf{z}}^l$,\\ $\hat{\mathbf{z}}^{l+1}=\text{SW-MSA}(\text{LN}(\mathbf{z}^l))+\mathbf{z}^l$,\\$\mathbf{z}^{l+1}=\text{MLP}(\text{LN}(\hat{\mathbf{z}}^{l+1}))+\hat{\mathbf{z}}^{l+1}$, $\hat{\mathbf{z}}^{l}$ and $\hat{\mathbf{z}}^{l+1}$ denote the output features of the W-MSA and SW-MSA‌ modules, respectively.}
		\label{fig:swin-block}
	\end{subfigure}
	\hfill
	\begin{subfigure}[t][][c]{0.3\textwidth}
		\includegraphics[width=\textwidth]{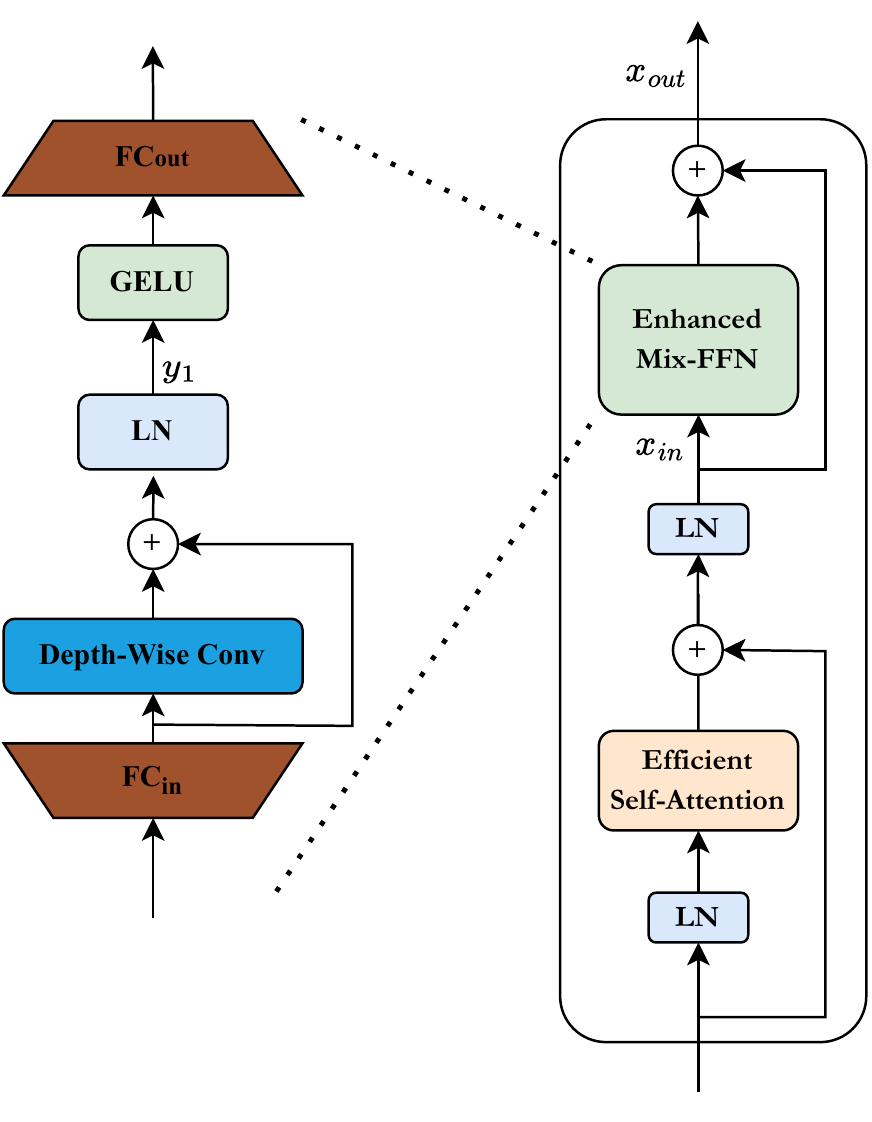}
		\caption{The Efficient Mix-FFN block applies the following operations: \\ {\tiny $y_1 = \text{LN}(\text{Conv}_{3\times 3}(\text{FC}(x_{in}))+\text{FC}(x_{in}))$}, \\ $x_{out}=\text{FC}(\text{GELU}(y_1)) + x_{in}$.}
		\label{fig:missformer-block}
	\end{subfigure}
	\hfill
	\begin{subfigure}[t][][c]{0.3\textwidth}
		\includegraphics[width=\textwidth]{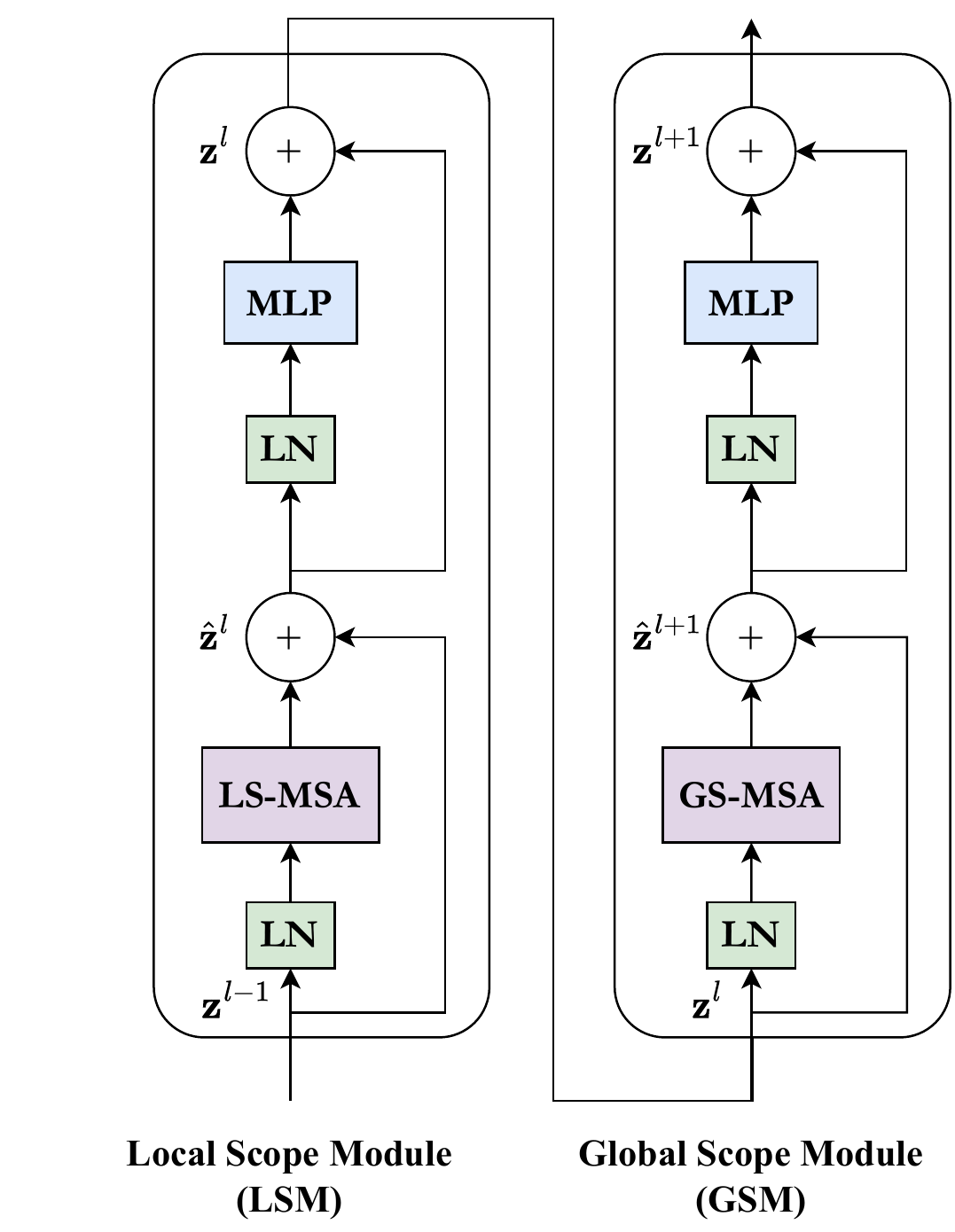}
		\caption{The LSM and GSM modules (main counterparts of the D-Former block) apply the following formulations:\\ $\hat{\mathbf{z}}^l = \text{LS-MSA}(\text{LN}(\mathbf{z}^{l-1}))+\mathbf{z}^{l-1}$,\\$\mathbf{z}^l=\text{MLP}(\text{LN}(\hat{\mathbf{z}}^l))+\hat{\mathbf{z}}^l$,\\ $\hat{\mathbf{z}}^{l+1}=\text{GS-MSA}(\text{LN}(\mathbf{z}^l))+\mathbf{z}^l$,\\$\mathbf{z}^{l+1}=\text{MLP}(\text{LN}(\hat{\mathbf{z}}^{l+1}))+\hat{\mathbf{z}}^{l+1}$, $\hat{\mathbf{z}}^{l}$ and $\hat{\mathbf{z}}^{l+1}$ denote the output features of the LS-MSA and GS-MSA modules, respectively.}
		\label{fig:dformer-block}
	\end{subfigure}
	\hfill
	\caption{Overview of main Transformer counterpart blocks mentioned in \Cref{sec:standalone-transformer}. (a) Swin Transformer block \cite{liu2021swin,cao2021swin}, (b) MISSFormer Enhanced Transformer block \cite{huang2021missformer}, (c) D-Former Transformer block \cite{wu2022d}. MLP, FC, and LN operations indicate the Multi-Layer Perceptron, Fully Connected, and Layer Normalization \cite{ba2016layer}, respectively.}
	\label{fig:attention-standalone-transformer}
\end{figure*}

\subsubsection{Standalone Transformer Backbone for U-Net Designs} \label{sec:standalone-transformer}
So far, multiple studies incorporating the Transformer concept and conventional CNN modules have been reviewed in \Cref{sec:hybrid-transformer}. In this section, we investigate the usage of Transformer as a standalone main counterpart for designing backbones for U-Net-like structures. One of the pioneering structures in this domain was proposed by Valanarasu et al. \cite{valanarasu2021medical}, namely MedT. Like most of the other networks, MedT plans to contribute to capturing long-range spatial context with pure Transformer rather than the CNN-based methods that partially broaden the hindered-receptive field of CNN, e.g., D-UNet \cite{jin2019dunet} with deformable convolution operations \cite{dai2017deformable}, ASPP-FC-DenseNet \cite{hai2019fully} with atrous convolution operations \cite{chen2017deeplab}, and H-DenseUNet \cite{li2018h} with successive convolution operations. However, Transformer's performance (also ViTs) has a strong bond with the fed data scale to the Transformer module \cite{dosovitskiy2020image}, which in the medical scale, could be degraded more, and a high amount of data could not be available. This lack of data is a considerable corner point in learning positional encoding as one of the preliminary steps of Transformers, which have shown their capacity to model images' spatial structure. Therefore MedT \cite{valanarasu2021medical} proposes a gated axial-attention mechanism to control the information flow by positional embeddings to query, key, and value \cite{wang2020axial} in a multi-axis attention operation \cite{ho2019axial}. In \cite{wang2020axial} the accurate relative positional encoding learned on large-scale datasets rather than small-scale datasets improves the performance, therefore MedT introduces a gating parameter to control the amount of positional bias in capturing non-local information in hindering non-accurate positional embedding. In addition, to effectively extract information, MedT utilizes a Local-Global (LoGo) training strategy to compensate for the Transformer's patch-wise technique weakness in capturing inter-patch pixel dependencies. To do so, MedT investigates two branches in its network diagram (see \Cref{fig:medt}), one as a global branch to work on the original resolution of the image and the local branch that operates on the patches of the image. Overall, MedT demonstrated vanguard results on Brain US \cite{valanarasu2020learning,wang2018automatic}, Glas \cite{sirinukunwattana2017gland}, and MoNuSeg \cite{kumar2017dataset,kumar2019multi} datasets in Dice and IoU metrics.

Transformers are well capable of capturing long-range dependencies through data, however, they suffer from severe and inevitable handicaps that impede them from their versatile use in vision tasks. These shortages commonly are correlated with chains to each other, e.g., Transformers computational complexity is quadratic \cite{dosovitskiy2020image,aghdam2022attention} and this restrains its usability in dense vision tasks such as segmentation and detection, which needs the neighboring pixel dependencies in the multi-scale and hierarchical pattern. Due to the fixed size non-overlapping tokenization step in the naive Transformer rather than the pixel-by-pixel calculation of attention to diminishing the mentioned computational burden, the Transformer is unworthy of extracting the local contextual dependencies in intra-path pixels. These constraints make the interest to provide efficient Transformers, Linear Transformers, with a significant amount of reduction in parameters and computational complexity \cite{tay2020efficient}. In the vision tasks, Swin Transformer \cite{liu2021swin} plays a critical role as an efficient and linear Transformer with the capability of supporting hierarchical architectures. A key design counterpart of the Swin Transformer is its shifted windowing scheme that makes the Transformer calculate the affinities for patches in the same window. Afterward, the window swipes on the patches, and the attention calculates among the patches in the same window. This successive shifting operation and capturing local contextual information within patches in windows can stack multiple times. Ultimately a patch merging layer is introduced to build CNN-like hierarchical feature maps by merging image patches in deeper layers. This intuition and the U-Net-like structure success emerged Swin-Unet \cite{cao2021swin} structure in the medical segmentation field. From \Cref{fig:swin-unet}, Cao et al. \cite{cao2021swin} used the Swin Transformer block as the main counterpart of their U-shaped network. 2D medical images split into non-overlapping patches, and each patch fed into the encoder path comprised of Swin blocks. The contextual features from the bottleneck output upsample in the decoder path with patch expanding layer (contrary to path merging layer) end couples with the multi-stage features from the encoder via skip connections to restore the spatial information. Swin-Unet presented SOTA results over the CNN-Transformer hybrid structures like TransUNet \cite{chen2021transunet} and demonstrated the robust generalization ability with the help of two multi-organ (Synapse) \cite{landman2015miccai} and cardiac (ACDC) \cite{bernard2018deep} segmentation datasets.

Due to some intrinsic features of medical images or, more specifically, the properties of human body organs, e.g., multi-scale and deformations, \cite{ronneberger2015u}, the necessity to capture long-range dependencies for accurate segmentation boundaries \cite{valanarasu2021medical} and inherence of generalizability of models even with no pre-training strategies and applicable for low-data regime \cite{maier2018rankings,prevedello2019challenges,varoquaux2022machine}, Huang et al. \cite{huang2021missformer} proposed MISSFormer a pure U-shaped Transformer network to address these apprehensions. \Cref{fig:missformer} displays the MISSFormer network with enhanced Transformer block as a primary entity in the network. One of the Transformer's drawbacks mentioned earlier is their un-suitability for capturing local context \cite{chu2021conditional,li2021localvit}, which comes with the solution for lessening the computational complexity with patching operation. However, the local contextual information plays a pivotal role in high-resolution vision tasks, therefore some studies in the vision Transfomer domain tackle this problem by embedding convolution operations in their attention module, e.g., PVTv1 \cite{wang2021pyramid}, PVTv2 \cite{wang2022pvt}, and Uformer \cite{wang2022uformer}. Huang et al. \cite{huang2021missformer} argues this methodology and brings up the point that direct usage of convolution layers in Transformer blocks limits the discrimination of features. For an input image, MISSFormer applies a $4\times 4$ convolutions with the stride size instead of $4$ (overlapping windows) for preserving local continuity in building the patches stage. The encoder path molds hierarchical representation (see \Cref{fig:missformer}) with the help of Enhanced Transformer Block, which accompanies the \Cref{fig:missformer-block} Transformer module. Enhanced Transformer block comprises an Efficient Self-Attention module to decrease the traditional attention calculation by downsampling the corresponding query, key, and value matrixes. Afterward, the Enhanced Mix-Feed Forward Network (FFN) sub-module (modified clone of Mix FFN from \cite{xie2021segformer}) aligns features and makes discriminant representation with $3 \times 3$ depth-wise convolutions for capturing the local context efficiently. Analogous to \cite{wang2022uctransnet}, MISSFormer rethinks the skip connection design, utilizes the Enhanced Transformer Context Bridge module for multi-scale information fusion, and hinders the gap between encoder and decoder feature maps. This module captures the local and global correlations between different scale features. For a brief review of this module, the context bridge (\Cref{fig:missformer}) flattens the output attention matrixes from different scales in the spatial dimension and reshapes it in a way to have a consistent channel dimension and concatenates the representations in flattened spatial dimension and feed to enhanced Transformer block to produce long-range dependencies and local context correlations. Ultimately the output of Enhanced Mix-FFN is split and restores to its original shape to get the discriminant fused hierarchical multi-scale representation. MISSFormer plotted high performances compared to TransUNet \cite{chen2021transunet} and Swin-Unet \cite{cao2021swin} over Synapse \cite{landman2015miccai} and ACDC \cite{bernard2018deep} datasets. 

Since most U-Net-shaped standalone Transformer utilized models incorporate 2D inputs, Wu et al. \cite{wu2022d} proposed a Dilated Transformer (D-Former) for 3D medical image segmentation with consideration of degrading the computational complexity. D-Former \cite{wu2022d}, \Cref{fig:d-former}, is a U-shaped hierarchical architecture with specialized D-former blocks which consist of Local Scope Modules (LSMs) and Global Scope Modules (GSMs) in an alternate order to capture local and global contextual information. Each D-Former block could repeat the successive LSM and GSM counterparts, however the original D-Former \cite{wu2022d} uses three successive sequences of LSM-GSM modules in the third and sixth D-Former blocks. LSM captures local self-attention by dividing a 3D feature map into non-overlapping 3D-volumetric units that consist of 3D patches and calculating the self-attention within these 3D units. As a complement to the locally fine-grained attention, the GSM module attains interactions through different units in a dilated manner to enlarge the attention's performance range. This intuition captures local and global interactions with fixed-size patches within sets of 3D volumetric units without any computation overflow. Also, D-Former takes advantage of positional encoding within Transformers with Dynamic Positional Encoding (DPE), which can be a vital cue in dense prediction tasks such as segmentation. Due to the 3D functionality and 3D contextual information, D-Former is a SOTA method on various 3D datasets such as Synapse \cite{landman2015miccai} and ACDC \cite{bernard2018deep} datasets with large margins in Dice score over CNN-hybrid or standalone Transformer designs.

Still, numerous studies out there, e.g., DS-TransUNet \cite{lin2021ds} utilize the Swin Transformer in the dual-path U-shaped structure for modeling the multi-scale patch size to lessen the deficiency of Transformers in capturing the local context. We reviewed multiple studies that utilized the Transformer in their U-Net pipeline in various methods. With such evidence and stunning growth in the Transformer field, we still hear about the outstanding collaboration between Transformers and U-Net-like networks so often.

\subsection{Rich Representation Enhancements} \label{sec:rich-representation}
To obtain a rich representation, the common approaches applied to medical image segmentation are multi-scale and multi-modal methods, e.g. \cite{wang2021sar,zhao2020scau,zhang2018road}. The key objective is to enhance the performance of the trained models by utilizing all available information from multi-modal or multi-scale images while retaining the most desirable and relevant features.

The multi-scale method, also referred to as the pyramid method originated from the Laplace pyramid method proposed by Burt et al. \cite{burt1987laplacian}. The approach converts the source input image by resizing it into a series of images with decreasing spatial resolutions. This scheme allows encoders of models to directly access the features of the enhanced images of different sizes and thus learn the respective features.

The study of the organs of interest requires their specific imaging modality to provide targeted information. However, each imaging technique has its limitations and can only reveal partial details about the organ, which may lead to inaccurate clinical analysis. Therefore, a fusion of images from various imaging modalities can be conducted to supplement each other's information by integrating complementary information retrieved from several input images.

The powerful structural design of the UNet network with the encoder and decoder allows the network to mine salient features at multiple input levels and enables effective feature fusion of different modalities.

Lachinov et. al. evaluate the performance of the Cascaded U-Net\cite{lachinov2018glioma} with multiple encoders processing each modality respectively to demonstrate the improvement due to the extraction multi-modal representation. The results indicate that the architecture taking the multiple modalities into account outperforms the network only relying on one single modality.

The following classifications will illustrate the modality fusion proposed to learn richer representations.

\subsubsection{Multi-scale Fusion}
Image pyramid input or side output layers are aggregated into U-Net structures to fuse the multi-scale information in the encoder or decoder stage.

Abraham et al. \cite{abraham2019novel} propose the Focal Tversky Attention U-Net with a generalized focal loss function that modulates the Tversky index \cite{hashemi2018asymmetric} to address the issue of data imbalance and improve precision and recall balance in medical image segmentation. Furthermore, they incorporate multi-scale image inputs into the attention U-Net model with deep supervision output layers \cite{oktay2018attention}. The novel architecture facilitates the extraction of richer feature representations and results in 3\% dice score improvement on multi-class CT abdominal segmentation task. Compared to the commonly used Dice loss, the Tversky similarity index introduces a specific weight for each class, which is inversely proportionate to the label occurrences. This index shown as follows alleviates precision and recall imbalance due to equal weighting of false positive (FP) and false negative (FN).

\begin{equation}
	\begin{aligned}
		TI_c &= \frac{\sum^N_{i=1}p_{ic}g_{ic}+\epsilon}{\sum^N_{i=1}p_{ic}g_{ic}+\alpha\mathbf{A}+\beta\mathbf{B}+\epsilon}
		\\
		\mathbf{A} &= \sum^N_{i=1}p_{i\bar{c}}g_{ic},\qquad
		\mathbf{B} =\sum^N_{i=1}p_{ic}g_{i\bar{c}}
	\end{aligned}
\end{equation}
where, $\mathbf{A}$ and $\mathbf{B}$ refer to FN and FP respectively. In the terms $\mathbf{A}$ and $\mathbf{B}$, $p_{ic}$ denotes the probability that the pixel $i$ is classified to the lession class $c$ and $p_{i\bar{c}}$ is the probability pixel $i$ belongs to the non-lesion class $\bar{c}$. $g_{ic}$ and $g_{i\bar{c}}$ are of the same form for ground truth pixels. Adjusting the weights $\alpha$ and $\beta$ to fine-tune the emphasis can enhance recall in case of significant class imbalance. 

The authors further develop a focal Tversky loss function (FTL) for better small regions-of-interest (ROIs) segmentation by forcing the function to shift the focus on less accurate and misclassified predictions.
\begin{equation}
	FTL_c = \sum(1-TI_c)^{1/\gamma}
\end{equation}

where $\gamma$ is set in the interval $[1,3]$ to enable the loss function to concentrate more on incorrectly classified predictions that are less accurate. The reason is that when $\gamma > 1$, the FTL is almost unaffected if a pixel is misclassified with a high Tversky index. And if a pixel is incorrectly classified with a small Tversky index, the FTL will be high and forces the model to focus on hard samples.
\begin{figure}
	\centering
	\includegraphics[width=\columnwidth]{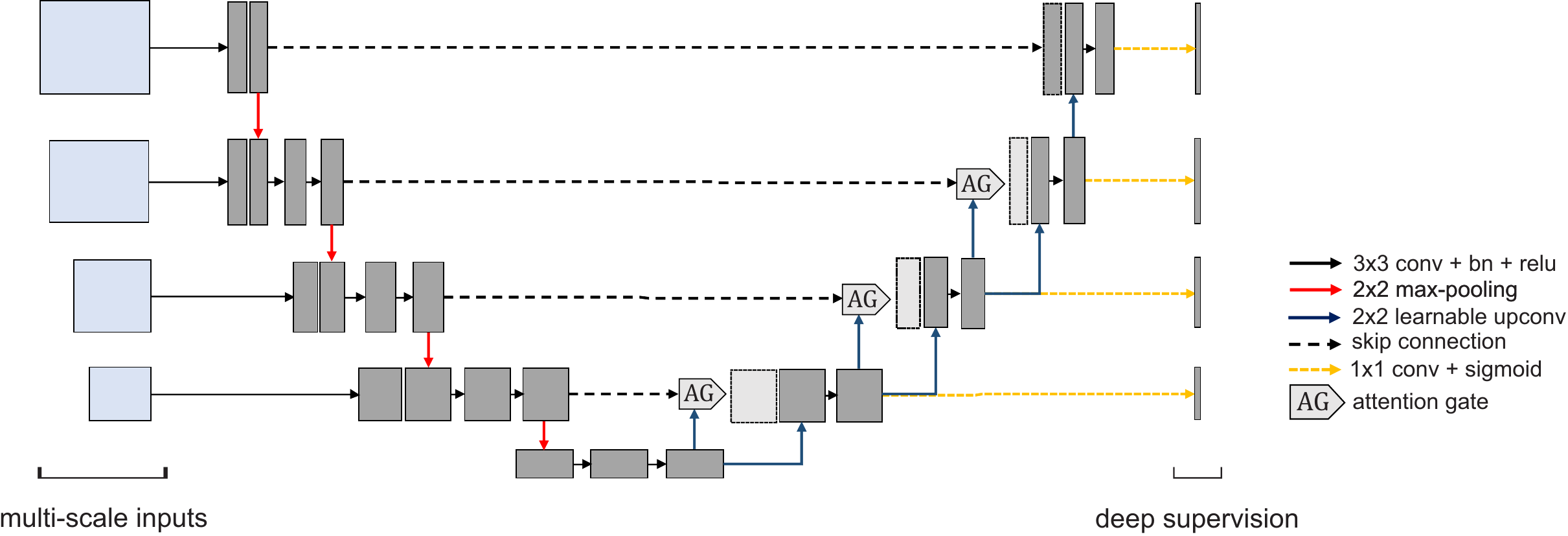}
	\caption{Architecture of the Focal Tversky Attention U-Net with multi-scale input and deep supervised output layers. AGs indicate soft attention gates which combine spatial information from low-level feature maps with coarser contextual information from skip connections. Figure from \cite{abraham2019novelFig}.}
	\label{fig:Focal_Tversky}
\end{figure}

As shown in \Cref{fig:Focal_Tversky}, they use Soft Attention Gates (AGs) to prune features and propagate only relevant spatial information to the decoding layers to enhance the balance between precision and recall at a structural level. In addition, an input image pyramid injected into each of the max pooling layers in the encoder and deep supervision module enriches feature learning at different scales.

To better segment the Optic Disc (OD) and Optic Cup (OC) for accurate diagnosis of glaucoma from fundus images, Fu et al. \cite{fu2018joint} introduce the polar transformation into the U-shape convolutional network with multi-scale input layers to build the Polar Transformation M-Net, aims to extract the richer context representation of the original image in the polar coordinate system. Compared to prior work, which treats OD and OC individually while ignoring their interdependency and overlap, the proposed Polar transformation M-Net considers both OD and OC simultaneously and formulates the segmentation as a multi-label task. Moreover, a novel loss function built on the Dice coefficient is employed to address the data imbalance between OD and OC in the fundus images. The model aggregates the side-output layers serving as early classifiers that generate prediction maps at different scales.

\begin{figure}
	\centering
	\includegraphics[width=\columnwidth]{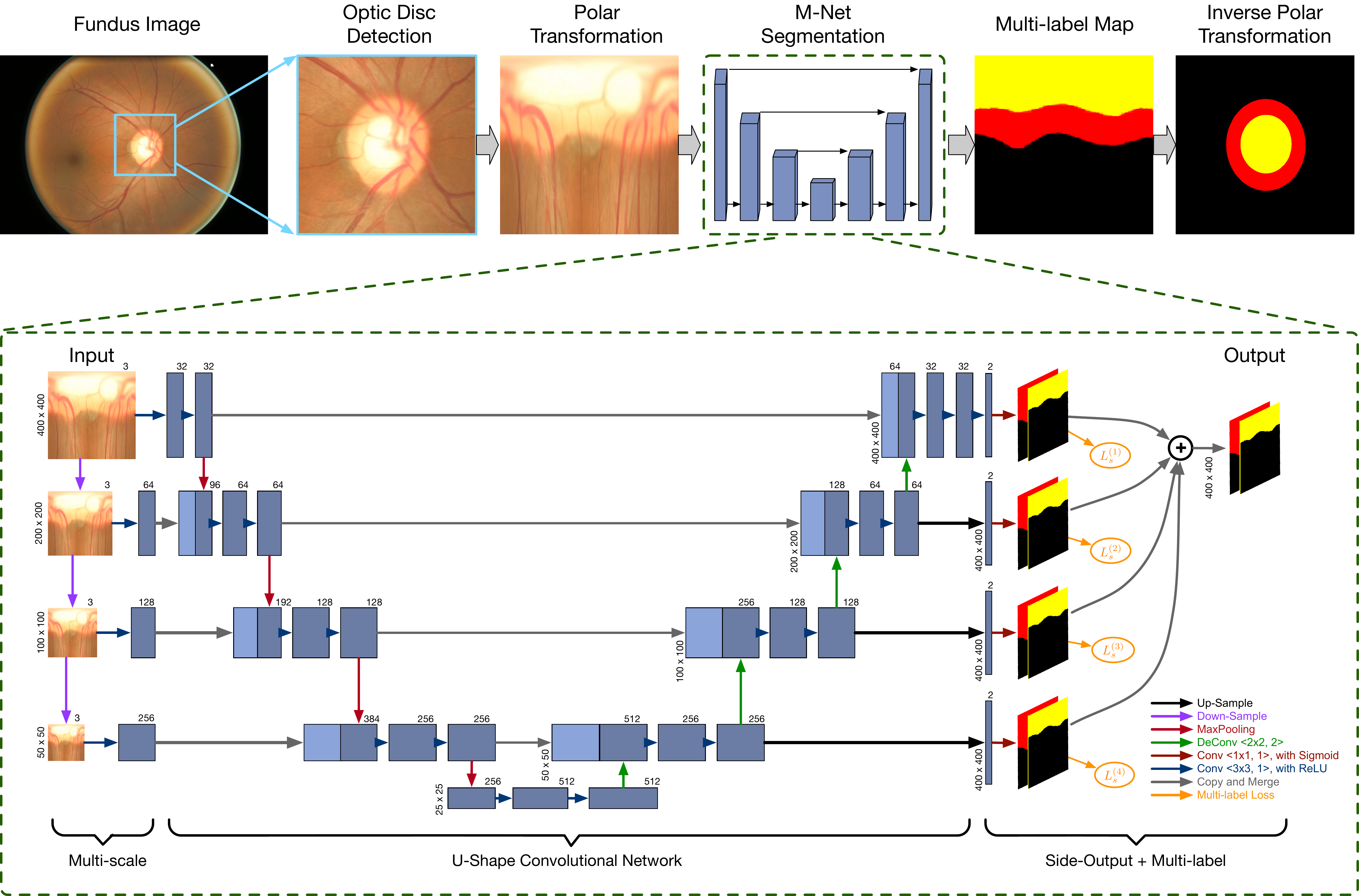}
	\caption{Illustration of the Polar Transformation M-Net method. The upper part of the figure shows the overall pipeline. The lower part demonstrates the architecture of M-Net. The method first localizes the disc center to extract the ROI, then applies the polar transformation to the detected patches of ROI. The transferred results are passed through the M-Net to obtain multi-label maps. The inverse polar transformation restored the final segmentation prediction to the Cartesian coordinate. The M-Net architecture is composed of an input image pyramid, U-shape convolutional network, and side-output layers. Figure from \cite{fu2018jointFig}.}
	\label{fig:PolarMUnet}
\end{figure}

\Cref{fig:PolarMUnet} demonstrates the overall structure of M-Net. It consists of a multi-scale input layer, U-shape convolutional network, side-output layer, and multi-label loss function. The initial multi-scale layer constructs the image input with descending resolutions for hierarchical representation learning. The processed image pyramid is passed through a U-shape network similar to the original U-Net architecture \cite{ronneberger2015u} whose encoder path and decoder path are concatenated by the skip connections. The output of each stage within the decoder path is fed to a side-output layer that produces a local output map. The superiority of the side-output layer is that the issue of gradient vanishing could be mitigated by the backpropagation of the side-output loss together with the final layer loss. Since the disc region overlays the cup pixels, the authors evaluate segmentation performance by the proposed multi-label loss function based on the Dice coefficient, which is defined as:
\begin{equation}
	L_s = 1 - \sum^{K}_k \frac{2w_k\sum^N_np_{(k,i)}g_{k,i}}{\sum^N_ip^2_{(k,i)}+\sum^N_ig^2_{(k,i)}}
\end{equation}
where $K$ is the total number of classes and $N$ denotes the number of pixels. $p_{(k,i)}$ and $g_{(k,i)}$ are the predicted probability and binary ground truth respectively. Class weights $w_k$ control the contribution of each class to the final results. Another contribution of the network is the pixel-wise polar transformation operated on the fundus image plane. The polar transformation can map the radial relationship that OC should be within the OD to the layer-like spatial structure, which makes the features easier to identify. Additionally, the interpolation during the polar transformation based on the OD center could expand the cup region. It helps relieve the heavily biased distribution of OC/background pixels. The experiments on the ORIGA dataset have demonstrated an increase in segmentation performance.

The original U-Net may not fully exploit all contributions of the semantic strength since it only generates output segmentation maps from the final layer of the decoder path.
Moreover, the output of the layers in different steps cannot be connected to one another, which blocks feature sharing and leads to redundant parameters. To address the mentioned issue, Moradi et al. proposed MFP-UNet which allows the output of all the blocks in different stages to be fed to the last layer \cite{moradi2019mfpFig}.
Their architecture is composed of two pathways, the ``bottom-up pathway'' and the ``top-down pathway''.
The encoder of the UNet with dilated convolution filter serves as the "bottom-up pathway". The dilated convolutional kernel can increase the receptive fields of the module by the dilation factor. Besides, the expansion path of U-Net acts as the FPN top-down pathway. Each step of the top-down pathway provides prediction maps where lower-resolution semantically stronger features can be processed for transfer to higher resolutions. Additional convolution layers are included for processing the feature maps at different scales to one fixed resolution compared to the decoder path of the original U-Net, which can boost the accuracy and improve the resolution of each stage. According to the experiment results, the novel model provides a robust and powerful architecture regarding the capabilities of feature representation in a pyramid, which shows robustness to large and rich training sets.

\subsubsection{Multi-modality Fusion}
In this section, we summarize the U-Net variant models with multimodal fusion modules, where a single encoder of U-Net is extended to multiple encoders to receive medical images in different modalities. The branches of encoders are connected by their respective strategies of aggregation, thus sharing information in different modalities, extracting richer representations, and complementing each other.

The novel architecture Dolz et al. \cite{dolz2018dense} propose in the Dense Multi-path U-Net enhances traditional U-Net models regarding rich representation learning on two key aspects: modality fusion and inception module extension. Two typical strategies are employed to deal with multi-modal image segmentation tasks. The early fusion merges the low-level features of inputs of multiple imaging modalities at the very early stage. As for the late fusion strategy, the CNN outputs of different modalities are fused at a later point. Nevertheless, these previous strategies cannot thoroughly model the highly complex relation of the image information across different paths of modalities. To alleviate the limitation, the proposed HyperDenseNet adopts the strategy where each stream receives inputs of image data of one modality, and the layers in the same and different paths are densely connected.
\begin{figure}
	\centering
	\includegraphics[width=\columnwidth]{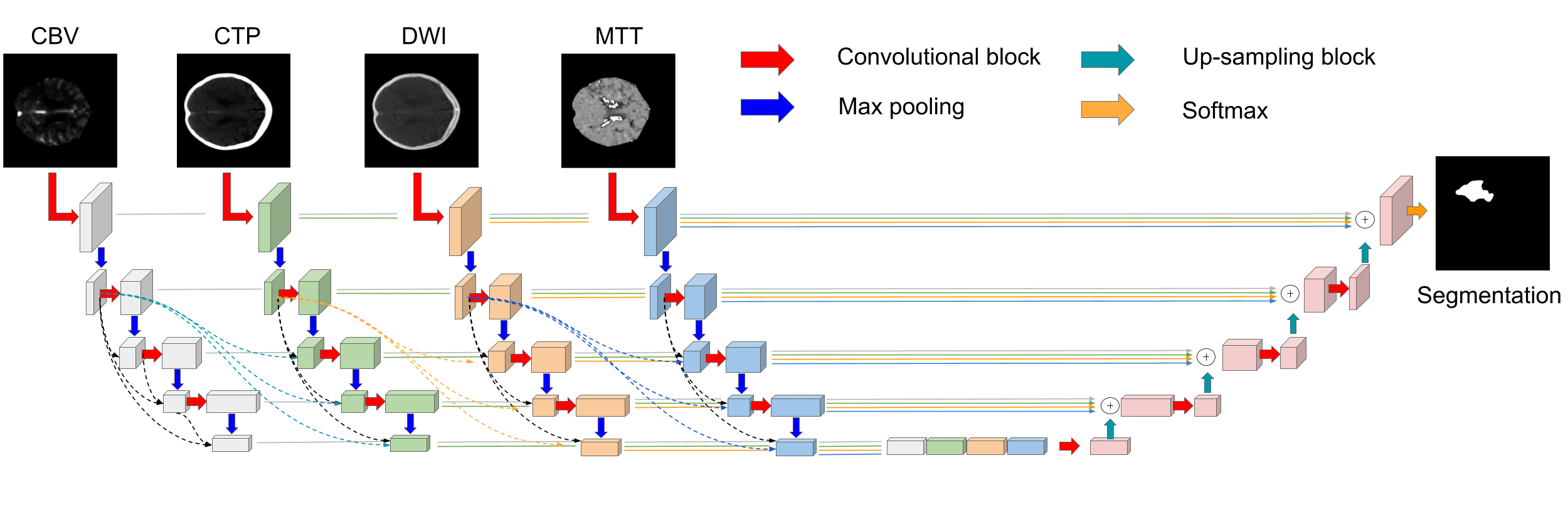}
	\caption{The architecture of network for segmenting ischemic stroke lesions across various imaging modalities. In the encoding path, each imaging modality is input into a single stream. All layers are directly connected to one another in a single stream, which facilitates the information flow of the network. The dashed lines depict the dense connections between the layer outputs in the streams of different modalities. Figure from \cite{dolz2018denseFig}.}
	\label{fig:DenseUnet}
\end{figure}

As shown in \Cref{fig:DenseUnet}, the encoding path contains N streams, each responsible for one imaging modality. The Dense Multi-path U-Net supports Hyper-dense connections both within a single path and between several paths. In a densely-connected network, the output of the $l^{th}$ layer is produced by the mapping $H_l$ of the concatenation of all the feature layers.
\begin{equation}
	x_l = H_l([x_{l-1},x_{l-2},...,x_0])
\end{equation}
HyperDenseNet integrates the outputs among different paths based on the densely-connected network to obtain richer feature representation from combined modalities. In addition, the permuting and interleaving operations are applied to the concatenation to improve performance. Considering the case of two modalities with streams 1 and 2 denoted as $x_l^1$ and $x_l^2$ respectively, the output of the $l^{th}$ layer of HyperDenseNet can be expressed as follows:

\begin{equation}
	x_l^s = H_l^s(\pi_l^s([x_{l-1}^1,x_{l-1}^2,x_{l-2}^1,x_{l-2}^2,...,x_0^1,x_0^2])),
\end{equation}
where $\pi^s_l$ represents the shuffling function acting on the feature maps.

Inspired by the Inception module in \cite{szegedy2016rethinking} which employs convolutions with multiple kernel sizes on the same level to capture both local and global information, the authors further expand the convolutional module of Inception with two additional convolutional blocks to facilitate the learning of multi-scale features. The two blocks exploit different dilation rates to enable multiple receptive fields larger than the original inception module. The $n \times n$ convolutions are replaced with the consequent $1 \times n$ and $n \times 1$ convolutions with the aim to be more efficient.

Lachinov et al. \cite{lachinov2018glioma} propose a deep cascaded variant of U-Net, Cascaded Unet, to process multi-modal input for better performance regarding brain tumor segmentation. Despite the feasibility of the original U-Net to handle multi-modal MRI image input, it fuses the feature information of all the modalities which is processed in an identical manner. Based on the original U-Net, the proposed Cascaded Unet employs multiple encoders in parallel for better exploiting feature representations for each specific modality.
\begin{figure}
	\centering
	\includegraphics[width=\columnwidth]{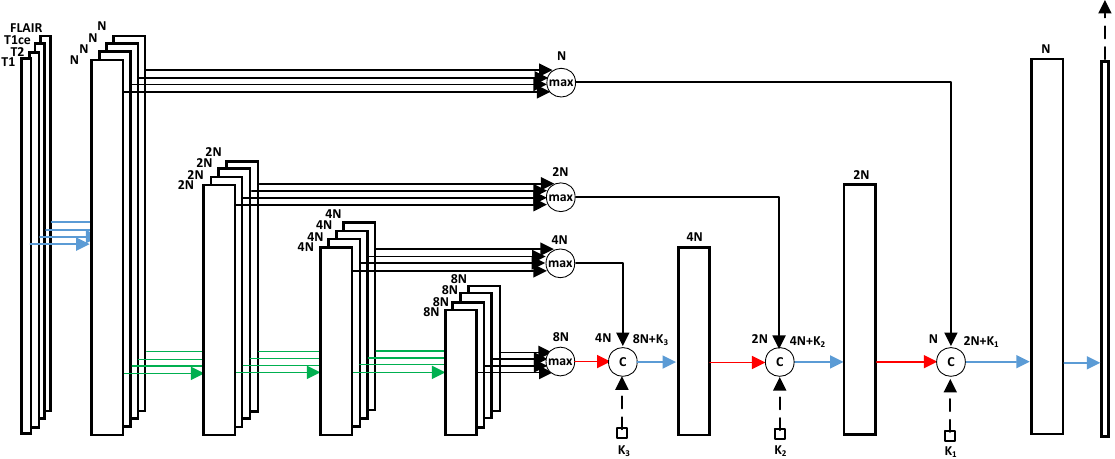}
	\caption{Architecture of Cascaded Unet with multiple encoders. T1, T2, T1CE, and FLAIR indicate several MRI modalities. N denotes the current number of filters. K is a number of filters in the context feature map produced from lower-scale models. At each stage of the encoding path across modalities, the resulting features are calculated as a maximum of the feature maps processed by encoders. In the decoder of the model, the output at each scale is obtained from the skip connections from the encoder, the decoder output from the previous stage, and the context connections. Figure from \cite{lachinov2018gliomaFig}.}
	\label{fig:CascadedUNet}
\end{figure}
The overall architecture of the Cascaded Unet model can be seen in \Cref{fig:CascadedUNet}.
The encoder path contains separate subpaths where every subpath utilizes a convolution group to process one input modality and generate feature maps. Then elementwise maximum operation acts on the multiple feature maps per stage to obtain the resulting features. The output of the feature map is afterward joined with the corresponding feature map of the larger-scale block, which boosts the information flow between the feature maps at different scales. The decoder of Cascaded UNet produces output at each level depending on the output at the same scale and the output of the decoder block at the previous stage. This strategy encourages the model to iteratively improve the results from earlier iterations.

\subsubsection{Leveraging Depth Information}
Some methods modify the U-Net into a 3D model and design modules to extract the information across channels in order to fully exploit the structural information of the third-dimension medical images. For improving automatic brain tumor prognosis, Islam et al. adapt the U-Net architecture to a 3D model and integrate the 3D attention strategy to perform image segmentation \cite{islam2019brain}. Compared to only skip connections, the introduced 3D attention model is aggregated into the decoder part of U-Net that includes channel and spatial attention in parallel with skip connections. The additional 3D attention layers encourage the module to encode richer spatial features from the original images.

\begin{figure}
	\centering
	\includegraphics[width=\columnwidth]{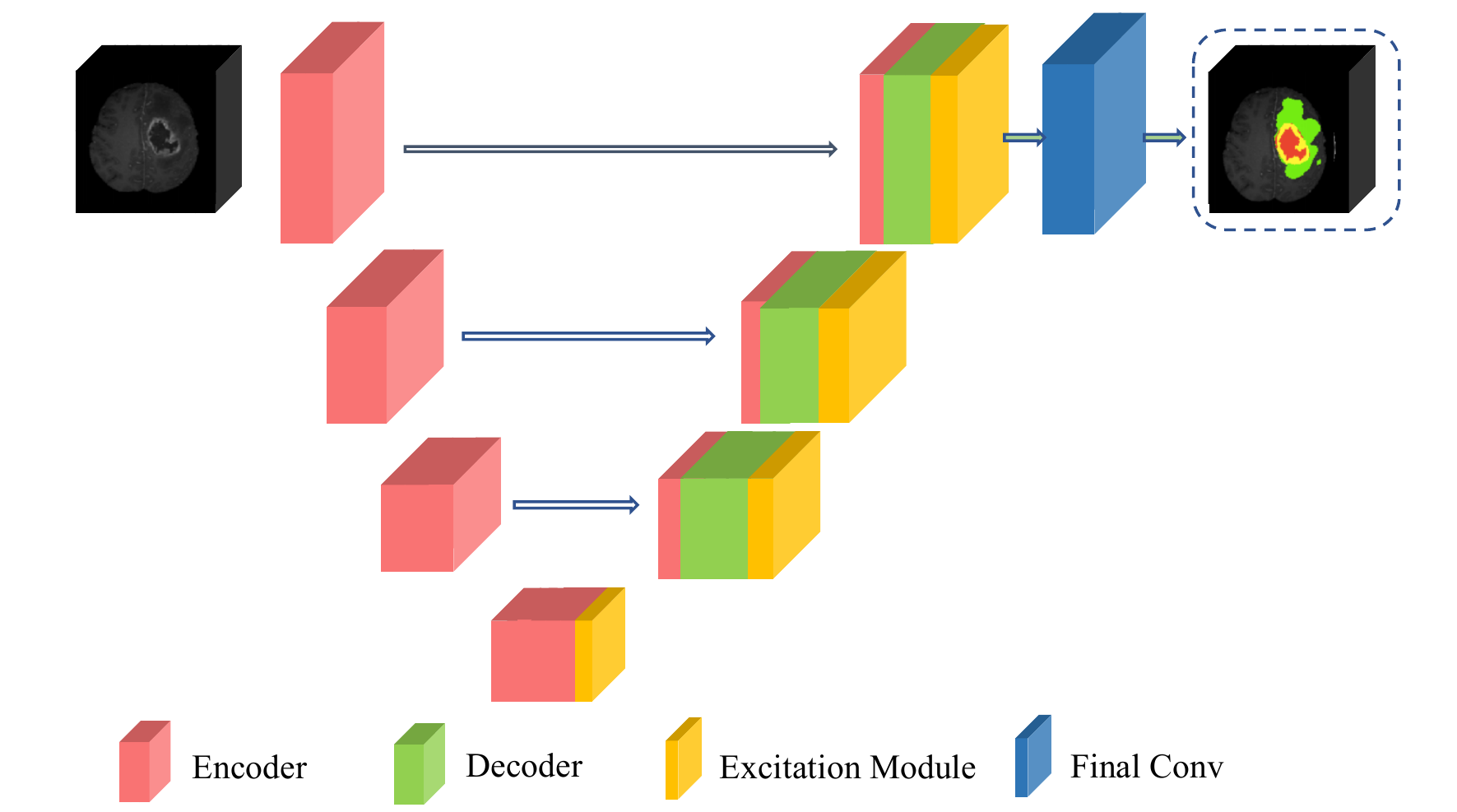}
	\caption{Architecture of 3D Attention U-Net with a channel and a spatial attention parallel to skip connections. Figure from \cite{islam2019brainFig}.}
	\label{fig:3d-attention-UNet}
\end{figure}

As shown in \Cref{fig:3d-attention-UNet}, the 3D attention U-Net is composed of a 3D encoder, the decoder, and skip connections combined with the channel and spatial attention mechanism.
In the path for 3D spatial attention, the authors perform $1 \times 1 \times C $ convolution on the input feature maps to obtain the result of the $H \times W \times 1$ dimension. In parallel, the input feature maps are passed through an average pooling and then fed to the fully-connected layer to get the $1 \times 1 \times C$ sequential channel correlation. Since the two paths capture features parallelly, the inconsistency and sparsity caused by the two excitations can be alleviated by fusing skip connections. Furthermore, the integration of skip connections can enhance the performance of segmentation prediction which can be inferred from the experiments on the BraTS 2019 dataset.

\subsection{Probabilistic Design} \label{sec:probablistic}
Another type of U-Net extension combines the classic U-Net with different types of probabilistic extensions. Depending on the task that should be achieved or the process that should be enhanced, different types of extensions from bayesian skip connections, over variational auto-encoders to Markov random fields are used, which are introduced in the following.

\subsubsection{Variational Auto Encoder (VAE) Regularization}
In medical image segmentation tasks, different graders often produce different segmentations. Most of these different segmentations are plausible as many medical images contain ambiguities that can not be resolved considering only the image at hand. Taking this into consideration, Kohl et al. learn a distribution over segmentations from an ambiguous input to produce an unlimited number of possible segmentations instead of just providing the most likely hypothesis \cite{kohl2018probabilistic}. In their approach, they combine a U-Net, for producing reliable segmentations, with a conditional variational autoencoder (CVAE), which can model complex distributions and encodes the segmentation variants in a low-dimensional latent space. The best results were obtained for a latent space of dimension 6.

\begin{figure}
	\centering
	\includegraphics[width=\columnwidth]{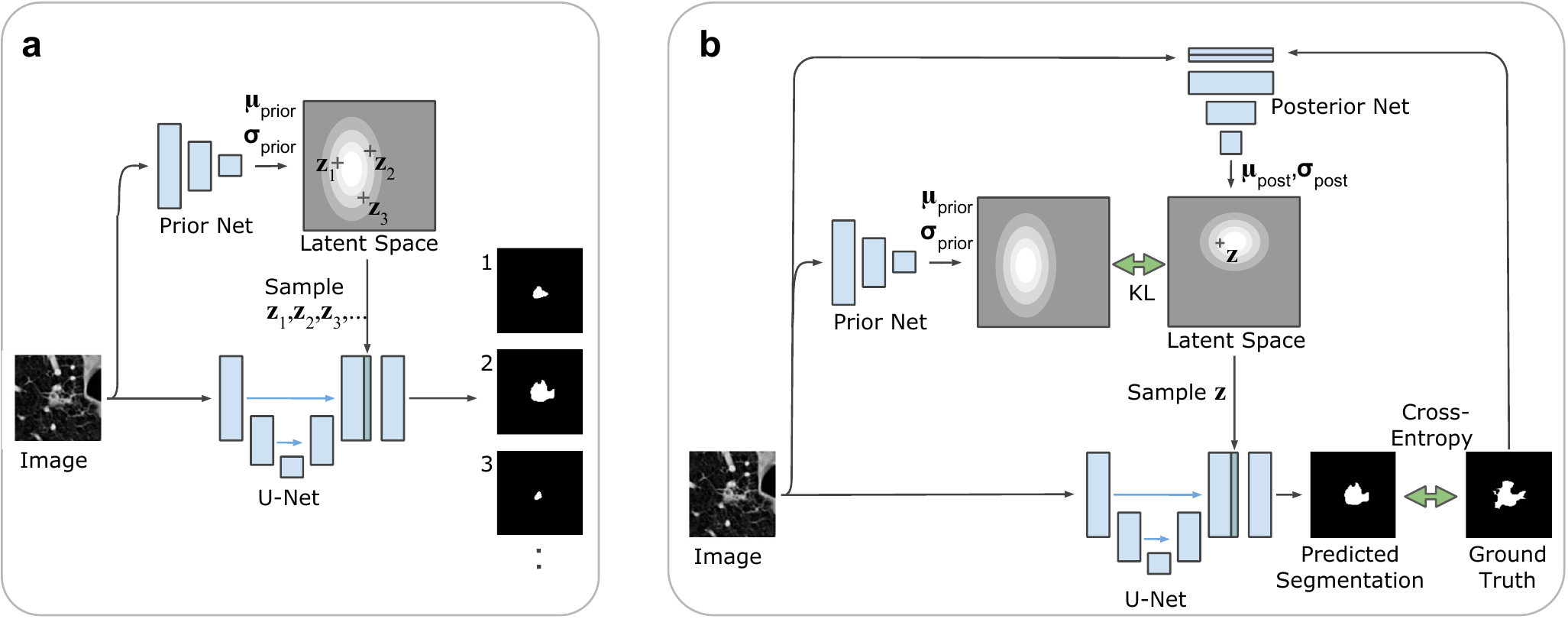}
	\caption{\textbf{(a)} Illustration of the sampling process of M segmentations with probabilistic U-Net. \textbf{(b)} Illustration of the training process of the probabilistic U-Net for one training sample. The green arrows represent loss functions. Figure from \cite{kohl2018probabilistic-arx}.}
	\label{fig:probabilisticUnet}
\end{figure}

\Cref{fig:probabilisticUnet} (a) shows the sampling process given a trained prior net and u-net as well as a low-dimensional latent space.
Each position in the latent space encodes a different segmentation variant. Passing the input image through the prior net, it will determine the probability of the encoded variants for the given input image.
For each possible segmentation to be predicted the network is applied to the same input image. A random sample from the prior probability distribution is drawn and broadcast to an N-channel feature map with the same shape as the segmentation map. It will then be concatenated with the final feature maps of the u-net and processed with successive $1\times1$ convolutions to produce the segmentation map corresponding to the drawn point from the latent space. Only the combination needs to be recalculated in each iteration, as the last feature maps of the U-Net and the output of the prior net can be reused for each hypothesis.\\
\Cref{fig:probabilisticUnet} (b) shows the training procedure of the probabilistic U-Net.

Apart from the standard training procedures for conditional VAEs and deterministic segmentation models, it has to be learned how to embed the segmentation variants in the latent space in a useful way. This is solved by the posterior net. It learns to recognize a segmentation variant and map it to a certain position in the latent space. A sample from its output posterior distribution combined with the activation map of the u-net must result in a segmentation identical to the ground truth segmentation. From this, it follows that the training data set must include a set of different but plausible segmentations for each input image.

Myronenko \cite{myronenko20183d} adds a VAE branch to a 3D U-Net architecture to address the problem of limited training data for brain tumor segmentation. In their architecture, the U-Net is used for the segmentation of the tumor, and the VAE is used for the reconstruction of the image sharing the same encoder. For the VAE, the output of the encoder is reduced to a lower dimensional space and a sample is drawn from the Gaussian distribution with the given mean and standard derivation (std). The sample is then reconstructed to the input image using an architecture similar to that of the U-Net decoder but without any skip connections. The total loss to be minimized during training is made up of three terms:
\begin{equation}
	\mathbf{L} = \mathbf{L}_\text{dice} + 0.1 \cdot \mathbf{L}_\text{L2} + 0.1 \cdot \mathbf{L}_\text{KL}
\end{equation}
$\mathbf{L}_\text{dice}$ is a soft dice loss between the predicted segmentation of the u-net and the GT segmentation.\\
$\mathbf{L}_\text{L2}$ and $\mathbf{L}_\text{KL}$ are the losses for the VAE where $\mathbf{L}_\text{L2}$ describes how well the reconstructed image matches the input image and $\mathbf{L}_\text{KL}$ is the Kullback-Leibler (KL) divergence between the estimates normal distribution and a prior distribution $\mathcal{N}(0,1)$. Using the VAE brach helps to better cluster the features at the end of the encoder. This helps to guide and regularize the shared encoder for small training set sizes. Adding the additional VAE branch, therefore, improved the performance and led to stable results for different random initializations of the network.

\subsubsection{Graphical Model Algorithm}
While the classic U-Net performs well on data from the same distribution as the training data, its accuracy decreases on out-of-distribution data.

\begin{figure}
	\centering
	\includegraphics[width=\columnwidth]{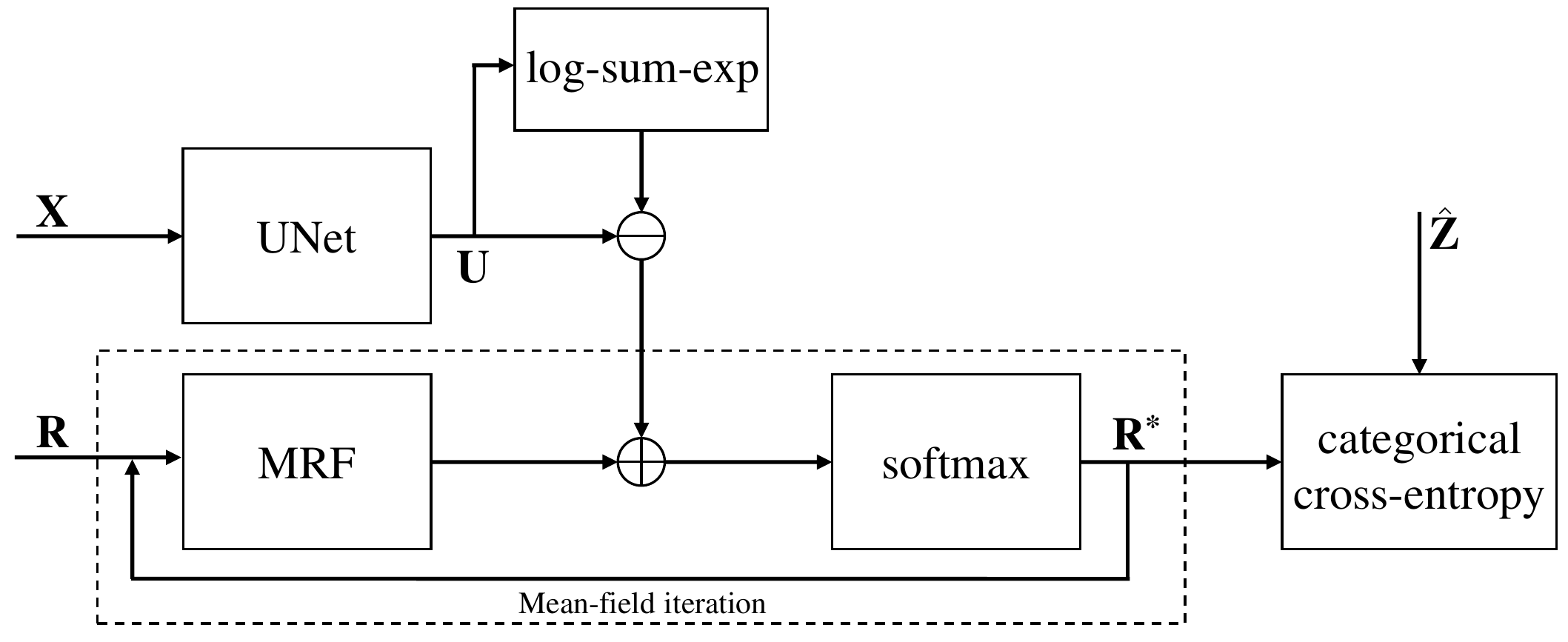}
	\caption{Schematic illustration of the MRF-UNet product. Figure from \cite{brudfors2021mrf-arx}.}
	\label{fig:mrfUnet}
\end{figure}

To address this problem, Brudfors et al. \cite{brudfors2021mrf} combine a U-Net with Markov random fields (MRFs) to form the MRF-Unet. The low-parameter, first-order MRFs are better at generalization because they encode simpler distributions which is an important quality to fit out of distribution data. The very accurate U-Net predictions make up for the fact that the MRFs are less flexible. The architecture of the proposed model can be seen in \Cref{fig:mrfUnet}. As the combination of the U-Net and MRF distribution is intractable by calculating the product of the two, an iterative mean-field approach is used to estimate the closest factorized distribution under the Kullback-Leibler divergence. A detailed mathematical derivation of the process can be found in the work by Brudfors et al. \cite{brudfors2021mrf}. Experiments showed that the combination of MRF and U-Net improved performance on in- and out-of-distribution data.
The lightweight MRF component, which does not add any additional parameters to the architecture, serves as a simple prior and therefore learns abstract label-specific features.

Klug et al. \cite{klug2020bayesian} use a Bayesian skip connection in an attention-gated 3D U-Net to allow a prior to bypass most of the network and be reintegrated at the final layer in their work to segment stroke lesions in perfusion CT images. The skip connection provides the prior to the final network layer and should reduce false-positive rates for small and patchy segmentations of varying shapes. As a prior, the segmentation of the ischemic core obtained by a standard thresholding method is used. Klug et al. \cite{klug2020bayesian} evaluated two ways to combine the prior and the output of the U-Net to calculate the final output segmentation: Addition and convolution of the two maps. Superior results were achieved by using convolution for combination in all experiments.

The input to the U-Net is the concatenation of the 3D perfusion CT image and the prior. When comparing a 3D attention gate U-Net to the same architecture with the bayesian skip connection additionally reintegrating the prior at the end of the network, the latter achieves a better performance in terms of dice score with faster convergence. It is worth mentioning that we have found excessive papers in which probabilistic design is integrated into the U-Net in applications such as for brain tumor \cite{savadikar2020brain}, and skin lesion segmentation \cite{chen2022medical}.

\begin{figure*}[!thb]
	\centering
	\includegraphics[width=\textwidth]{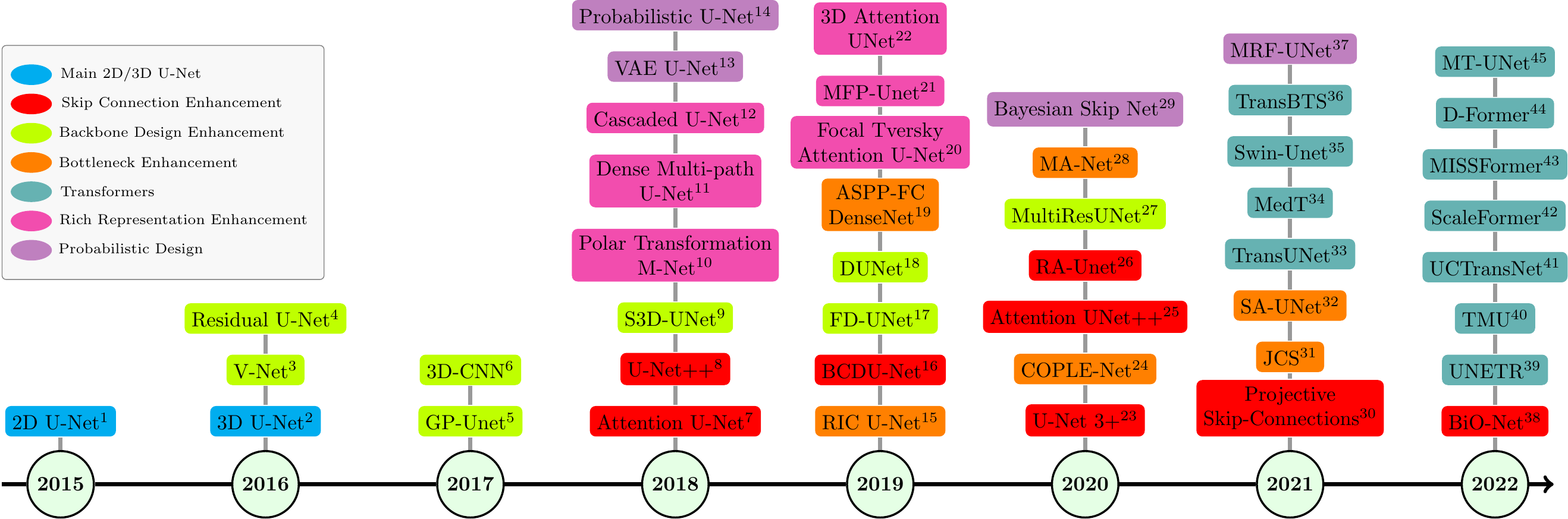}
	\caption{The timeline of prominent U-Net-based methods proposed in medical semantic segmentation literature, from 2015 to 2022. The superscripts in ascending order denote the 1. \cite{ronneberger2015u}, 2. \cite{cciccek20163d}, 3. \cite{milletari2016v}, 4. \cite{drozdzal2016importance}, 5. \cite{dubost2017gp}, 6. \cite{alakwaa2017lung}, 7. \cite{oktay2018attention}, 8. \cite{zhou2018unet++}, 9. \cite{chen2018s3d}, 10. \cite{fu2018joint}, 11. \cite{dolz2018dense}, 12. \cite{lachinov2018glioma}, 13. \cite{myronenko20183d}, 14. \cite{kohl2018probabilistic}, 15. \cite{zeng2019ric}, 16. \cite{azad2019bi}, 17. \cite{guan2019fully}, 18. \cite{jin2019dunet}, 19. \cite{hai2019fully}, 20. \cite{abraham2019novel}, 21. \cite{moradi2019mfp}, 22. \cite{islam2019brain}, 23. \cite{huang2020unet}, 24. \cite{wang2020noise}, 25. \cite{li2020attention}, 26. \cite{jin2020ra}, 27. \cite{ibtehaz2020multiresunet}, 28. \cite{fan2020ma}, 29. \cite{klug2020bayesian}, 30. \cite{lachinov2021projective}, 31. \cite{wu2021jcs}, 32. \cite{guo2021sa}, 33. \cite{chen2021transunet}, 34. \cite{valanarasu2021medical}, 35. \cite{cao2021swin}, 36. \cite{wang2021transbts}, 37. \cite{brudfors2021mrf}, 38. \cite{xiang2020bio}, 39. \cite{hatamizadeh2022unetr}, 40. \cite{reza2022contextual}, 41. \cite{wang2022uctransnet}, 42. \cite{huang2022scaleformer}, 43. \cite{huang2021missformer}, 44. \cite{wu2022d}, 45. \cite{wang2022mixed}, respectively.}
	\label{fig:timeline}
\end{figure*}

\begin{table*}[!thb]
	\centering
	\caption{The review of U-Net-like models for medical image segmentation based on the proposed taxonomy, \Cref{fig:unet-taxonomy}. \{\textit{SCE}, \textit{BDE}, \textit{BE}, \textit{T}, \textit{RRE} and \textit{PD}\} stands for \textit{Skip Connection Enhancement}, \textit{Backbone Design Enhancement}, \textit{Bottleneck Enhancement}, \textit{Transformer}, \textit{Rich Representation Enhancement} and \textit{Probabilistic Design}, respectively.} \label{tab:paperhighlights}
	\resizebox{\textwidth}{!}{
		\begin{tblr}{
				colspec={Q[l]Q[l,3cm]Q[l,7cm]Q[l,5.5cm]},
				rowspec={*{15}{Q[t]}},
				row{1}={font=\bfseries},
				cell{2-40}{1}={font=\itshape},
				cell{2-40}{1-4}={font=\tiny}
			}
			\toprule
			Strategy & Networks & Core Ideas & Practical Use Cases \\ \toprule
			SCE & {Attention U-Net \cite{oktay2018attention} \\ UNet++ \cite{zhou2018unet++,zhou2019unet++} \\ RA-Unet \cite{jin2020ra} \\BCDU-Net \cite{azad2019bi} \\ U-Net3+ \cite{huang2020unet} \\ BiO-Net \cite{xiang2020bio}} & {Skip connections are defined as connections in a neural network that do not connect two following layers but instead skip over at least one layer. This strategy initially aimed to encourage feature reusability and compensate for gradient vanishing in a deeper network. This modification introduces the feasibility of transferring high spatial localization features from the encoder to the decoder for better segmentation maps. In addition, some use cases used skip connections as hierarchical multi-scale fusion paths for feature enrichment in diverse U-Net stages. Furthermore, skip connection could efficiently decrease the semantic feature gaps between different layers and scales.} & {$\bullet$ Exploit multiscale features \cite{zhou2019unet++} \\$\bullet$ Robust boundary representation \cite{huang2020unet} \\$\bullet$ Bi-directional feature representation \cite{xiang2020bio} \\$\bullet$ Feature recalibration \cite{oktay2018attention} \\$\bullet$ Suppress irrelevant regions besides feature reusability \cite{jin2020ra} \\$\bullet$ Enrich semantic representation \cite{azad2019bi}} \\ \midrule
			BDE & {Residual U-Net \cite{he2016deep,drozdzal2016importance,milletari2016v} \\ Multi-Res U-Net \cite{ibtehaz2020multiresunet,szegedy2016rethinking} \\ Dense U-Net \cite{huang2017densely,guan2019fully} \\ H-DenseUNet \cite{li2018h} \\DUNet \cite{jin2019dunet,dai2017deformable} \\ S3D U-Net \cite{chen2018s3d}} & {The backbone defines how the layers in the encoder are arranged and its counterpart is therefore used to describe the decoder architecture. Ideally, the strong backbone design (e.g., inception model) with pre-trained weight can further improve the model generalization capability.} & {$\bullet$ Converges to lower loss \cite{li2018h} \\$\bullet$ Addressing gradient vanishing \cite{he2016deep} \\$\bullet$ Faster convergence rate \cite{drozdzal2016importance} \\$\bullet$ Feature reusability \cite{milletari2016v} \\$\bullet$ Multi-scale encoding \cite{ibtehaz2020multiresunet} \\$\bullet$ Better boundary representation \cite{szegedy2016rethinking} \\$\bullet$ Fine-grained feature set \cite{huang2017densely} \\$\bullet$ Cross-modality representation \cite{li2018h} \\$\bullet$ Reducing computation burden of multi-scale representation \cite{jin2019dunet} \\$\bullet$ Efficient multi-scale computation \cite{chen2018s3d}}\\ \midrule
			BE & {ASPP \cite{hai2019fully} \\ MA-Net \cite{fan2020ma} \\ COPLE-Net \cite{wang2020noise} \\ SA-UNet \cite{guo2021sa} \\ FRCU-Net \cite{azad2021deep} \\ JCS \cite{wu2021jcs} \\ MS-Net \cite{zhang2022multi}} & {The network bottleneck contains the compressed representation of the input data and provides necessary information (e.g., semantic, texture, shape features) to reconstruct the  segmentation map. Any improvement in the bottleneck design can further improve the prediction result.} & {$\bullet$ Frequency recalibration \cite{azad2021deep} \\$\bullet$ Spatial attention \cite{guo2021sa} \\$\bullet$ Feature pyramid \cite{hai2019fully,wang2020noise} \\$\bullet$ Imposing attention mechanism \cite{wu2021jcs}} \\ \midrule
			T & {TransUNet \cite{chen2021transunet} \\ TransBST \cite{wang2021transbts} \\ Swin-Unet \cite{cao2021swin} \\ UNETR \cite{hatamizadeh2022unetr} \\ TMU \cite{reza2022contextual} \\ UCTransNet \cite{wang2022uctransnet}} & {Transformers' critical point is to compensate for CNN's limited receptive field. Extracting long-range contextual information with intuition to look at the whole image at once is a promotional function of Transformers. Due to the 1D sequence mapping functionality of Transformers, they can use as a play-and-plug module at different parts of U-Net-like structures. However, due to the quadratic computational complexity nature, utilizing efficient Transformers' design even from the NLP field within vision architectures is beneficial. Since Transformers calculate the affinities between different parts of input data adaptively, utilizing them is a wise solution for multi-scale feature amalgamation.} & {$\bullet$ Improving CNN bottleneck's feature discriminancy \cite{chen2021transunet} \\$\bullet$ Capture inter-slice affinities from 3D data \cite{wang2021transbts} \\$\bullet$ Hierarchical Efficient Transformer-based design \cite{cao2021swin} \\$\bullet$ Modeling 3D volumetric data with global multi-scale information \cite{hatamizadeh2022unetr} \\$\bullet$ Feature re-calibration / degrading the boundary maps  erroneous \cite{reza2022contextual} \\$\bullet$ Decreasing the gap between multi-scale semantic features \cite{wang2022uctransnet}} \\ \midrule
			RRE & {Focal Tversky Attention U-Net \cite{abraham2019novel} \\ PT M-Net \cite{fu2018joint}\\Dense Multi-path U-Net \cite{dolz2018dense} \\MFP-Unet \cite{moradi2019mfp} \\ Cascaded Unet \cite{lachinov2018glioma}} &
			{The key objective is to enhance the performance of the trained models by utilizing all available information from multi-modal or multi-scale images while retaining the most desirable and relevant features. Some methods also operate directly on volumetric images to take full advantage of depth information.} & {$\bullet$ Improved precision and recall balance \cite{abraham2019novel} \\$\bullet$ Hierarchical representation learning \cite{fu2018joint} \\$\bullet$ Richer feature representation from combined modalities \cite{dolz2018dense} \\$\bullet$ Robust architecture regarding the capabilities of feature representation in a pyramid \cite{moradi2019mfp} \\$\bullet$ Boosted information flow between the different scales \cite{lachinov2018glioma}} \\
			\midrule
			PD & {Probabilistic U-Net \cite{kohl2018probabilistic} \\ MRF U-Net \cite{brudfors2021mrf} \\ VAE Regularization \cite{myronenko20183d} \\ Bayesian Skip \cite{klug2020bayesian}} & {In medical image segmentation tasks, different graders often produce different segmentations. Most of these different segmentations are plausible as many medical images contain ambiguities that can not be resolved considering only the shot at hand. Probabilistic models aim to model the nature of uncertainty in medical data.} & {$\bullet$ Modeling annotation uncertainty \cite{kohl2018probabilistic} \\$\bullet$ Addressing out-of-distribution \cite{brudfors2021mrf} \\$\bullet$ Imposing regularization to address data limitation \cite{myronenko20183d} \\$\bullet$ Encouraging feature-reusability and reduce FP rate \cite{klug2020bayesian}} \\ \bottomrule
		\end{tblr}
	}
\end{table*}

\subsection{Comparative Overview}
In this section, we briefly review the recent works regarding U-Net variants presented in \Cref{sec:skip-connection} to \ref{sec:probablistic} for medical image segmentation in \Cref{tab:paperhighlights}. It lists the related network list in each direction along with information about the core ideas and the practical use cases. As detailed in \Cref{tab:paperhighlights}, the modification dealing with skip connections is one of the directions for the extension of the U-Net structure. Some works redesign skip connections by increasing the number of forward skip connections or aggregating some modules within the skip connections for processing feature maps. Some methods also apply bi-directional LSTM to combine the feature maps from Encoder and Decoder. The novel skip connections in these mentioned methods enable the models to be more flexible and therefore explore local and semantic features more efficiently and from different scales. However, it also means a more complex design of the network architecture which leads to a larger number of parameters and more expensive computation.

Instead of altering skip connections, some proposed methods use other types of backbones apart from the original U-Net by Ronneberger et al. \cite{ronneberger2015u}. Residual mapping structure, inception modules, and dense-connections are incorporated into the architectures respectively. Such designs alleviate the vanishing gradient and degradation problems, which facilitates the faster convergence of the training process. Several approaches focus on adapting convolution operations such as using deformable convolutional kernels to make the network more general and robust. Nevertheless, a higher computational cost is needed due to the additional convolution layers.

In the third strategy, some approaches utilize other mechanisms in the bottleneck aiming at enhancing feature extraction and compressing more useful spatial information. Attention modules are employed to model long-range spatial dependencies between pixels in the bottle-neck feature maps. Atrous spatial pyramid pooling (ASPP) with different sampling rates can resample the compressed feature maps in the bottleneck separately, which helps to gain the output of bottleneck from different sizes of reception fields.

The rise of the Transformer, which is prevalent in the field of NLP, inspires the development of computer vision. The proposed ViT facilitates the new trend of the combination of U-Net and Transformer. The tokenized input images are passed through Transformer to extract global information that will supplement U-Net. Despite the state-of-the-art (SOTA) performance that the Transformer-based U-Net-like models have achieved, the large number of parameters of models sometimes cause a long time for convergence. Some models are also highly dependent on the pretrained weights.

The last direction combines the original U-Net with various types of probabilistic extension modules resulting in new types of variants. Probabilistic U-Net integrates the U-Net structure with a conditional variational autoencoder (CVAE) to generate an unlimited number of plausible prediction results when the inputs are obscure. Furthermore, the network with Markov Random Fields (MRF) has the advantage of preventing the model from overfitting with precise segmentations, which significantly improves the performance on out-of-distribution data. The methods in this direction demonstrate more or less robustness to defective input datasets, such as those of limited size or containing ambiguous images.

\Cref{fig:timeline} demonstrates the timeline of typical U-Net-based methods proposed in medical semantic segmentation literature from 2015 to 2022. As shown in the timeline, the U-Net structure has continued to be appealing in recent years regarding the task of medical image segmentation. The direction for the extension of U-Net is prominently influenced by Transformer after 2021 as a result of the emergence of ViT \cite{dosovitskiy2020image}.

\section{Quantitative Comparison} \label{sec:experiments}
In this section, we evaluate the performance of several of the
previously discussed U-Net variants on favored medical segmentation benchmarks. It is worth noting that although most models reported their performance on standard datasets and used standard metrics, some failed to do so, making across-the-board comparisons difficult. Furthermore, only a small percentage of publications provide additional information, such as hyper-parameters, execution time, and memory footprint, in a reproducible way, which is essential for industrial and real-world applications. To compensate for the gap between the solid foundation for comparison of the architectures, we conducted several studies on them in fair conditions to empower the equilibrium insights on their performance.

\subsection{Implementation details}\label{sec:implementation-detail}
In this section, we discuss our experiments and selected networks criteria. Before diving in deep, we should mention that all our implementation code is done in Python with the PyTorch library \cite{NEURIPS2019_9015}. We used a single Nvidia RTX 3090 GPU for training the networks. As a baseline network, we select the 2D U-Net \cite{ronneberger2015u} to build our comparison upon the naive U-Net and explore the effect of modifications to U-Net. Next will be the Attention U-Net \cite{oktay2018attention}, which is the pioneering method to integrate the attention mechanism with U-Net to enhance the feature reusability and a feature selection measure to make the features more discriminant. U-Net++ \cite{zhou2019unet++} brings the highly dense skip connection schema to U-Net to decrease the semantic gap between down-sampling and up-sampling paths. However, some methods tried to surpass the convolution operations of local receptive fields by using dense backbones to U-Net, but not only this procedure still lacks the real global context, but it also adds more parameters to the model, which is not desirable. To this end, Residual U-Net \cite{zhang2018road} brings a neighboring affinity recipe to hinder locality problems via RCNN blocks. Although this model originally applied to non-medical data, it demonstrated robust results on the medical images too. MultiResUNet \cite{ibtehaz2020multiresunet} by modeling the multi-modal information in the backbone is another intuition that we will see its contribution over the base U-Net, that which is more successful than the other in compensating the locality of CNN-based methods. Lastly, we selected three Transformer-utilized U-shaped networks, TransUNet \cite{chen2021transunet}, UCTransNet \cite{wang2021uctransnet} and MISSFormer \cite{huang2021missformer}, to demonstrate the Transformer evolution to the medical segmentation field by effectively capturing global contextual information. We should also note that no pretraining weights were utilized during the training process for any of the networks. Even though the pretraining weights on Imagenet might bring some advantages, we dropped the pretraining weight to provide a fair evaluation criterion. 

\subsection{Datasets}\label{sec:datasets}
To demonstrate a fair and productive comparison on the \cref{sec:implementation-detail} networks, we selected several datasets from the diverse modalities for the semantic segmentation task. In this respect, we consider \textbf{S}egmentation of \textbf{M}ultiple \textbf{M}yeloma \textbf{P}lasma \textbf{C}ells, \textit{SegPC} \cite{gupta2018pcseg,gupta2020gcti,gehlot2020ednfc}, which is a collection of 2D microscopic images for cancer screening to aid hematologists in better diagnosis. \textit{ISIC 2018} \cite{codella2019skin} is another 2D Dermoscopic dataset from the skin lesions for assisting dermatologists with an early-stage cancer diagnosis. Next is the multi-organ 3D CT \textit{Synapse} \cite{landman2015miccai} dataset covering the 13 organs' annotations which are published through the Synapse website \cite{synapse2015ct}. Here we want to clarify that the Synapse dataset is also known as \textbf{B}eyond the \textbf{C}ranial \textbf{V}ault (\textit{BCV} or \textit{BTCV}), so these two names using interchangeably, but there is a slight difference between the Synapse and the BCV datasets used in studies. Most of the studies used the Synapse dataset name in their works \cite{chen2021transunet} using the eight organ classes annotation; the rest used the number of classes for their report varies from eleven \cite{xie2021cotr} to twelve \cite{hatamizadeh2022unetr}. 

\subsubsection{SegPC}
For decades, automatic cell segmentation in microscopy images was studied, and various methods were developed \cite{bozorgpour2021multi,gupta2022segpc}. Multiple Myeloma (MM) is a type of blood cancer, specifically, a plasma cell cancer. The first stage of building an automated diagnostic tool for MM is the robust segmentation of cells. Segmenting plasma cells in these microscopic stained images is quite complex due to the diversity of situations. For instance, cells may appear in clusters or isolated, with varying nuclei and cytoplasm sizes, some of which touch each other with overlapping boundaries. The SegPC includes images captured from the bone marrow aspirate slides of MM patients. The current dataset has 775 images of MM plasma cells. In this study, we sort the training set according to the single entity's name and choose 70\% of the set as a train set, 10\% for the validation set, and 20\% for the test set. Also, we applied the resizing step to all images to a fixed size of $224 \times 224$. The SegPC dataset slides were stained using Jenner-Giemsa stain and contain the annotation for the Nucleus and Cytoplasm of Plasma cells. It should be noted that for this dataset, we only apply networks for the Cytoplasm segmentation task after cropping the Nucleus samples in each image.

\subsubsection{ISIC 2018}
Human skin tissue consists of three types, i.e., dermis, epidermis, and hypodermis. The epidermis is a susceptible tissue, which under severe solar radiation, could trigger the embedded melanocytes to produce melanin at a significant level \cite{azad2020attention}. Fatal skin cancer is a result of melanocyte growth, which is known as melanoma. In 2022, the American Cancer Society reported approximate melanoma skin cancer cases of 99,780, with death cases of 7,650, 7.66\% of all cases \cite{siegel2022cancer}. Early disease recognition plays a crucial role in medical diagnosis, where it reported that detection of melanoma in early phases could increase the relative survival rate to 92\%. However, robust skin lesion segmentation is a pretty challenging task due to the diverse lesion sizes, illumination changes, differences in texture, position, colors, and presence of unwanted objects like air bulbs, hair, or ruler markers. The ISIC 2018 \cite{codella2019skin} dataset was published by the \textit{International Skin Imaging Collaboration (ISIC)} as a large-scale dataset of dermoscopy images. It includes 2,594  RGB images of $700 \times 900$ pixels size. First, we resized all images to a $224 \times 224$ pixels size, and then like \cite{alom2019recurrent}, we used 1,815 images for training, 259 for validation, and 520 for the testing steps. The dataset consists of two class annotations: cancer or non-cancer lesions' heat map.

\subsubsection{Synapse}
The Synapse dataset is a multi-organ segmentation dataset \cite{landman2015miccai}, which was presented with the 30 abdominal CT scans provided in conjunction with the \textit{MICCAI 2015 Conference}, with 3,779 axial contrast-enhanced abdominal clinical CT images. Each CT volumetric data consists of $85 \sim 198$ slices of a consistent size $512 \times 512$ through whole samples. The spatial resolution of each voxel are $\left[0.54 \sim 0.54\right] \times \left[0.98 \sim 0.98\right] \times \left[2.5 \sim 5.0\right]$ mm$^{3}$ in each axis. In this report, we used the same preferences for data preparation analogous to \cite{chen2021transunet,azad2022transdeeplab,heidari2022hiformer}, randomly allocated 18 training cases and 12 cases for validation, and during the testing phase, we reported the final scores on the validation set. In addition, all slices resized to $224 \times 224$ to follow the same setting as \cite{chen2021transunet,cao2021swin}. We used eight classes annotation for our experiments with class names: Aorta, Gallbladder, Spleen, Kidney (L/R), Liver, Pancreas, Spleen, and Stomach. Our training process uses 2D data (similar to \cite{chen2021transunet,cao2021swin}) and reports the test results on the 3D volume.

\subsection{Loss Functions}\label{sec:loss-funcs}
Selecting the proper loss/objective function while designing the complex segmentation network is extremely important. There are a variety of loss functions to choose and train the network, but in this study, we determined the two well-known and traditional loss functions in the medical image segmentation domain: \textbf{C}ross \textbf{E}ntropy (\textit{CE}) and \textbf{D}ice \textbf{S}ørensen \textbf{C}oefficient (\textit{DSC}) loss functions.

\subsubsection{CE Loss}
CE \cite{zhang2018generalized} derives from the Kullback-Leibler (KL) divergence, a measure of dissimilarity between two distributions. In the segmentation task, CE is formulated as:
\begin{align}
	{L_{CE}} =  - \frac{1}{N}\sum\limits_{i = 1}^C {\sum\limits_{j = 1}^N {p_j^i\log q_j^i}},
\end{align}
where $p_j^i$ denotes the ground truth binary indicator of class label $i$ of voxel $j$, and $q_j^i$ is the corresponding predicted segmentation probability.

\subsubsection{Dice Loss}
\textbf{S}ørensen \textbf{D}ice \textbf{C}oefficient (\textit{DSC}) is typically used to evaluate the similarity between two samples, which will discuss in \Cref{sec:eval-metrics}. Based on this coefficient, \Cref{eq:dicecoeff}, the Dice loss was introduced in 2017 by Sudre et al. \cite{sudre2017generalised} and formulated as:
\begin{align}
	L_{Dice}(y,\hat{y})=1-\frac{2y\hat{y}+\alpha}{y+\hat{y}+\alpha},
\end{align}
where $y$ and $\hat{y}$ are actual and predicted values by model, respectively. $\alpha$ protects the function to not be undefined in edge case scenarios, e.g., $y=\hat{y}=0$.

\subsection{Evaluation Metrics}\label{sec:eval-metrics}
Unquestionably, a model should be examined from numerous aspects, such as several accuracy metrics, speed, and memory occupation efficiency. Nevertheless, most studies evaluate their model based on different accuracy metrics. This section brings brief keynotes on various accuracy metrics used in other research. Even though quantitative accuracy metrics are used to evaluate a segmentation algorithm on diverse benchmark datasets, since the ultimate goal of these approaches in computer vision is to apply them to real-world problems, the model's visual quality should also be considered.

\textbf{$\bullet$~Precision / Recall / F1 score ---} 
These are the most popular metrics for evaluating the accuracy of the segmentation models, either classical or deep learning-based methods resulting from the confusion matrix \cite{minaee2021image}. Precision and recall can be defined for each class or at the aggregate level as follows:
\begin{align}
	\text{Precision}= \frac{\text{TP}}{\text{TP} + \text{FP}}, \quad \text{Recall} = \frac{\text{TP}}{\text{TP}+ \text{FN}},
\end{align}
where TP stands for the true positive fraction, FP refers to the false-positive fraction and FN refers to the false-negative fraction. Recall in the segmentation context known also as sensitivity, which calculates the proportion of correctly labeled foreground pixels discarding background pixels. Usually, we are interested in a combined version of precision and recall rates, so a famous metric is called the F1 score, which is defined as the harmonic mean of precision and recall as follows:
\begin{align}
	F_1 =\frac{2 \text{Prec.}\: \text{Rec.}}{\text{Prec.}+\text{Rec.}}= \frac{2\text{TP}}{2\text{TP} + \text{FP} + \text{FN}} \label{eq:f1-score}
\end{align}

\textbf{$\bullet$~Accuracy ---}
Refers to as \textbf{Class Average Accuracy}, calculates the ratio of correct pixels to the whole mask based on each class. It is an updated version of pixel accuracy, which describes the percent of pixels in a predicted segmentation mask that is classified correctly. Accuracy is defined as:
\begin{align}
	Acc = \frac{1}{k} \sum^{k}_{j=1}\frac{p_{jj}}{g_j},
\end{align}
where $p_{jj}$ is the number of pixels that are classified in the correct class $j$, and $g_j$ is the total number of pixels of the ground truth for class $j$. However, the class imbalance prevalent in medical datasets will result in undesirable performance.

\textbf{$\bullet$~Intersection over Union (IoU) ---}
Also known as \textbf{Jaccard Index}, is a measure to describe the extent of overlap of predicted segmentation mask with ground truth. It is defined as the intersection area between the predicted segmentation and the ground truth, divided by the area of union between the predicted segmentation mask and the ground truth:
\begin{align}
	\text{IoU} = J(A, B) = \frac{\lvert A \cap B\rvert}{A \cup B} = \frac{\text{TP}}{\text{TP} + \text{FP} +  \text{FN}},
\end{align}
where A and B denote the ground truth and the predicted segmentation, respectively, it goes between $0$ and $1$.

\textbf{$\bullet$~Dice Coefficient ---}
This metric commonly applied in medical image analysis defines as the ratio of the twice overlapping region of ground truth ($G$) and prediction ($P$) maps to the total pixels of ground truth and predicted area. The Dice coefficient for a certain class can be formulated as:
\begin{align}
	Dice = \frac{2\vert G \cap P\vert}{\vert G\vert+\vert P\vert}, \label{eq:dicecoeff}
\end{align}
When Dice is used for binary segmentation maps, the Dice score is equivalent to the F1 score. From \Cref{eq:f1-score}, it is evident that this metric focuses on foreground pixels' accuracy and penalizes them for wrong prediction labels.

\textbf{$\bullet$~Hausdorff Distance ---}
This metric is one of the extensively used metrics to indicate the segmentation error. It computes the longest distance, Euclidean distance, from a point in the ground truth contour $\delta q$ to one point in segmented contour $\delta p$ as follows:
\begin{align}
    HD(X,Y)= max( hd(\delta p,\delta q) , hd(\delta q,\delta p)),
\end{align}
where $hd(\delta p,\delta q)$ and $hd(\delta q,\delta p)$ stand for the one-sided HD from $\delta p$ to $\delta q$ and from $\delta q$ to $\delta p$ respectively as follows:
\begin{align}
hd(\delta q,\delta p) &= max_{q \in \delta q}min_{p \in \delta p}||q-p||_2, \nonumber \\
hd(\delta p,\delta q) &= max_{p \in \delta p}min_{q \in \delta q}||q-p||_2.
\end{align}

\subsection{Experimental Results}
\label{sec:experimental-results}

\subsubsection{Skin lesion (ISIC 2018)}
Skin lesion segmentation is challenging due to various artifacts in dermoscopic images, such as ruler markers, water bubbles, dark corners, and strands of hair \cite{hasan2020dsnet}. \Cref{tab:er-isic2018} presents the comparison results for ISIC 2018 \cite{codella2019skin} dataset. It can be seen that the Dice score as a prominent metric is almost the same among CNN and Transformer based methods. The skin lesion usually appears in the texture and does not follow a specific shape or geometrical pattern. This might explain why transformer-based networks might not bring more advantages to texture-related patterns. It also can been seen that a network with multi-scale feature description capacity, e.g., U-Net++ \cite{zhou2018unet++} and UCTransNet \cite{wang2022uctransnet} are highly capable to localize abnormal regions comparing to other extensions. Overall, the model with multi-scale representation surpasses other extensions in both CNN and Transformer based approaches. 

\begin{figure*}[ht]
\captionsetup[subfigure]{labelformat=empty,}
\begin{subfigure}{.099\textwidth}
	\centering
	\includegraphics[width=\linewidth]{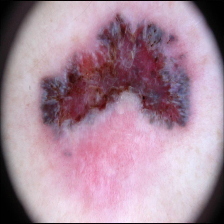}
	\caption{}
	\label{fig:er-isic-img-1}
\end{subfigure}\hfill
\begin{subfigure}{.099\textwidth}
	\centering
	\includegraphics[width=\linewidth]{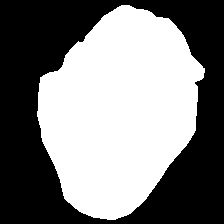}
	\caption{}
	\label{fig:er-isic-gt-1}
\end{subfigure}\hfill
\begin{subfigure}{.099\textwidth}
	\centering
	\includegraphics[width=\linewidth]{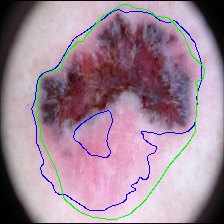}
	\caption{}
	\label{fig:er-isic-unet-1}
\end{subfigure}\hfill
\begin{subfigure}{.099\textwidth}
	\centering
	\includegraphics[width=\linewidth]{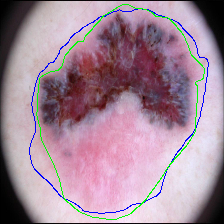}
	\caption{}
	\label{fig:er-isic-unetpp-1}
\end{subfigure}\hfill
\begin{subfigure}{.099\textwidth}
	\centering
	\includegraphics[width=\linewidth]{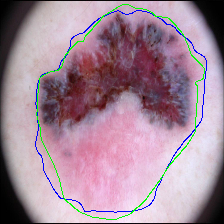}
	\caption{}
	\label{fig:er-isic-attunet-1}
\end{subfigure}\hfill
\begin{subfigure}{.099\textwidth}
	\centering
	\includegraphics[width=\linewidth]{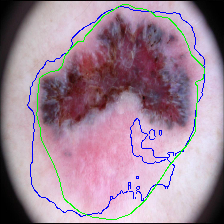}
	\caption{}
	\label{fig:er-isic-resunet-1}
\end{subfigure}\hfill
\begin{subfigure}{.099\textwidth}
	\centering
	\includegraphics[width=\linewidth]{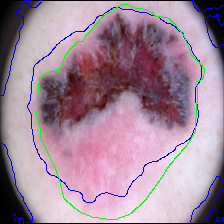}
	\caption{}
	\label{fig:er-isic-multiresunet-1}
\end{subfigure}\hfill
\begin{subfigure}{.099\textwidth}
	\centering
	\includegraphics[width=\linewidth]{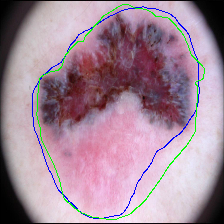}
	\caption{}
	\label{fig:er-isic-transunet-1}
\end{subfigure}\hfill
\begin{subfigure}{.099\textwidth}
	\centering
	\includegraphics[width=\linewidth]{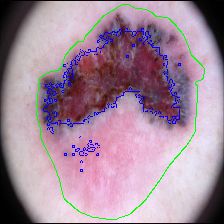}
	\caption{}
	\label{fig:er-isic-missformer-1}
\end{subfigure}\hfill
\begin{subfigure}{.099\textwidth}
	\centering
	\includegraphics[width=\linewidth]{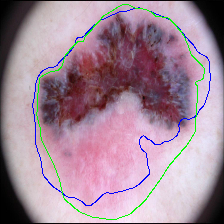}
	\caption{}
	\label{fig:er-isic-uctransnet-1}
\end{subfigure}\hfill
\\[-20.5pt]
\begin{subfigure}{.099\textwidth}
	\centering
	\includegraphics[width=\linewidth]{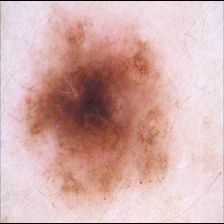}
	\caption{}
	\label{fig:er-isic-img-2}
\end{subfigure}\hfill
\begin{subfigure}{.099\textwidth}
	\centering
	\includegraphics[width=\linewidth]{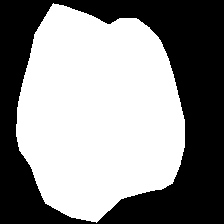}
	\caption{}
	\label{fig:er-isic-gt-2}
\end{subfigure}\hfill
\begin{subfigure}{.099\textwidth}
	\centering
	\includegraphics[width=\linewidth]{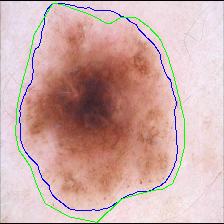}
	\caption{}
	\label{fig:er-isic-unet-2}
\end{subfigure}\hfill
\begin{subfigure}{.099\textwidth}
	\centering
	\includegraphics[width=\linewidth]{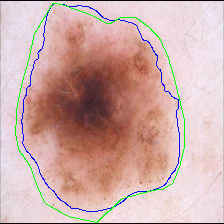}
	\caption{}
	\label{fig:er-isic-unetpp-2}
\end{subfigure}\hfill
\begin{subfigure}{.099\textwidth}
	\centering
	\includegraphics[width=\linewidth]{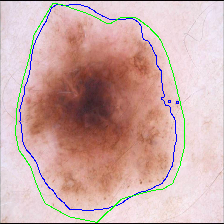}
	\caption{}
	\label{fig:er-isic-attunet-2}
\end{subfigure}\hfill
\begin{subfigure}{.099\textwidth}
	\centering
	\includegraphics[width=\linewidth]{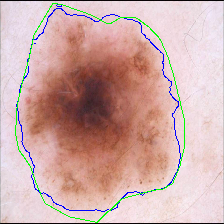}
	\caption{}
	\label{fig:er-isic-resunet-2}
\end{subfigure}\hfill
\begin{subfigure}{.099\textwidth}
	\centering
	\includegraphics[width=\linewidth]{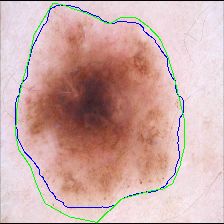}
	\caption{}
	\label{fig:er-isic-multiresunet-2}
\end{subfigure}\hfill
\begin{subfigure}{.099\textwidth}
	\centering
	\includegraphics[width=\linewidth]{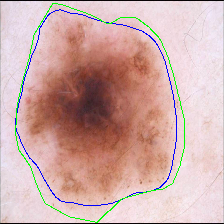}
	\caption{}
	\label{fig:er-isic-transunet-2}
\end{subfigure}\hfill
\begin{subfigure}{.099\textwidth}
	\centering
	\includegraphics[width=\linewidth]{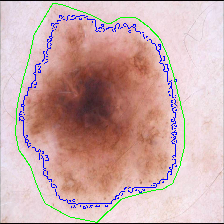}
	\caption{}
	\label{fig:er-isic-missformer-2}
\end{subfigure}\hfill
\begin{subfigure}{.099\textwidth}
	\centering
	\includegraphics[width=\linewidth]{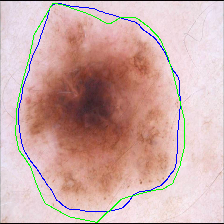}
	\caption{}
	\label{fig:er-isic-uctransnet-2}
\end{subfigure}\hfill
\\[-20.5pt]
\begin{subfigure}{.099\textwidth}
	\centering
	\includegraphics[width=\linewidth]{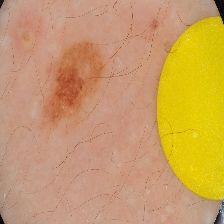}
	\caption{\scriptsize{Input Image}}
	\label{fig:er-isic-img-3}
\end{subfigure}\hfill
\begin{subfigure}{.099\textwidth}
	\centering
	\includegraphics[width=\linewidth]{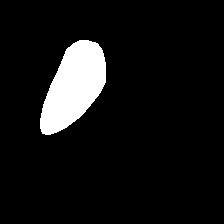}
	\caption{\scriptsize{Target (GT)}}
	\label{fig:er-isic-gt-3}
\end{subfigure}\hfill
\begin{subfigure}{.099\textwidth}
	\centering
	\includegraphics[width=\linewidth]{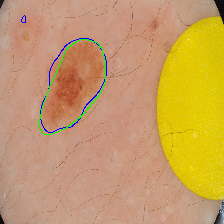}
	\caption{\scriptsize{UNet}}
	\label{fig:er-isic-unet-3}
\end{subfigure}\hfill
\begin{subfigure}{.099\textwidth}
	\centering
	\includegraphics[width=\linewidth]{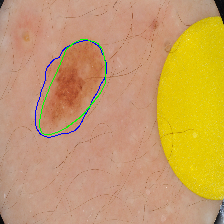}
	\caption{\scriptsize{UNet++}}
	\label{fig:er-isic-unetpp-3}
\end{subfigure}\hfill
\begin{subfigure}{.099\textwidth}
	\centering
	\includegraphics[width=\linewidth]{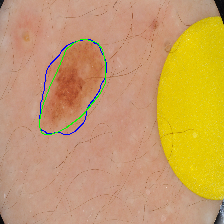}
	\caption{\scriptsize{AttUNet}}
	\label{fig:er-isic-attunet-3}
\end{subfigure}\hfill
\begin{subfigure}{.099\textwidth}
	\centering
	\includegraphics[width=\linewidth]{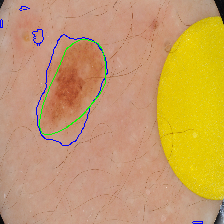}
	\caption{\scriptsize{ResUNet}}
	\label{fig:er-isic-resunet-3}
\end{subfigure}\hfill
\begin{subfigure}{.099\textwidth}
	\centering
	\includegraphics[width=\linewidth]{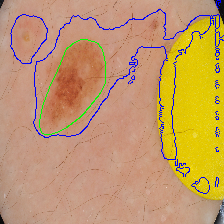}
	\caption{\scriptsize{MultiResUNet}}
	\label{fig:er-isic-multiresunet-3}
\end{subfigure}\hfill
\begin{subfigure}{.099\textwidth}
	\centering
	\includegraphics[width=\linewidth]{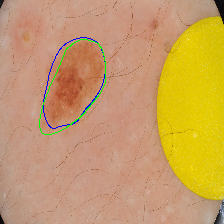}
	\caption{\scriptsize{TransUNet}}
	\label{fig:er-isic-transunet-3}
\end{subfigure}\hfill
\begin{subfigure}{.099\textwidth}
	\centering
	\includegraphics[width=\linewidth]{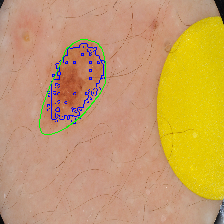}
	\caption{\scriptsize{MISSFormer}}
	\label{fig:er-isic-missformer-3}
\end{subfigure}\hfill
\begin{subfigure}{.099\textwidth}
	\centering
	\includegraphics[width=\linewidth]{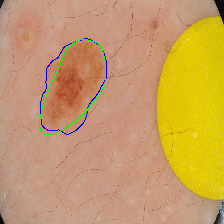}
	\caption{\scriptsize{UCTransNet}}
	\label{fig:er-isic-uctransnet-3}
\end{subfigure}\hfill
\caption{Visual comparisons of different methods on the \textit{ISIC 2018} skin lesion segmentation dataset. Ground truth boundaries are shown in \textcolor{green}{green}, and predicted boundaries are shown in \textcolor{blue}{blue}.}
\label{fig:er-isic2018}
\end{figure*}

For further investigation, we provided some segmentation results in \Cref{fig:er-isic2018}.
Although the same dice trend endorses the quantitative results among the CNN and Transformer models (\Cref{tab:er-isic2018}), the multi-scale representation derived from the global contextual modeling of the Transformer models enables these architectures to provide more smooth segmentation results even when the background pixels have a high overlap with the skin lesion class. In addition, the CNN-based networks generally are more likely to over-segment or under-segment the lesions comparing to the Transformer-based models. Since UCTransNet \cite{wang2022uctransnet} has the advantages of CNN hierarchical design and multi-scale feature fusion property within the CTrans module, it provides the SOTA results among the selected methods. The critical difference between the multi-scale representation learning with MISSFormer \cite{huang2021missformer} and UCTransNet \cite{wang2022uctransnet} is underlay the utilizing the down-sampling factor with in Efficient Transformer block, which degrades the feature representation (see \Cref{sec:standalone-transformer} and \Cref{fig:missformer-block} for more detail). However, UCTransNet utilizes the CNN-based feature representation instead of the Transformer-based backbone in a U-shaped structure. This representation makes the feature more discriminant, and the semantic gap between the encoder and decoder path successfully lessen by the CTrans module.

\begin{table*}[tb]
\caption{Performance comparison on \textit{ISIC 2018}, \textit{SegPC 2021} and \textit{Synapse} datasets (best results are bolded).}
\vspace{0.2\baselineskip}
\begin{subtable}[t]{0.48\textwidth}
	\centering
	\caption{\textit{ISIC 2018}} \label{tab:er-isic2018}
    \setlength{\tabcolsep}{4pt} %
    \renewcommand*{\arraystretch}{1.11}  %
    \vspace{-0.3\baselineskip}
	\scriptsize{
	\begin{tabular}{l | c c c c c c  }
		\hline
		\textbf{Methods} & 
		\textbf{AC} & 
		\textbf{PR} &  
		\textbf{SE} & 
		\textbf{SP} & 
		\textbf{Dice} & 
		\textbf{IoU} \\
		\hline
		U-Net \cite{ronneberger2015u} & 0.9446 & 0.8746 &  0.8603 & 0.9671 &  0.8674 & 0.8491 \\
		Att-UNet \cite{oktay2018attention} & 0.9516 & 0.9075 & 0.8579 & 0.9766 & 0.8820 & 0.8649 \\
		U-Net++ \cite{zhou2018unet++} & 0.9517 & 0.9067 & 0.8590 & 0.9764 & 0.8822 & 0.8651 \\
		MultiResUNet \cite{ibtehaz2020multiresunet} & 0.9473 & 0.8765 & 0.8689 & 0.9704 & 0.8694 & 0.8537 \\
		Residual U-Net \cite{zhang2018road} & 0.9468 &  0.8753 & 0.8659 & 0.9688 & 0.8689 & 0.8509 \\
		TransUNet \cite{chen2021transunet} & 0.9452 &  0.8823&  0.8578 &  0.9653 & 0.8499 &  0.8365\\
		UCTransNet \cite{wang2022uctransnet} & \textbf{0.9546} & \textbf{0.9100} & \textbf{0.8704} & \textbf{0.9770} & \textbf{0.8898} & \textbf{0.8729} \\
		MISSFormer \cite{huang2021missformer}  & 0.9453 & 0.8964 & 0.8371& 0.9742 & 0.8657 & 0.8484 \\
		\hline
	\end{tabular}
	}
\end{subtable}
\hfill
\begin{subtable}[t]{0.48\textwidth}
    \centering
    \caption{\textit{SegPC 2021}} \label{tab:er-segpc2021}
    \setlength{\tabcolsep}{4pt} %
    \renewcommand*{\arraystretch}{1.11}  %
    \vspace{-0.3\baselineskip}
    \scriptsize{
    \begin{tabular}{l | c c c c c c c }
    	\hline
    	\textbf{Methods} & 
    	\textbf{AC} & 
    	\textbf{PR} &  
    	\textbf{SE} & 
    	\textbf{SP} & 
    	\textbf{Dice} & 
    	\textbf{IoU} \\
    	\hline
    	U-Net \cite{ronneberger2015u} & 0.9795 & 0.9084 &  0.8548 & 0.9916 &  0.8808 & 0.8824 \\
    	Att-UNet \cite{oktay2018attention} & 0.9854 & 0.9360 & 0.8964 & 0.9940 & 0.9158 & 0.9144 \\
    	U-Net++ \cite{zhou2018unet++} & 0.9845 & 0.9328 & 0.8887 & 0.9938 & 0.9102 & 0.9092 \\
    	MultiResUNet \cite{ibtehaz2020multiresunet} & 0.9753 & 0.8391 & 0.8925 & 0.9834 & 0.8649 & 0.8676 \\
    	Residual U-Net \cite{zhang2018road} & 0.9743 &  0.8920 & 0.8080 & 0.9905 & 0.8479 & 0.8541 \\
    	TransUNet \cite{chen2021transunet} & 0.9702 & 0.8678 & 0.7831 & 0.9884 & 0.8233 & 0.8338 \\
    	UCTransNet \cite{wang2022uctransnet} & \textbf{0.9857} & \textbf{0.9365} & \textbf{0.8991} & \textbf{0.9941} & \textbf{0.9174} & \textbf{0.9159} \\
    	MISSFormer \cite{huang2021missformer} & 0.9663 & 0.8152 & 0.8014 & 0.9823 & 0.8082 & 0.8209\\
    	\hline
    \end{tabular}
    }
\end{subtable}
\hfill
\begin{subtable}[t]{\textwidth}
    \caption{\textit{Synapse}} \label{tab:er-synapse}
    \setlength{\tabcolsep}{8.7pt} %
    \renewcommand*{\arraystretch}{1.11}  %
    \vspace{-0.2\baselineskip}
    \scriptsize{
	\begin{tabular}{l|cc|*{8}c}
        \hline \textbf{Methods} & \textbf{DSC~$\uparrow$} & \textbf{HD~$\downarrow$} & \textbf{Aorta} & \textbf{Gallbladder} & \textbf{Kidney(L)} & \textbf{Kidney(R)} & \textbf{Liver} & \textbf{Pancreas} & \textbf{Spleen} & \textbf{Stomach} \\
        \hline 
        U-Net \cite{ronneberger2015u} & 76.85 & 39.70 & 89.07 & \textbf{69.72} & 77.77 & 68.60 & 93.43 & 53.98 & 86.67 & 75.58  \\
        Att-UNet \cite{oktay2018attention} & 77.77 & 36.02 & \textbf{89.55} & 68.88 & 77.98 & 71.11 & 93.57 & 58.04 & 87.30 & 75.75 \\
        U-Net++ \cite{zhou2018unet++} & 76.91 & 36.93 & 88.19 & 65.89 & 81.76 & 74.27 & 93.01 & 58.20 & 83.44 & 70.52 \\
        MultiResUNet \cite{ibtehaz2020multiresunet} & 77.42 & 36.84 & 87.73 & 65.67 & 82.08 & 70.43 & 93.49 & 60.09 & 85.23 & 74.66 \\
        Residual U-Net \cite{zhang2018road} & 76.95 & 38.44 & 87.06 & 66.05 & 83.43 & 76.83 & 93.99 & 51.86 & 85.25 & 70.13 \\
        TransUNet \cite{chen2021transunet} & 77.48 & 31.69 & 87.23 & 63.13 & 81.87 & 77.02 & 94.08 & 55.86 & 85.08 & 75.62 \\
        UCTransNet \cite{wang2022uctransnet} & 78.23 & 26.75 & 84.25 & 64.65 & 82.35 & 77.65 & 94.36 & 58.18 & 84.74 & 79.66\\
        MISSFormer \cite{huang2021missformer} & \textbf{81.96} & \textbf{18.20} & 86.99 & 68.65 & \textbf{85.21} & \textbf{82.00} & \textbf{94.41} & \textbf{65.67} & \textbf{91.92} & \textbf{80.81}\\
        \hline
    \end{tabular}
    }
\end{subtable}
\end{table*}

\subsubsection{Cell (SegPC 2021)}
\begin{figure*}[ht]
\captionsetup[subfigure]{labelformat=empty}
\begin{subfigure}{.099\textwidth}
	\centering
	\includegraphics[width=\linewidth]{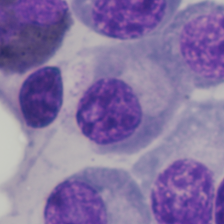}
	\caption{}
	\label{fig:er-segpc-img-1}
\end{subfigure}\hfill
\begin{subfigure}{.099\textwidth}
	\centering
	\includegraphics[width=\linewidth]{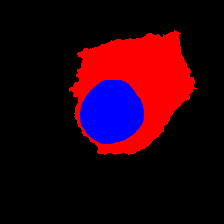}
	\caption{}
	\label{fig:er-segpc-gt-1}
\end{subfigure}\hfill
\begin{subfigure}{.099\textwidth}
	\centering
	\includegraphics[width=\linewidth]{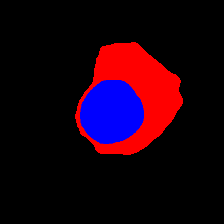}
	\caption{}
	\label{fig:er-segpc-unet-1}
\end{subfigure}\hfill
\begin{subfigure}{.099\textwidth}
	\centering
	\includegraphics[width=\linewidth]{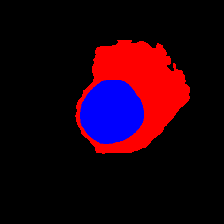}
	\caption{}
	\label{fig:er-segpc-unetpp-1}
\end{subfigure}\hfill
\begin{subfigure}{.099\textwidth}
	\centering
	\includegraphics[width=\linewidth]{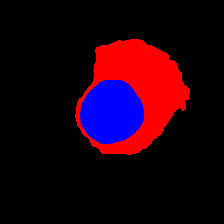}
	\caption{}
	\label{fig:er-segpc-attunet-1}
\end{subfigure}\hfill
\begin{subfigure}{.099\textwidth}
	\centering
	\includegraphics[width=\linewidth]{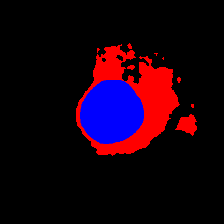}
	\caption{}
	\label{fig:er-segpc-resunet-1}
\end{subfigure}\hfill
\begin{subfigure}{.099\textwidth}
	\centering
	\includegraphics[width=\linewidth]{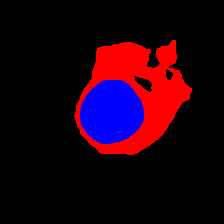}
	\caption{}
	\label{fig:er-segpc-multiresunet-1}
\end{subfigure}\hfill
\begin{subfigure}{.099\textwidth}
	\centering
	\includegraphics[width=\linewidth]{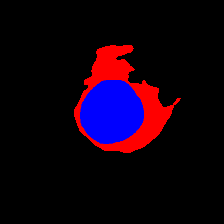}
	\caption{}
	\label{fig:er-segpc-transunet-1}
\end{subfigure}\hfill
\begin{subfigure}{.099\textwidth}
	\centering
	\includegraphics[width=\linewidth]{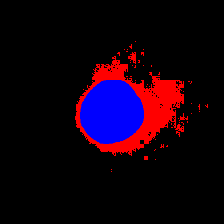}
	\caption{}
	\label{fig:er-segpc-missformer-1}
\end{subfigure}\hfill
\begin{subfigure}{.099\textwidth}
	\centering
	\includegraphics[width=\linewidth]{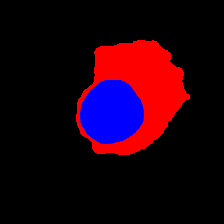}
	\caption{}
	\label{fig:er-segpc-uctransnet-1}
\end{subfigure}\hfill
\\[-20.5pt]
\begin{subfigure}{.099\textwidth}
	\centering
	\includegraphics[width=\linewidth]{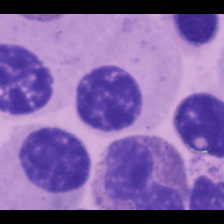}
	\caption{}
	\label{fig:er-segpc-img-2}
\end{subfigure}\hfill
\begin{subfigure}{.099\textwidth}
	\centering
	\includegraphics[width=\linewidth]{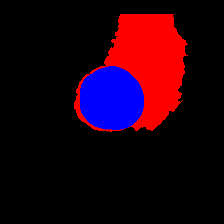}
	\caption{}
	\label{fig:er-segpc-gt-2}
\end{subfigure}\hfill
\begin{subfigure}{.099\textwidth}
	\centering
	\includegraphics[width=\linewidth]{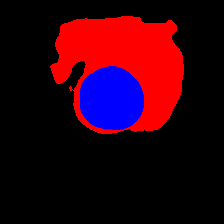}
	\caption{}
	\label{fig:er-segpc-unet-2}
\end{subfigure}\hfill
\begin{subfigure}{.099\textwidth}
	\centering
	\includegraphics[width=\linewidth]{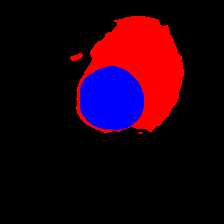}
	\caption{}
	\label{fig:er-segpc-unetpp-2}
\end{subfigure}\hfill
\begin{subfigure}{.099\textwidth}
	\centering
	\includegraphics[width=\linewidth]{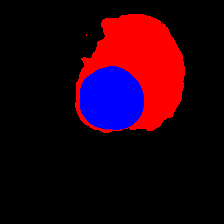}
	\caption{}
	\label{fig:er-segpc-attunet-2}
\end{subfigure}\hfill
\begin{subfigure}{.099\textwidth}
	\centering
	\includegraphics[width=\linewidth]{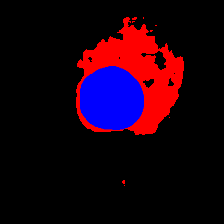}
	\caption{}
	\label{fig:er-segpc-resunet-2}
\end{subfigure}\hfill
\begin{subfigure}{.099\textwidth}
	\centering
	\includegraphics[width=\linewidth]{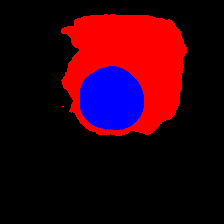}
	\caption{}
	\label{fig:er-segpc-multiresunet-2}
\end{subfigure}\hfill
\begin{subfigure}{.099\textwidth}
	\centering
	\includegraphics[width=\linewidth]{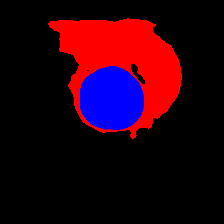}
	\caption{}
	\label{fig:er-segpc-transunet-2}
\end{subfigure}\hfill
\begin{subfigure}{.099\textwidth}
	\centering
	\includegraphics[width=\linewidth]{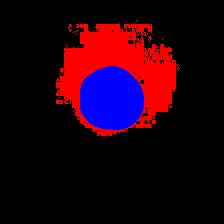}
	\caption{}
	\label{fig:er-segpc-missformer-2}
\end{subfigure}\hfill
\begin{subfigure}{.099\textwidth}
	\centering
	\includegraphics[width=\linewidth]{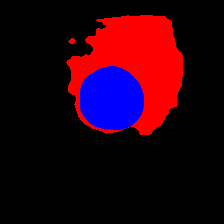}
	\caption{}
	\label{fig:er-segpc-uctransnet-2}
\end{subfigure}\hfill
\\[-20.5pt]
\begin{subfigure}{.099\textwidth}
	\centering
	\includegraphics[width=\linewidth]{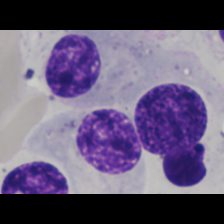}
	\caption{\scriptsize{Input Image}}
	\label{fig:er-segpc-img-3}
\end{subfigure}\hfill
\begin{subfigure}{.099\textwidth}
	\centering
	\includegraphics[width=\linewidth]{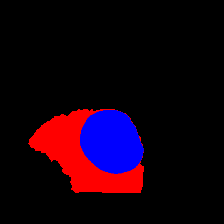}
	\caption{\scriptsize{Target (GT)}}
	\label{fig:er-segpc-gt-3}
\end{subfigure}\hfill
\begin{subfigure}{.099\textwidth}
	\centering
	\includegraphics[width=\linewidth]{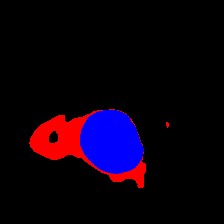}
	\caption{\scriptsize{UNet}}
	\label{fig:er-segpc-unet-3}
\end{subfigure}\hfill
\begin{subfigure}{.099\textwidth}
	\centering
	\includegraphics[width=\linewidth]{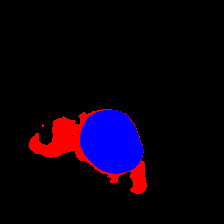}
	\caption{\scriptsize{UNet++}}
	\label{fig:er-segpc-unetpp-3}
\end{subfigure}\hfill
\begin{subfigure}{.099\textwidth}
	\centering
	\includegraphics[width=\linewidth]{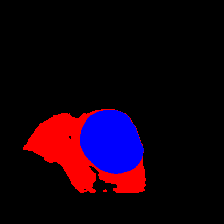}
	\caption{\scriptsize{AttUNet}}
	\label{fig:er-segpc-attunet-3}
\end{subfigure}\hfill
\begin{subfigure}{.099\textwidth}
	\centering
	\includegraphics[width=\linewidth]{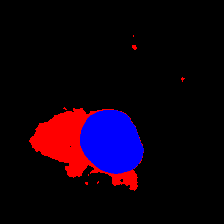}
	\caption{\scriptsize{ResUNet}}
	\label{fig:er-segpc-resunet-3}
\end{subfigure}\hfill
\begin{subfigure}{.099\textwidth}
	\centering
	\includegraphics[width=\linewidth]{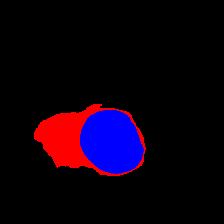}
	\caption{\scriptsize{MultiResUNet}}
	\label{fig:er-segpc-multiresunet-3}
\end{subfigure}\hfill
\begin{subfigure}{.099\textwidth}
	\centering
	\includegraphics[width=\linewidth]{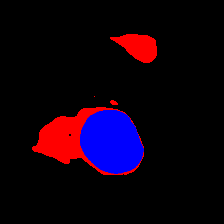}
	\caption{\scriptsize{TransUNet}}
	\label{fig:er-segpc-transunet-3}
\end{subfigure}\hfill
\begin{subfigure}{.099\textwidth}
	\centering
	\includegraphics[width=\linewidth]{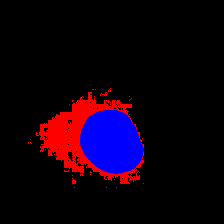}
	\caption{\scriptsize{MISSFormer}}
	\label{fig:er-segpc-missformer-3}
\end{subfigure}\hfill
\begin{subfigure}{.099\textwidth}
	\centering
	\includegraphics[width=\linewidth]{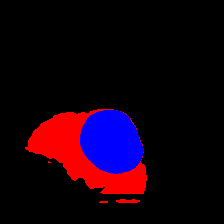}
	\caption{\scriptsize{UCTransNet}}
	\label{fig:er-segpc-uctransnet-3}
\end{subfigure}\hfill
\caption{Visual comparisons of different methods on the \textit{SegPC 2021} cell segmentation dataset. \textcolor{red}{Red} region indicates the Cytoplasm and \textcolor{blue}{blue} denotes the Nucleus area of cell.}
\label{fig:er-segpc2021}
\end{figure*}

Visual segmentation results presented in \Cref{fig:er-segpc2021} for SegPC dataset. Due to the overlapping nature of multiple myeloma and the spiky ground truth labels, segmenting is laborious. To this end, besides the Dice score comparison in \Cref{tab:er-segpc2021}, the mIoU metric plays a critical role in voting for the networks' performance. It is worth mentioning that the Transformer-based methods provide more softer contour of segmentation maps than the CNN-based studies, which in our opinion, reflects that global long-range contextual information helped the network to perceive the actual shape of cells. On the contrary, due to the locality of convolution counterparts, the CNN-based modules are more error-prone in boundaries. The multi-scale representation power of the UCTransNet \cite{wang2022uctransnet} once again shows the effectiveness of this approach in generating precise segmentation maps for cells with varying scales and backgrounds.

\subsubsection{Multi organ (Synapse)}
\begin{figure*}
\centering
\includegraphics[width=\textwidth]{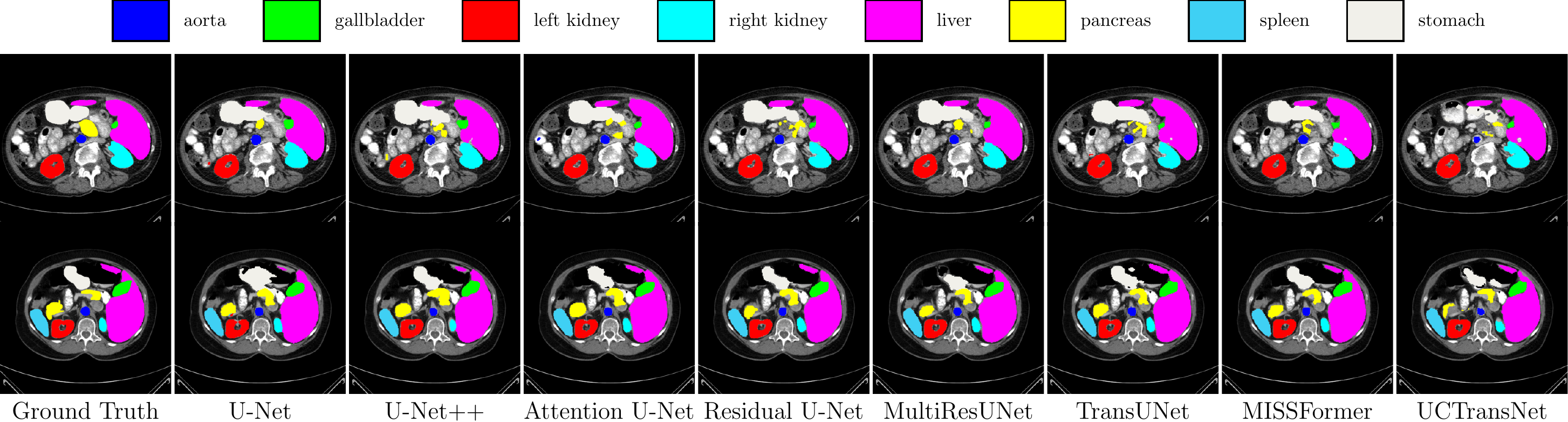}
\caption{Visual comparisons of different methods on the \textit{Synapse} multi-organ segmentation dataset.}
\label{fig:er-synapse}
\end{figure*}

In \Cref{tab:er-synapse}, we presented the quantitative results based on the Dice score and Hausdorff distance metrics. In addition, \Cref{fig:er-synapse} we depicted the visual qualitative results over the synapse multi-organ dataset. Due to the increasing class numbers and the multi-scale innate of the synapse dataset, it is a challenging task. Overall, as a baseline method, U-Net \cite{ronneberger2015u} performs well in comparison with parameters number and a basic design but suffers from several visual defects, e.g., miss-labeling, over-segmentation, and under-segmentation. U-Net++ \cite{zhou2019unet++}, MultiResUNet \cite{ibtehaz2020multiresunet}, and Residual U-Net \cite{zhang2018road} principally follow the same intuition within creating rich semantic representations with dense connections and bigger convolution kernels for capturing contextual information to increase performance, but still, they could not hit the target and depicted a minimal performance boost in Dice score with respect to U-Net. Attention U-Net \cite{oktay2018attUnet} utilized an attention mechanism within skip connections to recalibrate the feature reusability and demonstrates almost 1\% improvement in Dice score compared with U-Net and presents smoother segmentation boundaries over the mentioned methods. Finally, Transformer-based methods represent better results in mean Dice score with respect to CNN-based studies. It is predominately in bond with ViT's advantage in capturing global dependencies that are very important in multi-scale segmentation tasks. MISSFormer \cite{huang2021missformer}, due to its hierarchical and multi-scale design, performs as a SOTA strategy among our selected models. This result suggests that in multi-scale segmentation tasks, the determinative approach should follow the hierarchical design besides leveraging contextual information for superior performances.

\begin{table}
\centering
\caption{Comparison of the number of parameters in Million (M) scale and 
	Giga Floating-point Operations Per Second (GFLOPS).} \label{tab:parameter-flops}
	\begin{tabular}{l| *2{c}}
		\hline
		\textbf{Methods} & \textbf{\# Parameters (M)} & \textbf{GFLOPS} \\
		\hline
		MISSFormer \cite{huang2021missformer} & 42.46 & 7.25 \\
		U-Net \cite{ronneberger2015u} & 1.95 & 8.2 \\
		MultiResUNet \cite{ibtehaz2020multiresunet} & 7.82 & 15.74 \\
		U-Net++ \cite{zhou2018unet++} & 9.16 & 26.72 \\
		UCTransNet \cite{wang2022uctransnet} & 66.43 & 32.99 \\
		Att-UNet \cite{oktay2018attention} & 34.88 & 51.02 \\
		Residual U-Net \cite{zhang2018road} & 13.04 & 62.06 \\
		TransUNet \cite{chen2021transunet} &  105.28 & 80.68 \\		
		\hline
	\end{tabular}
\end{table}

\subsection{Discussion}\label{sec:discussion}
In this section, we analyze the models' performances from \Cref{sec:experimental-results} in more detail. Observing the experimental results, we find that the performance gain achieved by all the extensions of U-Net models compared to the original U-Net. This fact reveals the effectiveness of this contribution to the U-Net model. Although the U-Net acts as a baseline method, in a binary segmentation task such as skin lesion segmentation, it demonstrates a good performance compared to its extensions. However, its performance in multi-organ segmentation tasks, specifically the overlapped objects, seems less promising, evidenced by the over-segmentation or under-segmentation results presented in the qualitative Figures (see \Cref{fig:er-isic2018,fig:er-segpc2021,fig:er-synapse}).

On the contrary, attention-based strategies like Attention U-Net \cite{oktay2018attention} performs well in the case of overlapped objects. Furthermore, the hierarchy of the skip connections included in the U-Net++ \cite{zhou2019unet++} model seems successful in boosting the performance, which might explain the contribution of the nested structure. In addition, a hierarchical attention mechanism in Attention U-Net \cite{oktay2018attention} helps the naive U-Net compensate for the locality issue for more expansive capturing view in feature extraction to perform accurate segmentation over the multi-scale objects such as multi-organ segmentation. Even though each extension of the U-Net model in a CNN-based strategy aims to enhance the feature representation, the locality nature of the convolution operation usually limits the representation power for modeling global anatomical and structural representation. Residual U-Net \cite{zhang2018road}, even with the Residual enhancements, is no exception to this rule. On the other hand, Transformer models, e.g., MISSFormer \cite{huang2021missformer} (as a hybrid model), and UCTransNet \cite{wang2022uctransnet} seem highly capable of modeling the global representation as can be witnessed from the quantitative results on Synapse dataset. 

\begin{figure}
\centering
\includegraphics[width=\columnwidth]{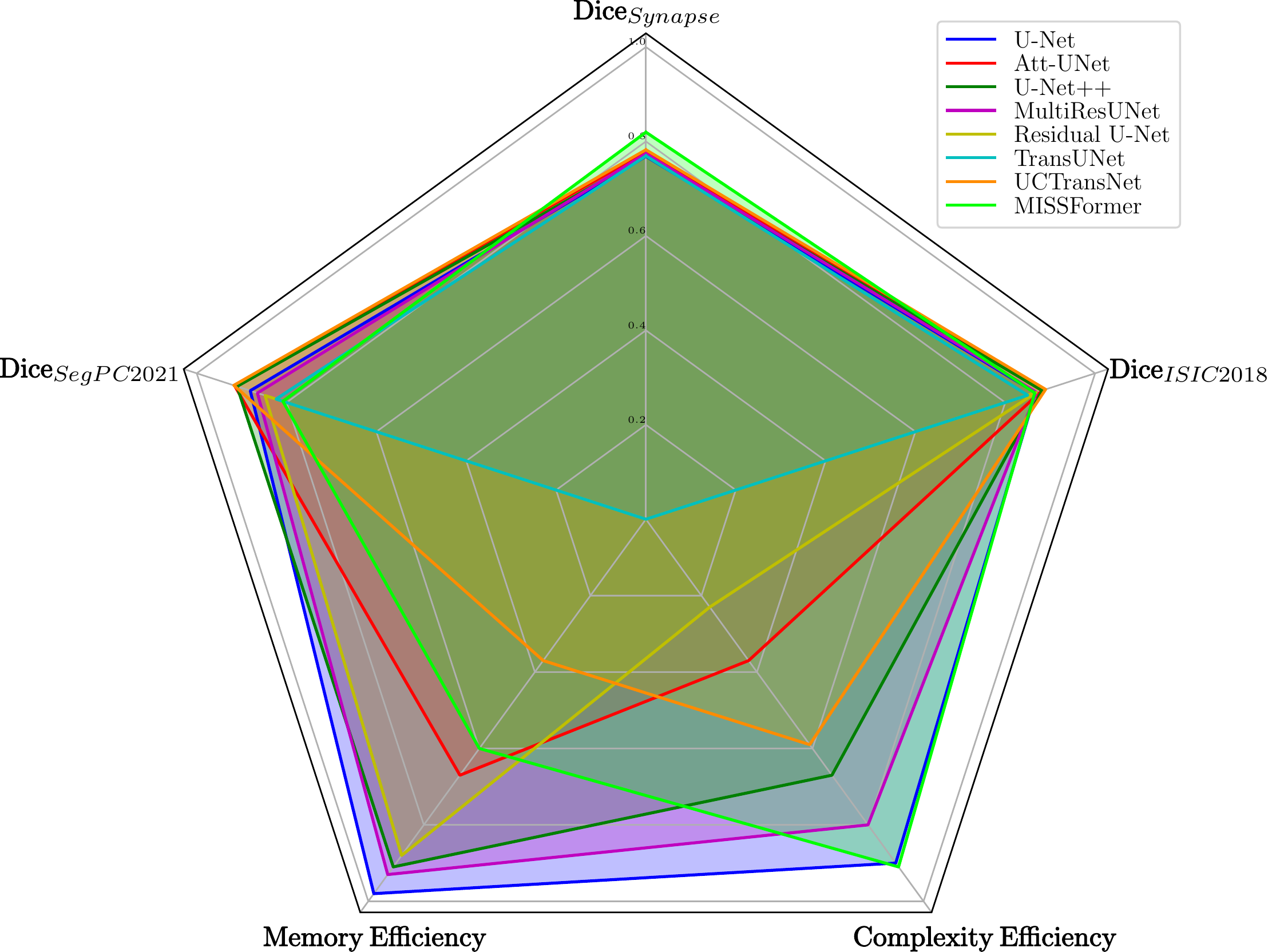}
\caption{Comparison of experimental models with respect to their Dice score over studied datasets, memory, and complexity factors.}
\label{fig:comparison-methods}
\end{figure}

The computational complexity, more specifically, the Giga Floating Point Operations (GFLOPS) and the number of trainable parameters, is another critical aspect that needs to be considered for the model assessment. In this respect, we provided \Cref{tab:parameter-flops} to show the number of GFLOPS and trainable parameters for each U-Net extension. Observing \Cref{tab:parameter-flops},  we realize that the modifications in naive U-Net with the integration of CNN-based plug-and-play modules increase the parameter numbers and GFLOPS remarkably. In addition, the extra FLOPS rate further increases with the Transformer-based models due to the quadratic computational complexity. On the contrary, MISSFormer, as a standalone Transfomer-based U-shaped network, benefits from the efficient Transformer design, less affected by the parameter numbers in comparison with TransUNet \cite{chen2021transunet}. Overall, the underlying trade-off between performance and efficiency needs to be considered for the practical use of these methods. To this end, firstly, we define the normalized computational efficiency of a particular network as follows:

\begin{align}
\text{Compexity}_{\text{net}}&=1-\left(\frac{\text{GFLOPS}(\text{net})}{max(\text{GFLOPS(net)}}\right), \nonumber \\
& s.t. \; \text{net} \in \text{Nets}
\end{align}

\noindent where the max function calculates the maximum number of GFLOPS from all models to find the upper bound, similarly, we define the normalized memory efficiency as:

\begin{align}
\text{Memory}_{\text{net}}&=1-\left(\frac{\#\text{Parameters}(\text{net})}{max(\# \text{Parameters}(\text{net}))}\right). \nonumber \\
& s.t. \; \text{net} \in \text{Nets}
\end{align}

\noindent Moreover, finally, we show the model's performance on a particular dataset using the dice score in \Cref{fig:comparison-methods}. The computational efficiency along with the performance gain demonstrated in \Cref{fig:comparison-methods} can be a good source for analyzing the trade-off between performance and efficiency and finally choosing an appropriate clinically-applicable model.

\section{Challenges and Opportunities } \label{sec:challenges}
Over the past few years, deep-learning-based methods, especially the U-Net family, have been utilized considerably in medical image fields to address clinical demands. In the following, future perspectives associated with medical image segmentation using the U-Net family will be introduced in favor of improvement in this field.

\subsection{Memory Efficient Models}
The primary purpose of almost all the approaches mentioned in this paper was to identify the limitation of the original U-Net model and design add-on modules to enhance feature reusability and enrich feature representation to bring more performance boost. However, including more parameters in the model usually results in large memory requirements, which makes the model unsuitable for clinical applications with limited computational devices \cite{fitzke2021oncopetnet}. To overcome this issue, one might consider the efficient design of the network or use a more efficient way of reducing unnecessary parameters. Deep model compression techniques such as pruning \cite{han2015deep}, quantization \cite{han2015deep}, low-rank approximation \cite{yu2017compressing}, knowledge distillation \cite{gou2021knowledge}, and neural architecture search \cite{wistuba2019survey} to new a few, is an area for advancement that researcher might consider in their approach.

\subsection{Balance Between Accuracy and Efficiency}
It is always appealing for the deep model to be accurate and efficient. However, model efficiency usually requires meeting some restrictions (e.g., low computational constraint). Such a scenario potentially leads to a poor network design and consequently results in less accurate predictions. Hence, there is always a trade-off between the model's accuracy and efficiency. From a clinical perspective, it is highly desirable to take into account the balance between efficiency and accuracy. Therefore, one might consider the trade-off between the accuracy and efficiency of U-Net extension.  

\subsection{Model scaling and complexity}
As mentioned in the previous sections, evident is the fact that some parameters such as running time, computational complexity, and memory are important for clinics with limited computing resources. In addition to the memory shortage, inference time and consequently computational complexity also have key roles in real-time applications that need to be considered. To do so, well-known metrics evaluate the model's complexity such as Model parameters, Floating-Point Operations (FLOPS), Runtime, and Frame Per Second (FPS) should be considered in the model assessment. The two given metrics, model parameters, and FLOPS are independent of the implementation environment. Hence, a larger value ends up with lower implementation efficiency. Metrics such as run-time and FPS may seem to be less desirable compared to model parameters and FLOPs, due to their reliance on hardware and implementation environment. However, these metrics are also imperative for a real-world application and need to be taken into account as well.

\subsection{Interpretable Models}
It is always a question for clinical, how deep learning models learns specific patterns to recognize certain cancer. From a clinical point of view, understanding the object recognition process by deep models is significantly appropriate. As a result, the radiologists can interpret the deep models (e.g., feature visualization), to find out how the diagnosis process happens inside the deep model and how they can include prior knowledge (e.g.,  pathology assumption) to further improve model performance.  Hence, model interpretation not only assists the radiologist to understand the hidden markers in medical data but also provides a way to further scrutinize the network architecture for possible improvement with modeling clinical assumptions. In this respect, future direction for U-Net extension might consider the importance of interpretable characteristics in their network design. 

\subsection{Federated Learning}
In a medical domain data, privacy provides a set of regulations to confirm data confidentiality and security. In most clinical applications, whole data acquisition usually happens in one center, which limits the data diversity and results in less generalizable deep neural networks. In contrast, a multi-central dataset with authorized data privacy can provide a more realistic dataset for clinical purposes \cite{rieke2020future}. The future network design might consider a federated learning strategy in their optimization process to provide a more generalizable model for clinical use.

\subsection{Software Ecosystems}
With the rapid development of Artificial Intelligence (AI) systems, these methods are integrated and adapted in almost all domains to facilitate the data processing strategy and provide assistive tools to meet consumer demands. A software ecosystem coordinates this process by defining a set of actors (e.g., developers, users, etc.) in conjunction with the appropriate interaction strategy to secure consumers' demands. The software ecosystem can be considered as a three-step pipeline (see \Cref{fig:unet-pipeline}). The first step (\Cref{fig:unet-pipeline}) is data acquisition and preparation. This step aims to recognize the hardware requirements (e.g., MRI scan, metadata processing) for the task-specific goal and prepare the data in a standard format for the next step. The deep learning model design and optimization process happens in the second step to learn the task-specific objective. Finally, the last step provides comprehensive tools to analyze and evaluate the extracted results for the clinical applications. Several software prototypes already exist in the literature, nnUNet \cite{isensee2021nnu}, Ivadomed \cite{gros2020ivadomed}, and etc., to provide open access code for both clinical/research communities. Similarly, our open-source software at GitHub\footnote{\url{https://github.com/NITR098/Awesome-U-Net}} provides an implementation of the U-Net family for several public datasets. The future research direction might consider contributing to an open-source library to further support software development. 


\subsection{Data Driven Decisions}
In a supervised learning task, the neural network uses the ground truth mask to learn a discriminative space, which enables the network to distinguish the target class from the background regions (e.g., segmenting tumor regions in bran images). All the extensions of the U-Net models presented in this review were based on the supervised learning paradigm. Such a training strategy imposes a mask bias on the training process and prevents the network from learning data-driven features. This might explain why supervised trained networks have less capacity to generalize for unseen objects. To address this issue, one might consider including unsupervised data in their training process for a richer data representation, like \cite{feyjie2020semi}.  As it is clear from the name, the unsupervised technique does not require any annotation and gives the model more freedom to infer from the data itself without imposing any prior knowledge (e.g., mask information). Although training and reasoning from the image itself is a complex task, modeling such behavior might open up new potential for the U-Net family to learn more generic and data-representative features. The future direction of the U-Net family might consider this point in their potential design for a further performance boost.

\section{Conclusion}
In this study, we presented a thorough review of the literature on U-Net and its variants, the emergence of which has continued to rise in the medical image segmentation field over the years.
We examined the area from its main taxonomy, the extensions to the variants, and a benchmark of performance and speed.
To structure the wide variety of U-Net extensions, the different extensions and variants are grouped depending on which type of change was made to the architecture.
Adaptions to the skip connections are presented in \Cref{sec:skip-connection} and comprise methods that increase the number of skip connections, apply additional processing to the feature maps in the skip connections to focus on the areas of interest and improve the fusion of the encoder and decoder feature maps combined through the skip connections.
\Cref{sec:backbone-design} introduces different types of backbones used in the U-Net architecture e.g. deeper network architectures, processing of 3D images or multi-resolution feature extraction for high inter-patient variability in terms of size of the object(s) of interest.
A variety of extensions of the bottleneck of the original U-Net is examined in \Cref{sec:bottleneck}.
The different approaches adapt the bottleneck for multi-scale representation of the bottleneck feature maps or position-wise attention to model spatial dependencies between pixels in the bottleneck.
The transformer variants of the U-Net architecture, introduced in \Cref{sec:transformers}, enable the networks to capture inter-pixel long-range dependencies and to compensate for the otherwise limited receptive field of the convolutions in the original U-Net.
Approaches presented in \Cref{sec:rich-representation} adapt the U-Net architecture to use information from multiple modalities and/or multiple scales for a rich representation of features. Finally, \Cref{sec:probablistic} presented to model the uncertainty in annotations of medical data or out-of-distribution samples.

Furthermore, a detailed evaluation of several of the U-Net variants on different medical datasets was conducted in \Cref{sec:experimental-results}. In our accuracy and complexity experiments, the transformer variants, the UCTransNet \cite{wang2021uctransnet} and MISSFormer \cite{huang2021missformer} achieved superior performance (2\% increases in terms of dice scores) compared to the CNN extension on all datasets. Eventually, our experimental results along with the computational and memory complexity provided a picture for the reader to consider a trade-off between the performance and efficiency (e.g., MISSFormer \cite{huang2021missformer} with 42M parameters and 0.81 DSC score on the Synapse dataset against U-Net with 1.9M parameters but 0.79 DSC score) for choosing the desired network for the problem at hand. We hope our study can assist researchers in further extending the U-Net model for both clinical and industry applications.

\bibliographystyle{IEEEtran}
\bibliography{ref}
\end{document}